\documentclass[a4paper,11pt]{article}
\usepackage{jcappub}
\usepackage{graphicx,epsfig,color,times,wasysym}
\usepackage[numbers]{natbib}
\usepackage{multirow}

\def \<{\langle}
\def \>{\rangle}

\def\prd{Phys. Rev. D}
\def\prl{Phys. Rev. Lett.}
\def\apj{Astrophys. J.}
\def\apjl{Astrophys. J. Lett.}
\def\apjs{Astrophys. J.Suppl.}
\def\mnras{Mon. Not. R. Astr. Soc.}
\def\aap{Astr. Astrophys.}

\def\aj{Astr. J.}
\def\nat{Nature}
\def\jcap{JCAP}

\def\aapr{\ref@jnl{A\&A~Rev.}}          

\def\lsim{\raise0.3ex\hbox{$<$\kern-0.75em\raise-1.1ex\hbox{$\sim$}}}
\def\gsim{\raise0.3ex\hbox{$>$\kern-0.75em\raise-1.1ex\hbox{$\sim$}}}

\def\beq{\begin{equation}} \def\eeq{\end{equation}} \def\bea{\begin{eqnarray}}
\def\eea{\end{eqnarray}}  
\def\be{\begin{equation}} \def\ee{\end{equation}} \def\ba{\begin{eqnarray}}
\def\ea{\end{eqnarray}} 
\def\dalemb#1#2{{\vbox{\hrule height.#2pt \hbox{\vrule width.#2pt height#1pt
\kern#1pt \vrule width.#2pt} \hrule height.#2pt}}}

\def\ba{\begin{eqnarray}} \def\ea{\end{eqnarray}} \def\be{\begin{equation}}
\def\ee{\end{equation}} 
\def\gtorder{\mathrel{\raise.3ex\hbox{$>$}\mkern-14mu
\lower0.6ex\hbox{$\sim$}}}
\def\ltorder{\mathrel{\raise.3ex\hbox{$<$}\mkern-14mu
\lower0.6ex\hbox{$\sim$}}}

\newcommand{\muK}{\mu  {\rm K}}

\usepackage{amssymb,amsbsy,amsmath,amsfonts,amssymb,amscd}
\usepackage{setspace}

\newcommand{\Planck}{\textit{\negthinspace Planck\/}} \newcommand{\planck}{\textit{\negthinspace Planck\/}}
\newcommand{\Spitzer}{\negthinspace \textit{Spitzer\/}}
\newcommand{\Herschel}{\textit{\negthinspace Herschel\/}}
\newcommand{\mission}{\textit{\negthinspace PRISM\/}}
\newcommand{\Euclid}{\negthinspace \textit{Euclid\/}}

\newcommand{\SPICA}{\negthinspace \textit{SPICA\/}}

\def\m{\ifmmode $m$\else \,m\fi}

\def\st{\ifmmode ^{\mathrm{st}} \else $^{\mathrm{st}}$\fi}
\def\nd{\ifmmode ^{\mathrm{nd}} \else $^{\mathrm{nd}}$\fi}
\def\rd{\ifmmode ^{\mathrm{rd}} \else $^{\mathrm{rd}}$\fi}
\def\th{\ifmmode ^{\mathrm{th}} \else $^{\mathrm{th}}$\fi}

\newcommand\ltsima{$\; \buildrel < \over \sim \;$}
\newcommand\simlt{\lower.5ex\hbox{\ltsima}}
\newcommand\gtsima{$\; \buildrel > \over \sim \;$}
\newcommand\simgt{\lower.5ex\hbox{\gtsima}}
\newcommand\simprop{\lower.5ex\hbox{$\; \buildrel \propto \over \sim \;$}}
\newcommand\mypar[1]{\smallskip\par\noindent\textbf{#1}}

\newcommand{\Msolar}{M_\odot}

\begin{document}
\title{PRISM (Polarized Radiation Imaging and Spectroscopy Mission): An Extended White Paper}
\author{\small
\mission\ collaboration:\footnote{www.prism-mission.org~ A full list of names appears on the next page.
This is an extended version of the White paper submitted to ESA in June 2013 (astro-ph arXiv:1306.2259)
as part of a competition to select the science themes for the next two ESA L-class (large) missions
as part of the ESA Cosmic Vision Programme. The approximate launch dates
announced for these two slots, known as L3 and L4, are 2028 and 2034,
respectively.}:}
\author[1]{Philippe Andr\'e}
\author[2]{Carlo Baccigalupi} 
\author[3,4]{Anthony Banday}
\author[5]{Domingos Barbosa}
\author[6]{Belen Barreiro}
\author[7,8]{James Bartlett}
\author[9,10]{Nicola Bartolo}
\author[11]{Elia Battistelli}
\author[12]{Richard Battye}
% ...


\abstract{\small
\noindent
\mission\ (Polarized Radiation Imaging and Spectroscopy Mission) was proposed to ESA in May 2013
as a large-class mission for investigating within the framework of the ESA Cosmic Vision
program a set of important scientific questions that require high resolution, high sensitivity, full-sky
observations of the sky emission
at wavelengths ranging from millimeter-wave to the far-infrared.
\mission's main objective is to explore the distant universe, probing cosmic
history from very early times until now as well as the structures, distribution of
matter, and velocity flows throughout our Hubble volume. \mission\  will
survey the full sky in a large number of frequency bands in both
intensity and polarization and will measure the absolute spectrum of sky emission more than
three orders of magnitude better than {\it COBE} FIRAS. The data obtained will allow us
to precisely measure the absolute sky brightness and polarization of all the components of the
sky emission in the observed frequency range,
separating the primordial and extragalactic components cleanly from the galactic and zodiacal light emissions.
The aim of this Extended White Paper is to provide a more detailed overview of the highlights
of the new science that will be made possible by \mission, which include: (1) the ultimate
galaxy cluster survey using the Sunyaev-Zeldovich (SZ) effect, detecting approximately $10^6$
clusters extending to large redshift, including a characterization of the gas temperature of
the brightest ones (through the relativistic corrections to the classic SZ template) as well
as a peculiar velocity survey using the kinetic SZ effect that comprises our entire Hubble
volume; (2) a detailed characterization of the properties and evolution of dusty galaxies, where
the most of the star formation in the universe took place, the faintest population of which
constitute the diffuse CIB (Cosmic Infrared Background); (3) a characterization of the B modes from
primordial gravity waves generated during inflation and from gravitational lensing, as well
as the ultimate search for primordial non-Gaussianity using CMB polarization, which is less
contaminated by foregrounds on small scales than the temperature anisotropies; (4) a search
for distortions from a perfect blackbody spectrum, which include some nearly certain signals
and others that are more speculative but more informative; and (5) a study of the role of the
magnetic field in star formation and its interaction with other components of the
interstellar medium of our Galaxy. These are but a few of the highlights presented here
along with a description of the proposed instrument.}

\maketitle
\flushbottom

{\small

\vspace{6mm}
\noindent
{\Large \bf {\textit{\textbf{PRISM}}} Science Case and White Paper Coordination}

\vspace{3mm}
\noindent
{\textit {\textbf{The preparation of the science case submitted to ESA has been coordinated by:}}}
\vspace{1mm}

\noindent
James Bartlett, Fran\c cois Bouchet, Fran\c cois  Boulanger, Martin Bucher, Anthony Challinor, Jens Chluba, 
Paolo de Bernardis (spokesperson), Gianfranco de Zotti, Jacques Delabrouille (coordinator), Pedro Ferreira, Bruno Maffei

\vspace{3mm}

\noindent
{\textit {\textbf{The writing of the extended version of the PRISM white paper has been coordinated by:}}}
\vspace{1mm}

\noindent
Martin Bucher, Jacques Delabrouille

\vspace{6mm}
\noindent
{\Large \bf {\textit{\textbf{PRISM}}} Steering Committee}
\vspace{3mm}

\noindent
{\bf France:} Fran\c cois Bouchet, Martin Bucher, Jacques Delabrouille, Martin Giard

\noindent
{\bf Germany:} Jens Chluba, Rashid Sunyaev

\noindent
{\bf Ireland:} Anthony Murphy

\noindent
{\bf Italy:} Marco Bersanelli, Carlo Burigana, Paolo de Bernardis

\noindent
{\bf Netherlands:} Rien van de Weijgaert

\noindent
{\bf Portugal:} Carlos Martins

\noindent
{\bf Spain:} Enrique Mart\'inez-Gonz\'alez, Jos\'e Alberto Rubi\~no-Mart\'in, Licia Verde

\noindent
{\bf Switzerland:} Martin Kunz

\noindent
{\bf United Kingdom:} Anthony Challinor, Joanna Dunkley, Bruno Maffei

%
%
%

}
\vfill\eject
\setcounter{tocdepth}{2}
{\small

\begin{spacing}{0.8}
\tableofcontents
\end{spacing}
}

\newpage
\setcounter{page}{1}
\pagenumbering{arabic}

\section{Executive summary}
The Polarized Radiation Imaging and Spectroscopy Mission (\mission) is a proposed large-class mission 
designed to carry out the ultimate survey of the microwave to far-infrared sky in both intensity and 
polarization as well as to measure its absolute emission spectrum. \mission\ will 
consist of two instruments: 

\begin{enumerate} 

\item A high angular resolution polarimetric imager with a 
3.5$\,$m usable diameter telescope, cooled to below 10K in order to maximally reduce the 
photon noise due to the thermal emission of the mirrors. This instrument will map the intensity and 
polarization of the complete sky in 32 broad frequency bands between 30~GHz (1~cm) and 6~THz (50 
microns) with unprecedented sensitivity and with an angular resolution ranging from about 17 arcminutes to 
about 6 arcseconds. 

\item A lower angular resolution ($1.4^\circ$) spectrometer that will compare the 
sky frequency spectrum to a nearly perfect reference blackbody and measure the \emph{absolute} sky 
emission over the same frequency range. 

\end{enumerate} 

The data from these two instruments will enable \mission\ to carry out breakthrough science
by answering key questions in many diverse areas of astrophysics and fundamental science 
as well as providing the astronomical community with high quality, full sky
observations of emission over a large frequency range. 

Highlights of the new science from \mission\ include:

\subsection{Ultimate galaxy cluster survey and census of structures in our Hubble volume} 

Because of the tight correlation between integrated $y$-distortion and cluster mass as well as the redshift independence
of the surface brightness,
the Sunyaev-Zeldovich (SZ) effect offers the best method for assembling a catalog of clusters at high 
redshift to be used for cosmology.  When \mission\ is launched (foreseen for 2028 or later), 
all-sky cluster samples (e.g., from {\it eROSITA,} \Euclid) will likely comprise some $10^5$ objects, mostly 
at $z\!<\!1.$ \mission\ will find 10 times more clusters, which importantly
will extend to deeper redshifts, with many 
thousands beyond $z=2$. In fact, \mission\ will detect \emph{all} clusters of mass larger than $5 \times 10^{13}\Msolar$ 
in the universe. Owing to its 
broad spectral coverage, high angular resolution, and exquisite sensitivity, 
using the kinetic SZ effect, \mission\ will measure the peculiar velocity of hundreds of thousands of clusters.
PRISM thus will carry out a complete survey of the large-scale velocity field throughout our entire Hubble 
volume.  \mission\ will also measure the gas temperature of more than ten thousand massive galaxy 
clusters using the relativistic corrections to the classic SZ spectral distortion spectrum.

By cross-correlating cluster detections with measurements of gravitational lensing of the CMB, 
\mission\ will measure cluster masses and calibrate the scaling laws crucial for cosmological 
studies.  This will be possible not only for \mission\ clusters, but for all cluster 
samples including X-ray and optically selected clusters, and even candidate high redshift proto-clusters 
detected in the far infrared. The \mission\ cluster catalogue will provide an exceptionally powerful probe of 
dark energy, modified gravity, and structure formation throughout the Hubble volume.

\mission\ will provide a comprehensive view of both the dark matter and the baryonic matter 
distributions, as well as their relation and its evolution over cosmic time.  With ultra-precise measurement of 
gravitational lensing of the CMB, maps of the dark matter distribution will be obtained,
which can be correlated with various baryonic tracers. These include the hot gas observed through the 
thermal SZ effect, optical and IR galaxies, extragalactic radio sources catalogues, and the 
cosmic infrared background (CIB) arising from primordial galaxies reaching redshifts $z>5.$
Cross-correlations with upcoming cosmic shear surveys such as Euclid will
also be extremely interesting. In this way, invaluable insight into the 
transformation of baryons from primordial gas into stars within their dark matter halo hosts will be 
obtained as well as a multi-component view of the distribution of mass and its tracers in the 
entire observable universe. Because the CMB is the most distant source plane, \mission\ 
maps the density contrasts at much higher redshifts than traditional gravitational lensing surveys.

\subsection{Understanding the Cosmic Infrared Background and star formation across cosmic time} Most star 
formation in the universe took place at high redshift. Hidden from optical observations by shrouds of 
dust in distant galaxies, star formation at large redshift is visible in the far infrared. 
Emission from these dusty galaxies constitutes the cosmic infrared background (CIB), which \mission\   
owing to its high sensitivity and angular resolution in the far infrared is uniquely situated to investigate. 
The survey will sharpen and extend to higher redshifts the determination of the bolometric luminosity 
function and the clustering properties of star-forming galaxies. Tens of thousands of easily 
recognizable, bright, strongly lensed galaxies and hundreds of the very rare maximum starburst galaxies
beyond $z>6$ will be detected, providing unique information on the history of star formation, the physics 
of the interstellar medium under a variety of conditions including the most extreme, and the growth of 
large-scale structure, including proto-clusters of star-forming galaxies. The survey will also probe the 
evolution of radio sources at (sub-)mm wavelengths and provide measurements of the spectral energy 
distribution (SED) of many thousands of radio sources over a poorly explored but crucial frequency 
range.

The broad frequency coverage and many frequency bands enable us to model the SEDs of 
individual distant sources as a superposition of emission from enshrouded star formation and 
emission powered by the active nucleus. \mission\ will thus provide insight into the 
interplay between these two components and their claimed co-evolution. The measurement of the source 
SEDs provide photometric redshifts, enabling optimized follow-up with high resolution 
ground-based spectroscopic observations.

In addition, the many frequency bands of \mission\ will enable us to decompose the 
CIB of unresolved galaxies into sub-components
originating from different redshift shells and thus 
constrain models of structure and star formation 
throughout cosmic history. By measuring both the total absolute CIB emission and its 
fluctuations over the CIB emission spectrum bump as well as resolving many 
individual high redshift galaxies, \mission\ will provide a comprehensive view of the far infrared background.

\subsection{Detecting inflationary gravity waves and constraining inflationary scenarios} Present 
precision measurements of cosmic microwave background (CMB) temperature anisotropies lend considerable 
support to simple models of inflation. However the most spectacular prediction of inflation---the 
generation of gravitational waves with wavelengths as large as our present horizon---remains unconfirmed. 
Several initiatives from the ground and from stratospheric balloons are currently underway to attempt to 
detect these gravitational waves through the B-mode spectrum of the CMB polarization. However
these initiatives
suffer severe handicaps including limited frequency coverage due to atmospheric opacity and 
emission, unstable seeing conditions, and far sidelobe pickup of ground emission. Only from space 
can one reliably detect the very low-$\ell $ B-modes due to the re-ionization bump. Because 
of its broad frequency coverage and extreme stability, \mission\ will be able to detect B-modes at 
$5\sigma $ for $r=5\times 10^{-4},$ even under pessimistic assumptions concerning the complexity of the 
astrophysical foreground emissions that must be reliably removed. Moreover, \mission\ will be able to 
separate and filter out the majority of the lensing signal due to gravitational deflections. Hence 
\mission\ will perform the best possible measurement of primordial CMB B-mode polarization, and hence the 
best possible measurement of the corresponding inflationary tensor perturbations achievable through CMB 
observations.

\subsection{Probing new physics through CMB spectral distortions} The excellent agreement between the 
microwave sky emission and the perfect blackbody observed by the {\it COBE FIRAS} instrument is rightfully 
highlighted as a crucial confirmation of Big Bang cosmology. However theory predicts that at higher 
sensitivity this agreement breaks down.
Energy injection at any redshift $z\ltorder ({\rm few})\times 10^6$ superimposes
spectral distortions relative to a perfect CMB blackbody spectrum. Some of the predicted 
deviations are nearly sure bets. Others provide powerful probes 
of possible new physics. The \mission\ absolute spectrometer will measure the spectrum more than three 
orders of magnitude better than FIRAS. $y$-distortions from the re-ionized gas as well as from hot 
clusters constitute a certain detection. However $\mu$-distortions and more general spectral distortions 
have the potential to uncover decaying dark matter and to probe the primordial power spectrum on very 
small scales that cannot be measured by other means, being contaminated by the nonlinearity of 
gravitational clustering at late times.

\subsection{Probing Galactic astrophysics} 

\mission\ will have a major impact on Galactic astrophysics by providing a unique set of all-sky maps. 
The \mission\ data will extend \Herschel\ dust observations to the whole sky, map the absolute sky intensity 
and polarized emission with high sensitivity in more than 20 different frequency bands, and also map 
emission lines key to quantifying physical processes in the interstellar medium. The survey 
will have the sensitivity and angular resolution required to map dust polarization down to sub-arcminute 
resolution even at the Galactic poles. No other project provides a comparable perspective on interstellar 
components over such a wide range of scales.

The \mission\ data will provide unique clues to study the interstellar medium, the Galactic magnetic 
field, and star formation, and will address three fundamental questions in Galactic astrophysics: 
What are the processes that structure the interstellar medium? 
What role does the magnetic field play in star 
formation? What are the processes that determine the composition and evolution of interstellar dust?

\section{Legacy archive}
The science themes outlined above sample but a few of the highlights of the rich and diverse physics 
and astrophysics that \mission\ will be able to carry out. These are the main drivers for 
the mission definition. However a secondary but also important objective is the collection of high 
quality survey full-sky data for a large variety of applications by the astronomical community.

The hundreds of intensity and polarization maps of \mission\  spanning over two decades in frequency will 
constitute a legacy archive useful for almost all branches of astronomy for decades to come. Combining 
low resolution spectrometer data and high resolution images from the imager, \mission\ will deliver a 
well calibrated and fully characterized spectro-polarimetric survey of the full sky from 50$\, \mu$m to 
1$\,$cm. The spectral resolution will range from about 0.5$\,$GHz to 15$\,$GHz at 1.4$^\circ$ angular 
resolution, and from $\delta \nu/\nu \approx 0.025$ to $0.25$ at the diffraction limit of a 3.5$\,$m 
telescope (from $\sim 6''$ to $17'$). It is not possible to foresee all the ways in 
which this extraordinary data set can be exploited.

The \mission\ collaboration will make public full-sky maps of the absolute emission of the sky (in 
intensity and polarization) in many frequency channels, as well as maps of absolute temperature of the 
CMB and its polarization (at a resolution of about 2 arcminutes with a sensitivity of order 1$\,\mu$K 
or better per resolution element). Similarly maps of the intensity and polarization of all main emission 
processes will be made available. These include (1) emission of all galactic components in absolute 
intensity and polarization (including main spectral lines such as lines of CO, C-I, C-II, N-II, O-II, HCN, 
HCO+), 
(2) maps of thermal (tSZ) and kinetic (kSZ) Sunyaev-Zel'dovich effect, 
(3) maps of CIB emission in various 
redshift shells, (4) maps of the lensing potential and of the matter density in different
redshift shells, and (5) several catalogues of 
various galactic and extragalactic objects including a catalogue of about a million galaxy clusters 
and large groups up to redshift $z>3$, and (6) catalogues of thousands of very high redshift dusty 
galaxies, strongly lensed galaxies, and high redshift proto-clusters.


\section{Probing the Universe with galaxy clusters}
\label{sec:clusters}
\begin{figure}
\begin{center}
\includegraphics[width=0.49\textwidth]{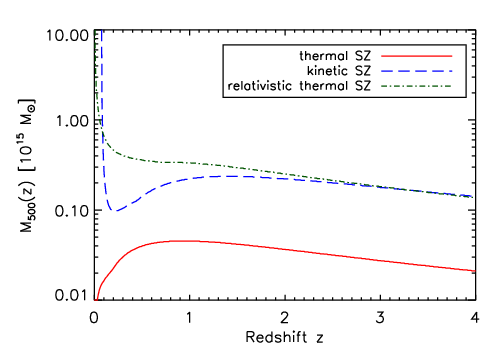}
\includegraphics[width=0.49\textwidth]{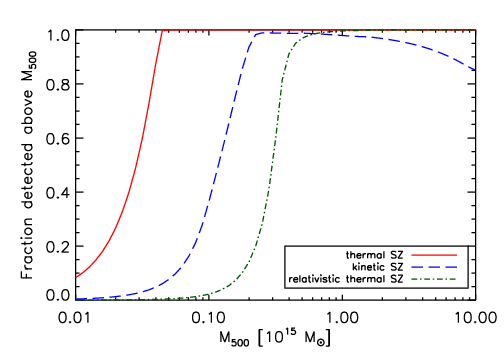}
\caption{{\small Left panel: Lower mass limits for detection of the labeled SZ effects at signal-to-noise $S/N>5$ 
as a function of redshift. Right panel: The completeness of the detection for all three effects, as a function of mass. Objects more massive than $4 \times 10^{13}\Msolar$ are detected at more than 
5$\sigma$ at all redshifts, a limit extending well down into the relatively unexplored group range (red line).  
\mission\ is able to measure peculiar velocities over most of the cluster range, i.e., for $M>2 \times 10^{14}\Msolar$ (dashed blue line), and relativistic effects, giving access to cluster temperature, for the more massive systems (dot-dashed green line), both out to high redshift. Note that in practice, the detection depth will be lower close to the galactic plane, where a fraction of the clusters will inevitably be missed (although this should be only a very small fraction of the total).}}
\label{fig:clusterdetmasses}
\end{center}
\end{figure}

Galaxy clusters are an invaluable and proven cosmological tool. The SZ effect is playing an increasingly
major role in the exploration of our Universe at large redshift 
\citep[e.g.,][]{barbosa1996, dasilva2001, bartlett2004, benson2013, reichardt2013, hasselfield2013, planck2013XX}.
Using the thermal Sunyaev-Zeldovich effect, \mission\ will produce the largest galaxy cluster catalog ever, with far
more objects than any past or planned cluster survey and, more importantly, extending to higher 
redshifts.  Cluster and group systems will be detected by \mission\ throughout our Hubble volume from the moment they 
first emerge.  With the kinetic SZ effect, \mission\ will map the cosmic peculiar velocity 
field throughout the entire observable universe, an objective that cannot be attained by any other means and which
will provide a new cosmological probe.
\mission\ will furthermore provide cluster mass determinations out to high redshift through 
gravitational lensing of the CMB in both temperature \citep{seljak2000} and polarization \citep{lewis2006}. 
\mission's high angular resolution and frequency coverage in bands unreachable from the ground bring these
objectives within reach.  The \Planck, ACT, and SPT experiments demonstrated the potential of 
the SZ effect for studying galaxy clusters and using them to constrain cosmological models.  
\mission\ will transform SZ cluster studies into arguably our most powerful probe of cosmic large-scale 
structure and its evolution.

\subsection{Characterizing the PRISM cluster catalog} 

We forecast the content of the \mission\ SZ catalog by applying a multi-frequency matched filter 
\citep[MMF,][]{melin2006} to a simulated typical field at intermediate Galactic latitude.  Our detection mass 
remains below $5 \times 10^{13}\Msolar$ at {\em all redshifts} (Fig.~\ref{fig:clusterdetmasses}), and extrapolating 
from the observed \Planck\ counts, we predict more than $10^{6}$ clusters, with many thousands at $z>2$.  The 
primary source of uncertainty in this prediction lies in the adopted SZ signal--mass scaling relation, $Y$--$M$.
Fortunately, this is more clear than prior to \Planck. We now know from \Planck\ SZ observations 
\citep{planck2011-pepXII,planck2012-pipXI} that the SZ signal scales with mass according to our adopted relation  
down to much smaller masses in the local universe, leaving as our main uncertainty the presently poor knowledge 
of its redshift dependence.  \mission\  will determine this evolution, through its mass measurement
capabilities (see below), and constrain cosmology from the observed cluster evolution.

One of the notable strong points of \mission\ is its ability to measure cluster masses at all redshifts through gravitational lensing 
of the CMB anisotropies.  Using detailed lensing simulations, we estimate that a cluster mass of $2\times 10^{14}\Msolar$ will be detected by
\mission\ at all redshifts with a signal-to-noise of unity (Melin \& Bartlett 2013, in preparation).  
We will therefore be able to measure individual masses for the more massive systems.  More
importantly however, we can measure the mean mass of clusters through binning as a function of SZ signal (or other observable) and redshift.  
This gives us the essential capability to determine the scaling relations with cluster mass necessary for cosmological 
interpretation of the cluster counts and for structure formations studies.  It is important to note that \mission\ is self-sufficient in 
this critical aspect of cluster science.

\mission\ will surpass all current and planned cluster 
surveys, including {\it eROSITA} and Euclid---not just in total number, but most importantly in number of 
objects at $z>1.5$.  Because optical/NIR cluster searches suffer from high contamination rates, especially at redshifts 
beyond unity, cluster identification  and the construction of a precise selection
function will be vastly more robust for \mission\ than for Euclid. 
Only \mission\ can find a significant number of clusters in the 
range $2<z<3$, the critical epoch that current observations identify as the emergence of the characteristic 
cluster galaxy population on the red sequence.  
\mission\ will also enable us to explore the abundance of the 
intra-cluster medium (ICM) through the $Y$--$M$ relation and its relation to the galaxy population at these 
key redshifts.

At the time of operation, large imaging (e.g., {\it DES, LSST, HSC}) and spectroscopic surveys (e.g., {\it 
4MOST, PFS, WEAVE, BigBOSS/MS-DESI, SKA}) will have covered the entire extragalactic sky.  We can therefore
easily obtain redshifts, spectroscopic or photometric, for all objects up to $z=2$, and the two micron cutoff 
of Euclid's IR photometric survey (H band) will be sufficient to detect the $4000\AA $ break in brighter 
cluster galaxies at higher redshifts. Follow-up with ALMA or CCAT can provide additional information 
(e.g., redshift, structure, object population) of the most interesting objects.

\subsection{Thermal SZ maps}

\begin{figure}[tb]
\includegraphics[width=0.195\textwidth]{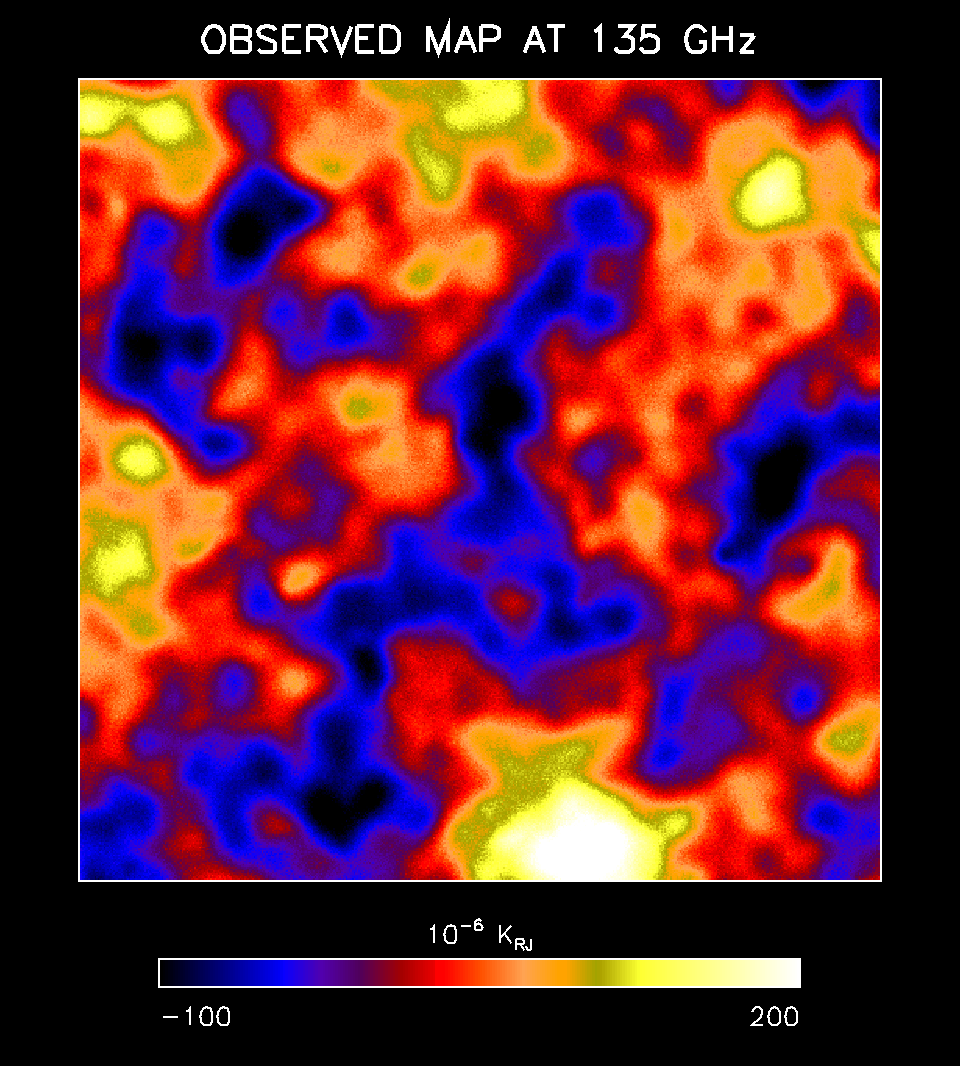}
\includegraphics[width=0.195\textwidth]{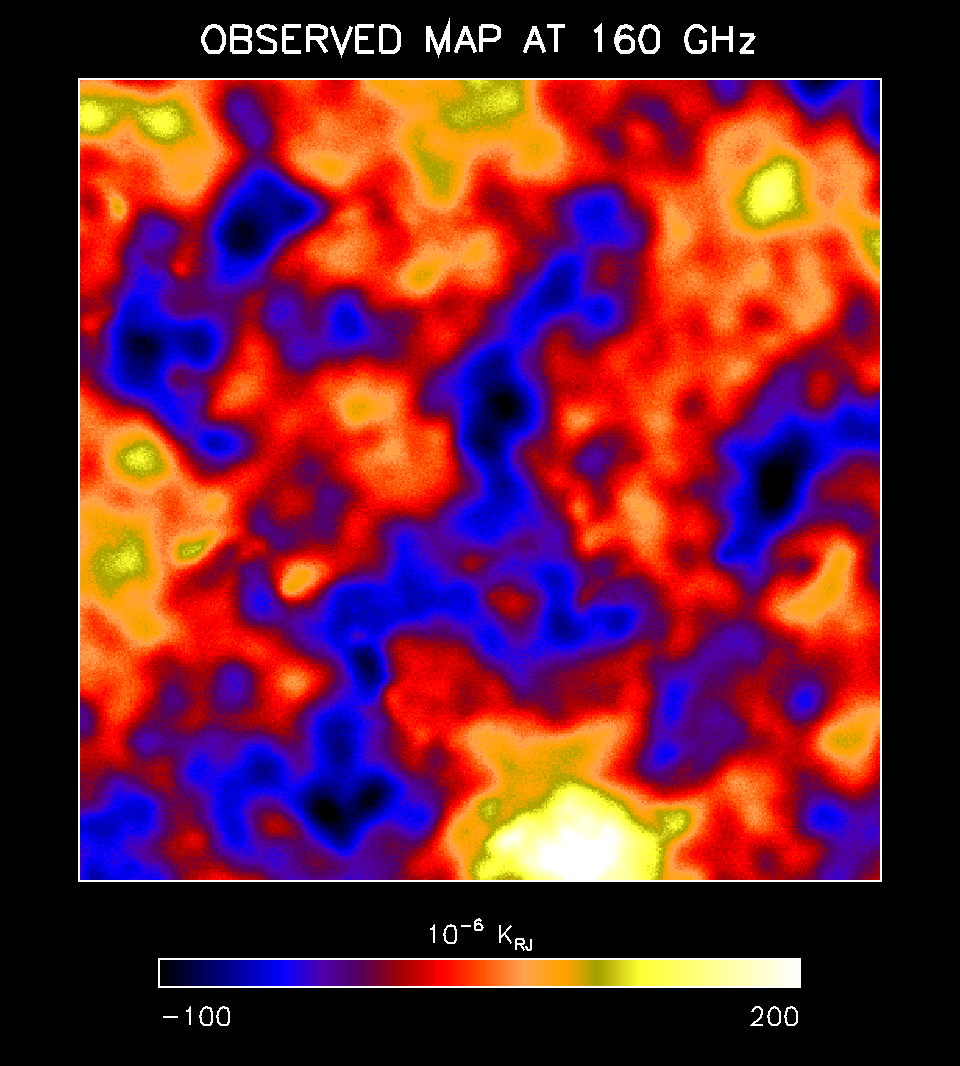}
\includegraphics[width=0.195\textwidth]{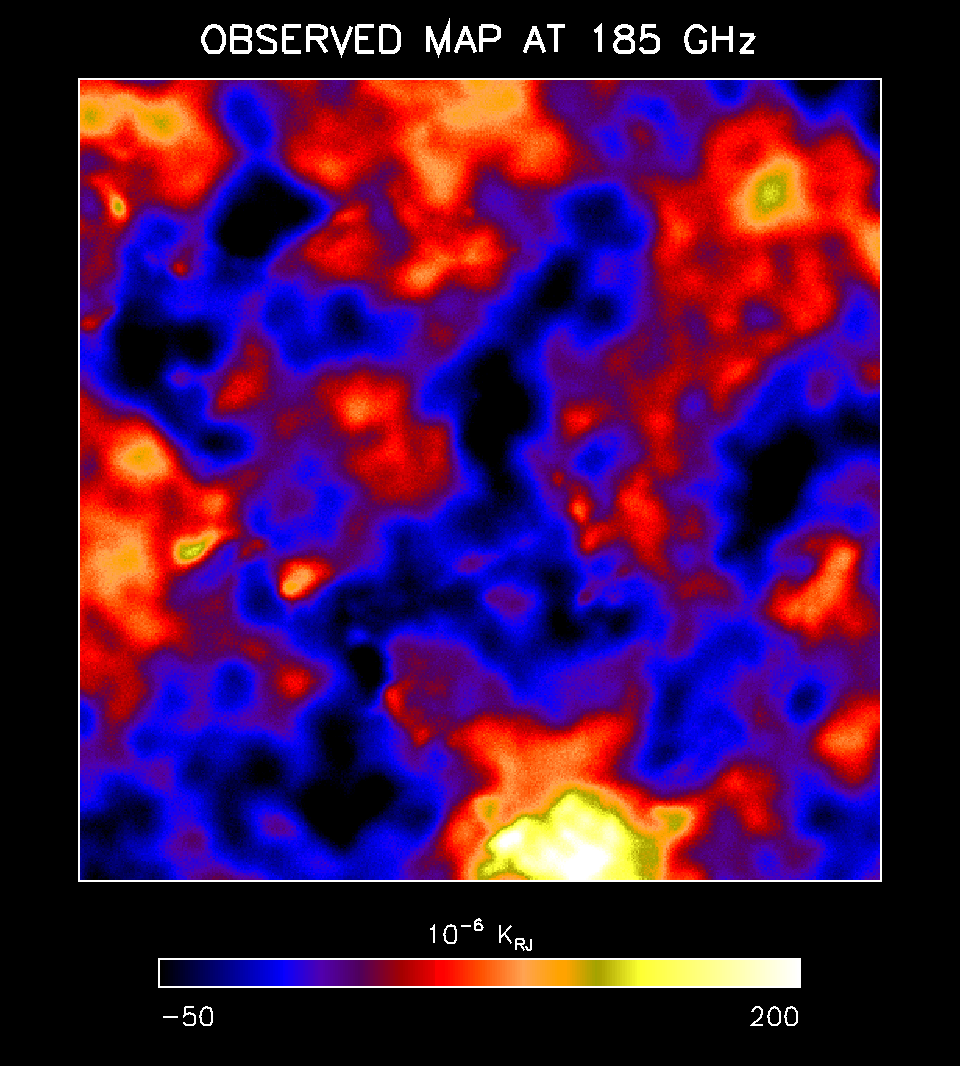}
\includegraphics[width=0.195\textwidth]{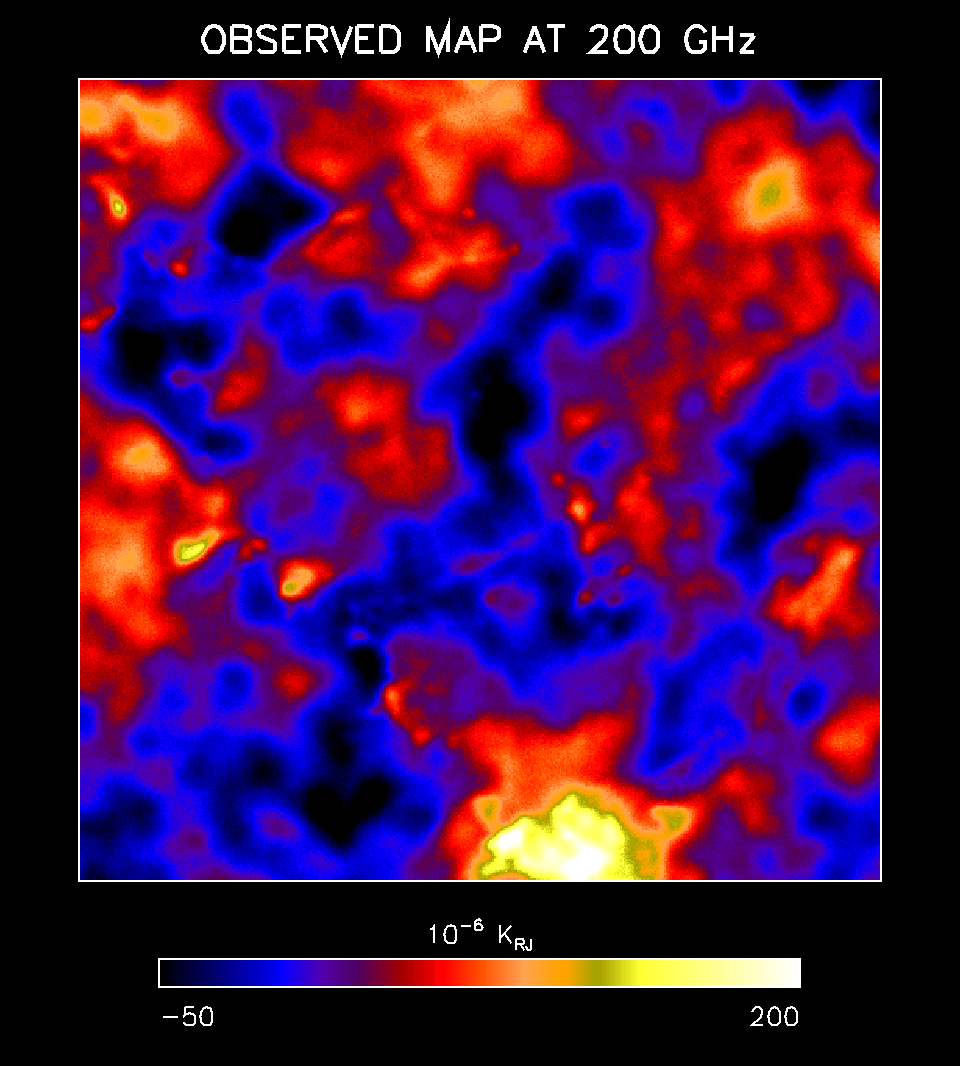}
\includegraphics[width=0.195\textwidth]{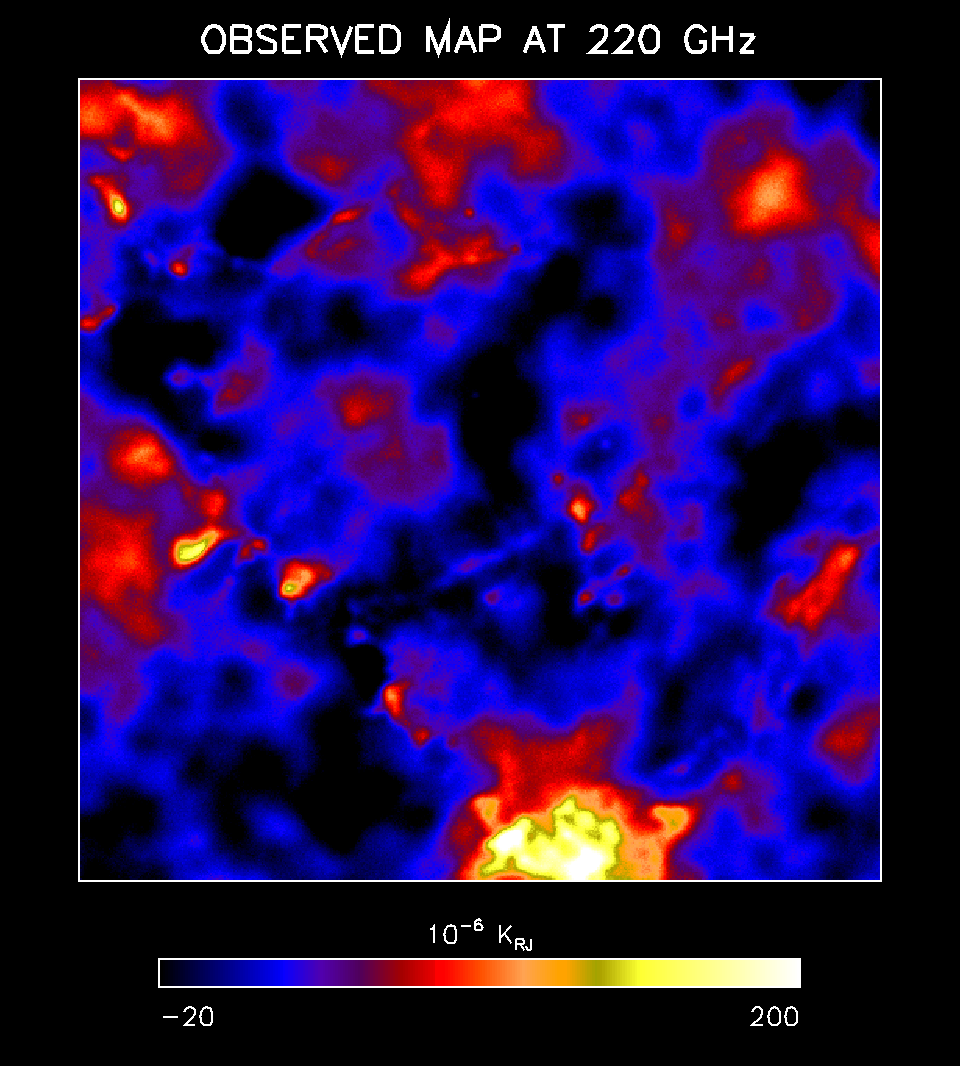}\\
\vskip 0.4cm
\includegraphics[width=0.195\textwidth]{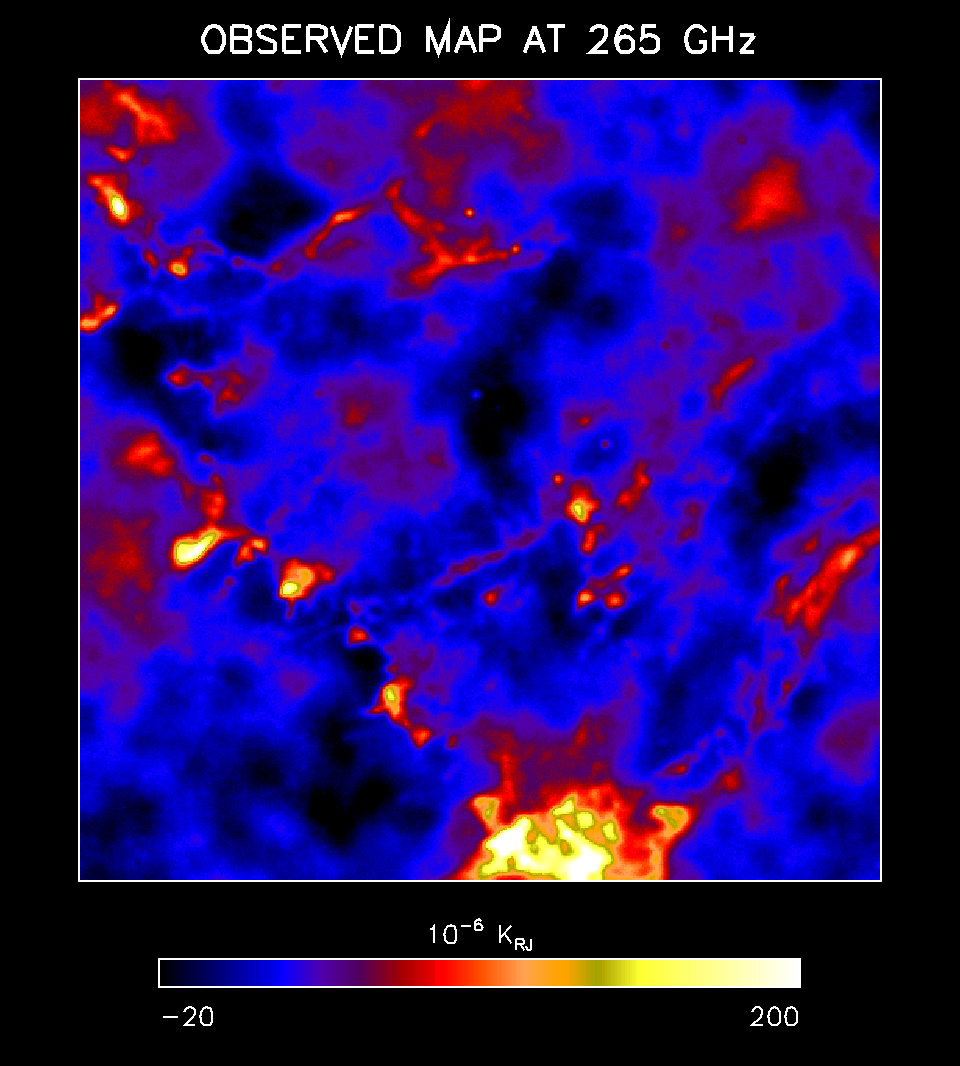}
\includegraphics[width=0.195\textwidth]{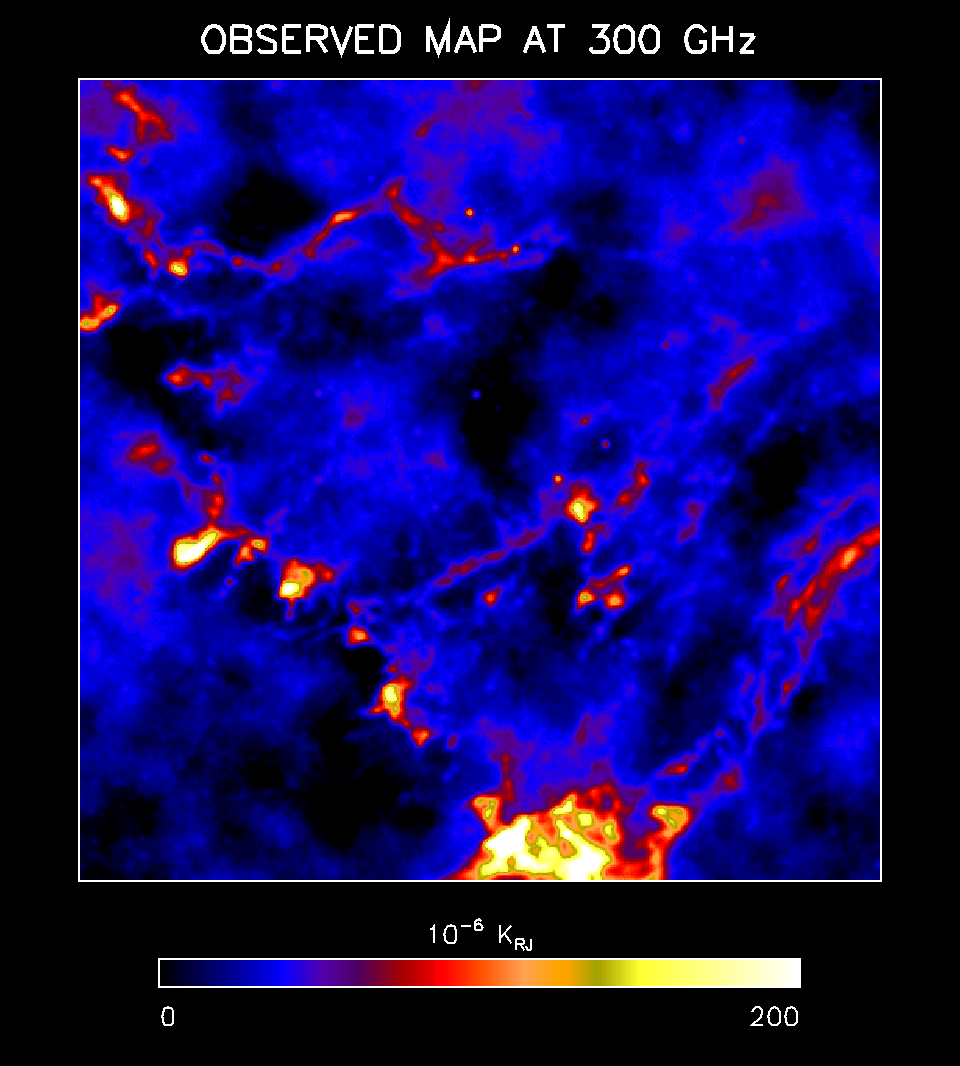}
\includegraphics[width=0.195\textwidth]{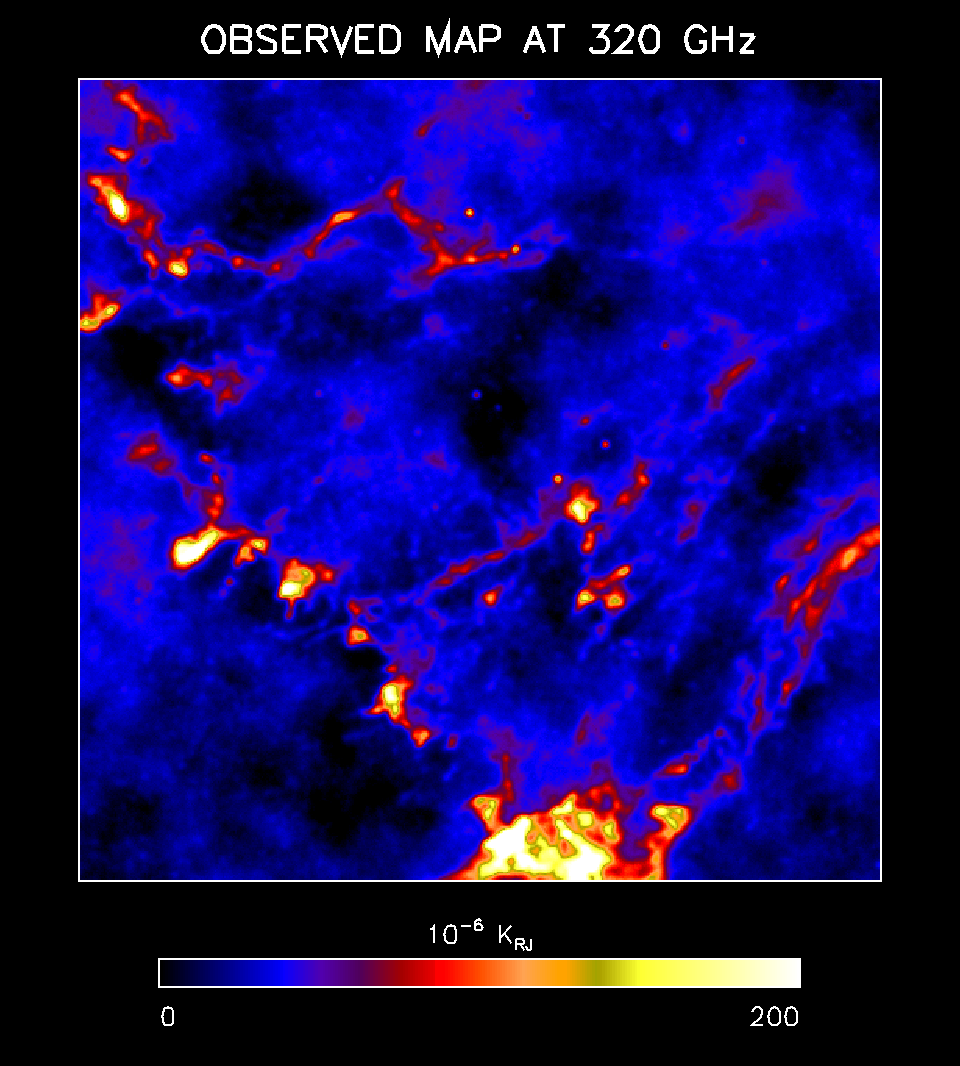}
\includegraphics[width=0.195\textwidth]{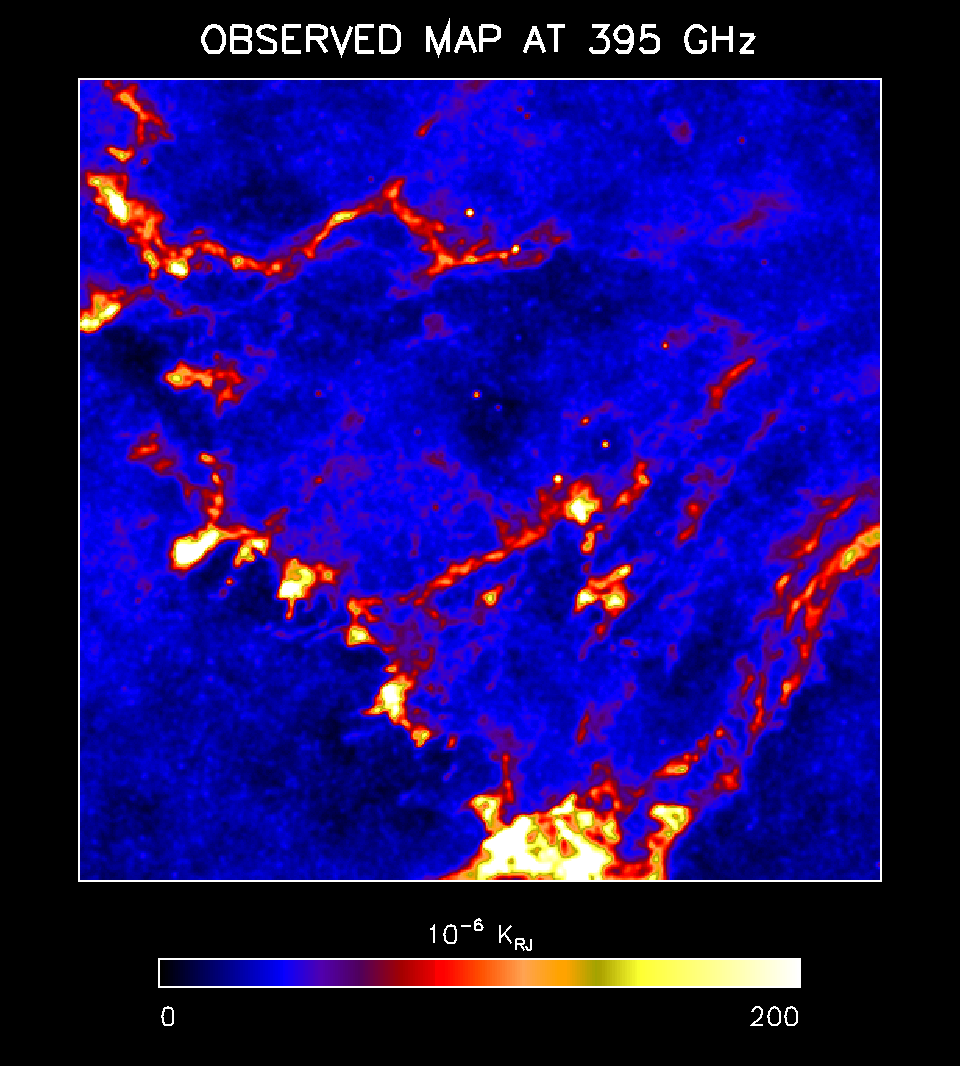}
\includegraphics[width=0.195\textwidth]{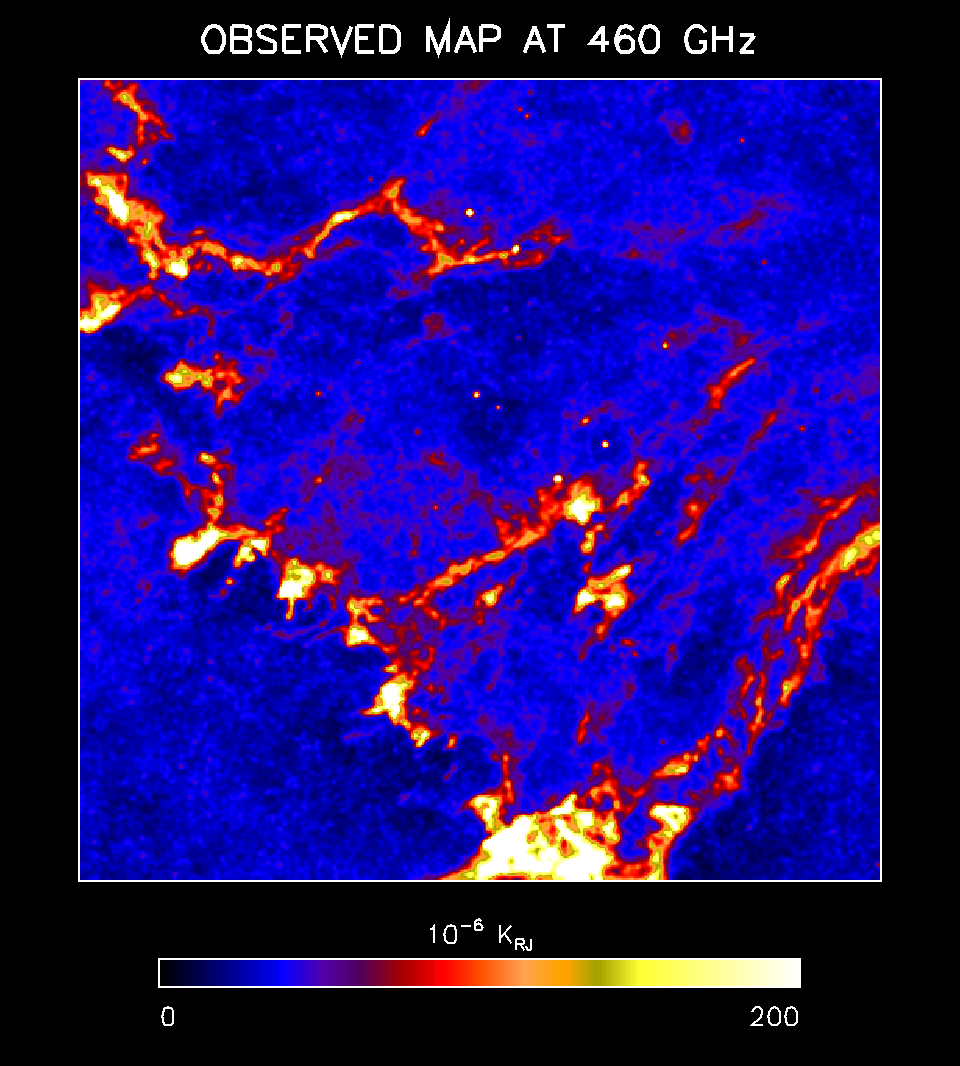}\\
\vskip 0.4cm
\includegraphics[width=0.195\textwidth]{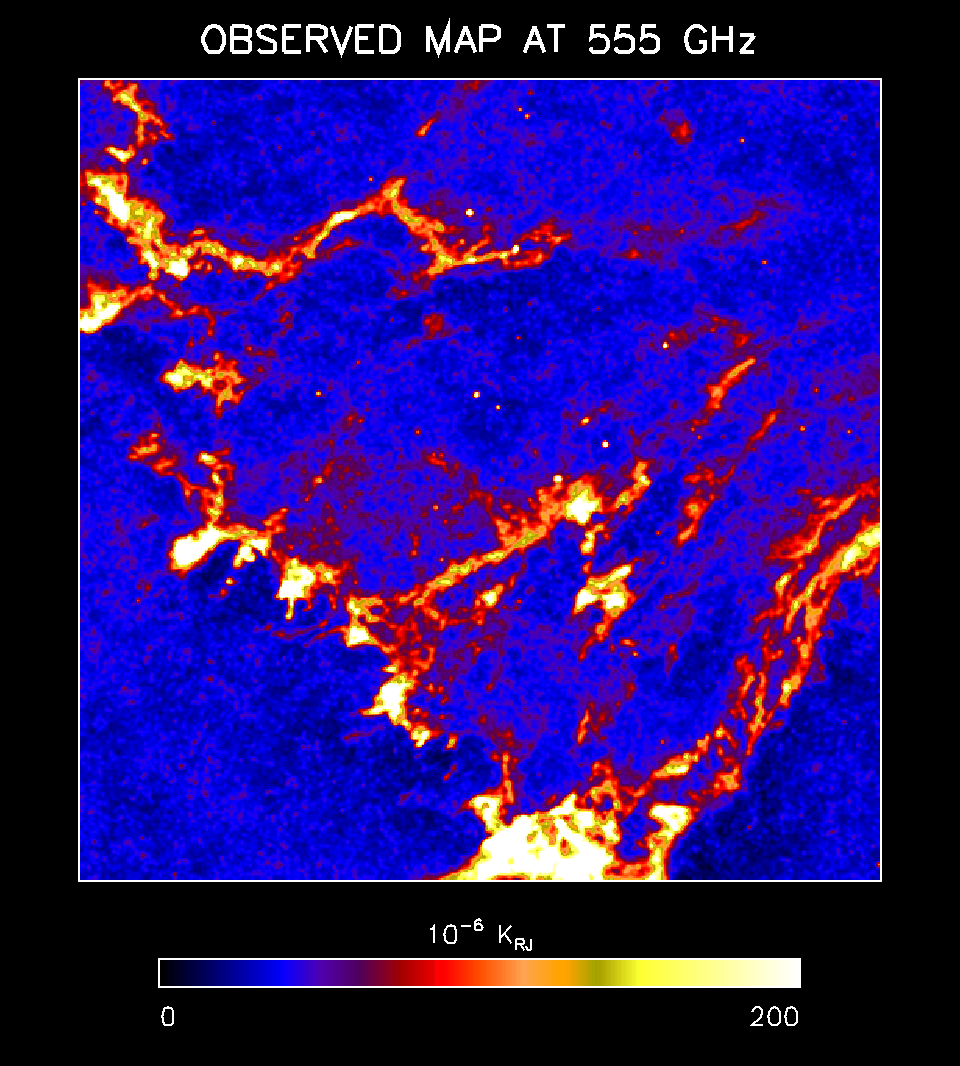}
\includegraphics[width=0.195\textwidth]{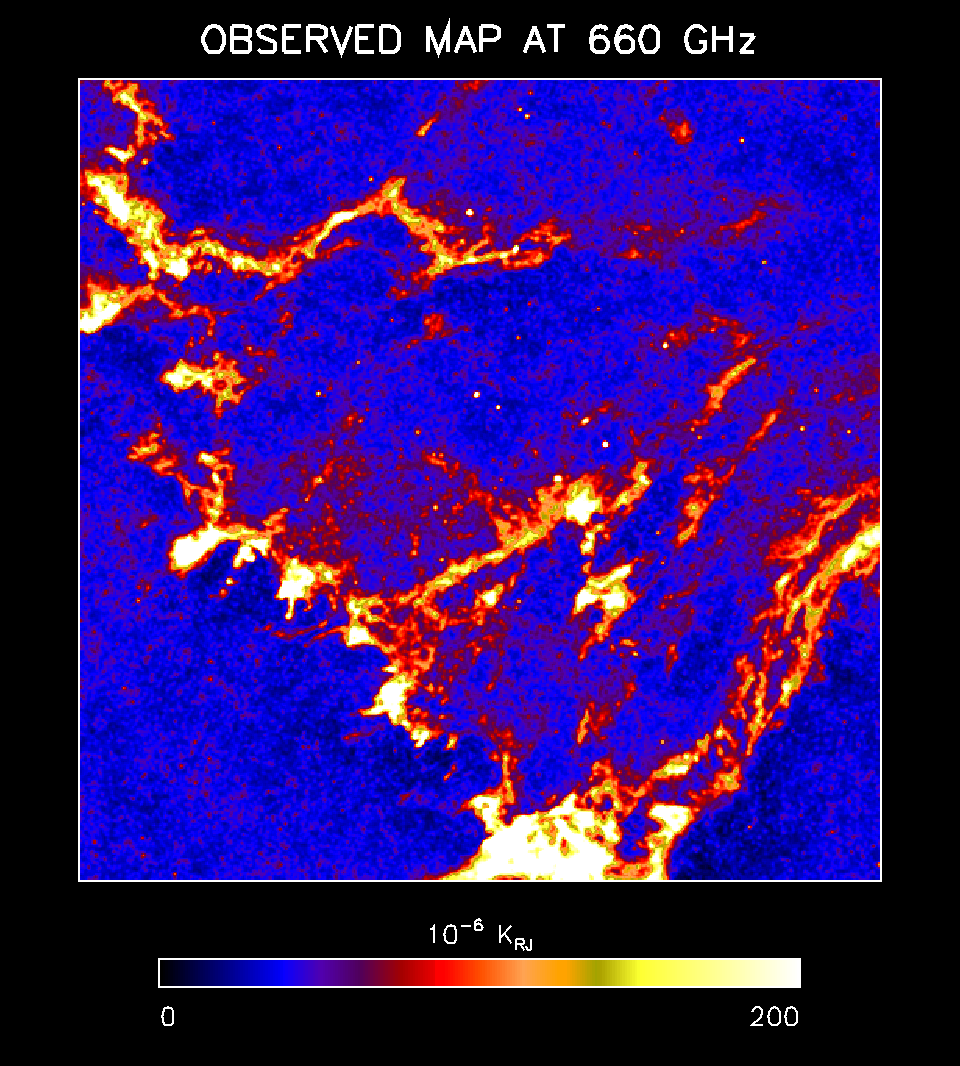}
\includegraphics[width=0.195\textwidth]{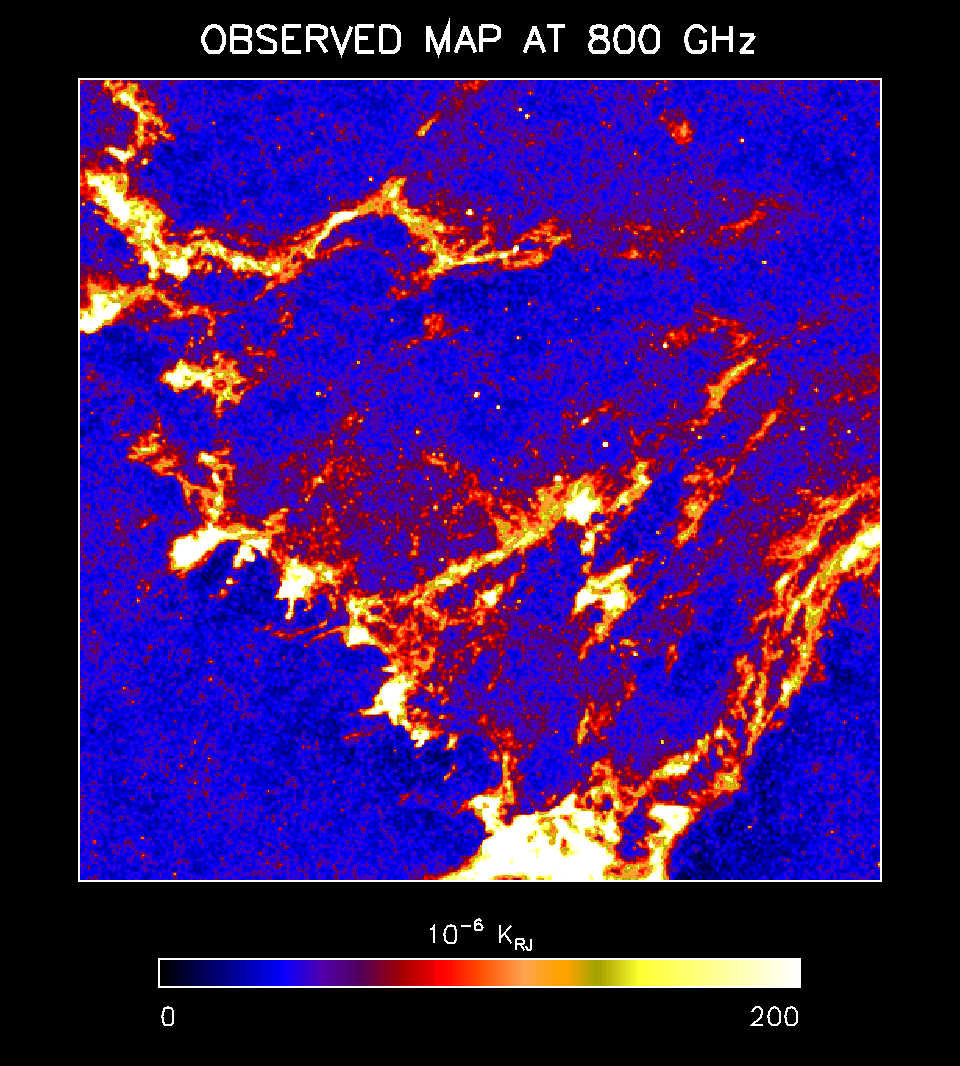}
\includegraphics[width=0.195\textwidth]{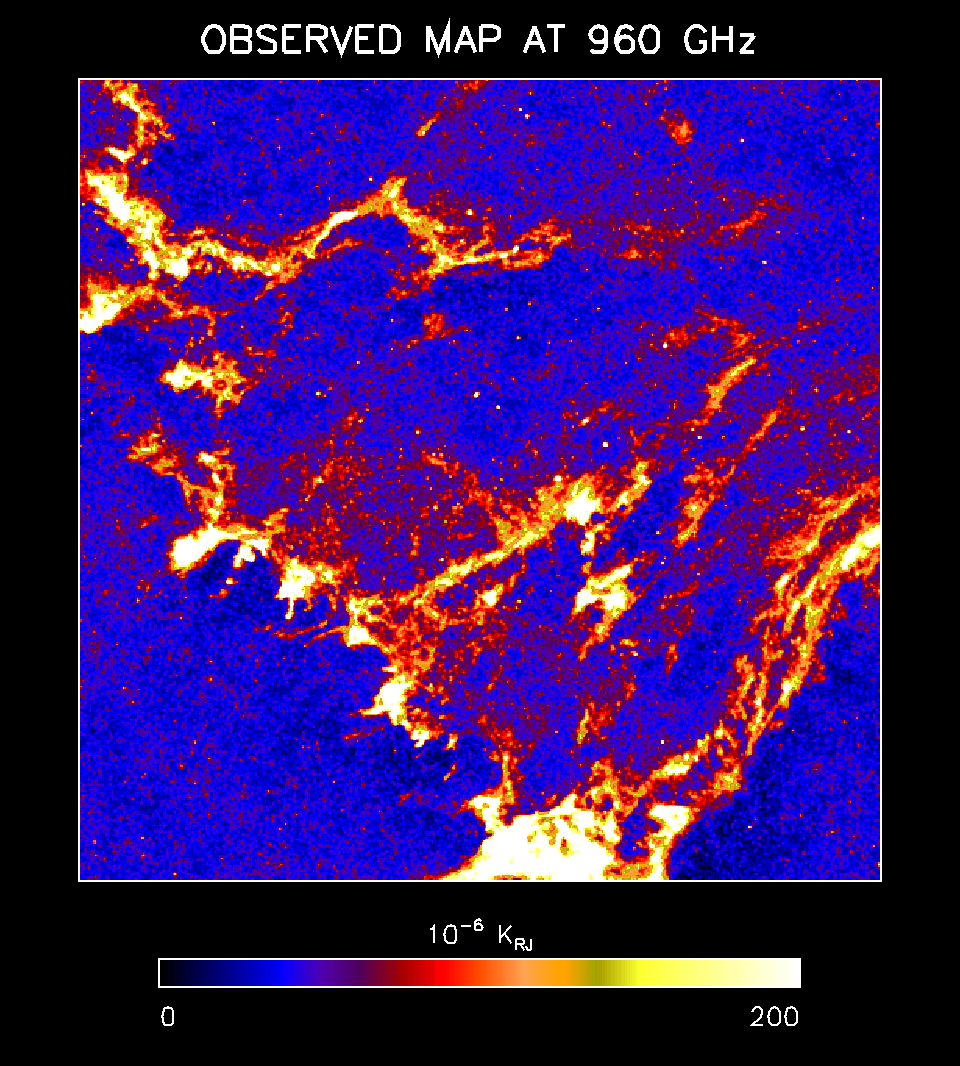}
\includegraphics[width=0.195\textwidth]{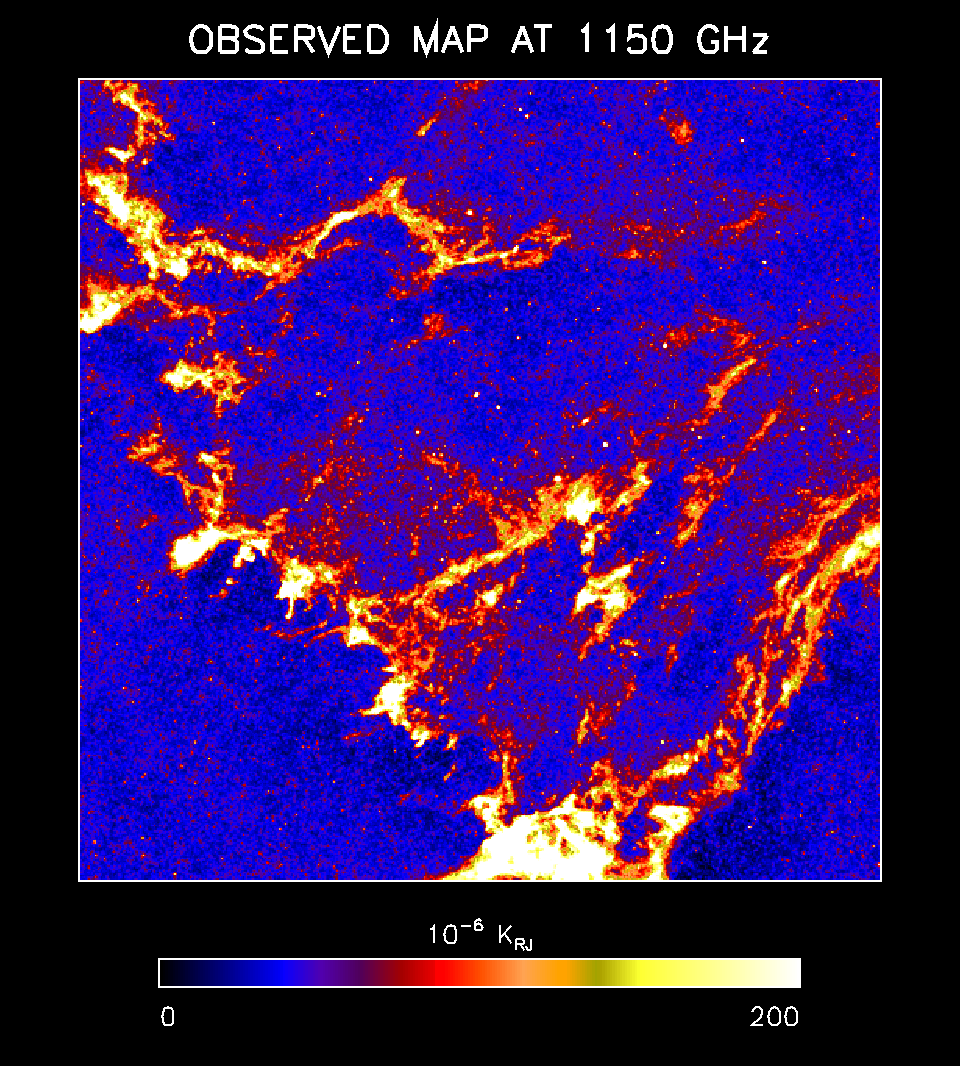}
\caption{{\small Simulated \mission\ observations in 15 broadband channels from 135 to 1150 GHz. The top left map, at 135 GHz, is dominated by CMB temperature anisotropies, and is observed with 3.8' angular resolution. At 1150 GHz (bottom right), the angular resolution is 27'' and the emission is dominated by Galactic dust, nearby galaxies, and CIB emission. The thermal SZ emission is sub-dominant at all frequencies.}}
\label{fig:prism-observations}
\end{figure}

The impressive performance of \mission\ for cluster detection throughout our Hubble volume is 
illustrated by a realistic simulation of \mission\ observations on a small patch of  sky of about 10 square degrees 
($3.4 \times 3.4$ degrees) through a region of low dust emission in the Draco constellation.  Figure~\ref{fig:prism-observations} shows
simulated observations in 15 frequency channels ranging from 135 to 1150 GHz in intensity only. 
The simulation comprises thermal dust and infrared sources scaled from \Herschel\ observations 
using the most recent dust model being developed within the \Planck\ collaboration, a model for the cosmic infrared background 
generated by a population of high redshift proto-spheroidal galaxies \citep{Xia2012,2013ApJ...768...21C}, as implemented in the 
Planck Sky Model \citep{2013A&A...553A..96D}, thermal SZ emission from a population of clusters and large groups matching \planck\ 
number counts and scaling relations \citep{planck2013XX}, kinematic SZ signal 
generated following the approach of Delabrouille, Melin and Bartlett \citep{2002ASPC..257...81D}, and CMB anisotropies 
matching the recent \planck\ power spectrum and cosmological constraints.  Other Galactic foreground emission (synchrotron, free-free, spinning dust, and 
molecular lines) and radio sources will not contaminate SZ cluster detection significantly after being subtracted out using \mission's lower frequency data and narrow band observations. They are therefore neglected for this simulation.

The SZ thermal emission signal (tSZ) is recovered from the maps using the so-called \emph{Internal Linear Combination} 
(ILC) method over a set of spectral windows, which define a set of scales adjusted to match the resolution of the 
different observed maps. A thermal SZ map is first obtained at 3.8' resolution using all 15 bands, each 
smoothed to 3.8'. Fourteen maps at 3.2' resolution (corresponding to that of the 160 
GHz channel) are then computed from all maps at frequencies above 160 GHz. We subtract from these 14 maps the 
corresponding 3.8' resolution maps, to obtain 14 maps with structure on scales ranging from 3.2 to 3.8 
arc-minutes. An ILC on these 14 maps is performed to produce a map of thermal SZ emission on scales 3.2-3.8 
arc-minutes, which is added to the lower resolution tSZ map. The process is iterated using each time only 
observations in the highest frequency channels until a tSZ map at 1.3' resolution, corresponding to the beam 
of the 395 GHz channel, is obtained. Considering that the process is localized in pixel space by being applied 
to a small patch of sky, this implements an ILC that is localized in both pixel space and across scales, similar to
the needlet LC method successfully used for CMB extraction on WMAP data 
\citep{2009A&A...493..835D,2012MNRAS.419.1163B,2013MNRAS.435...18B}.

\begin{figure}[tb]
\includegraphics[width=0.495\textwidth]{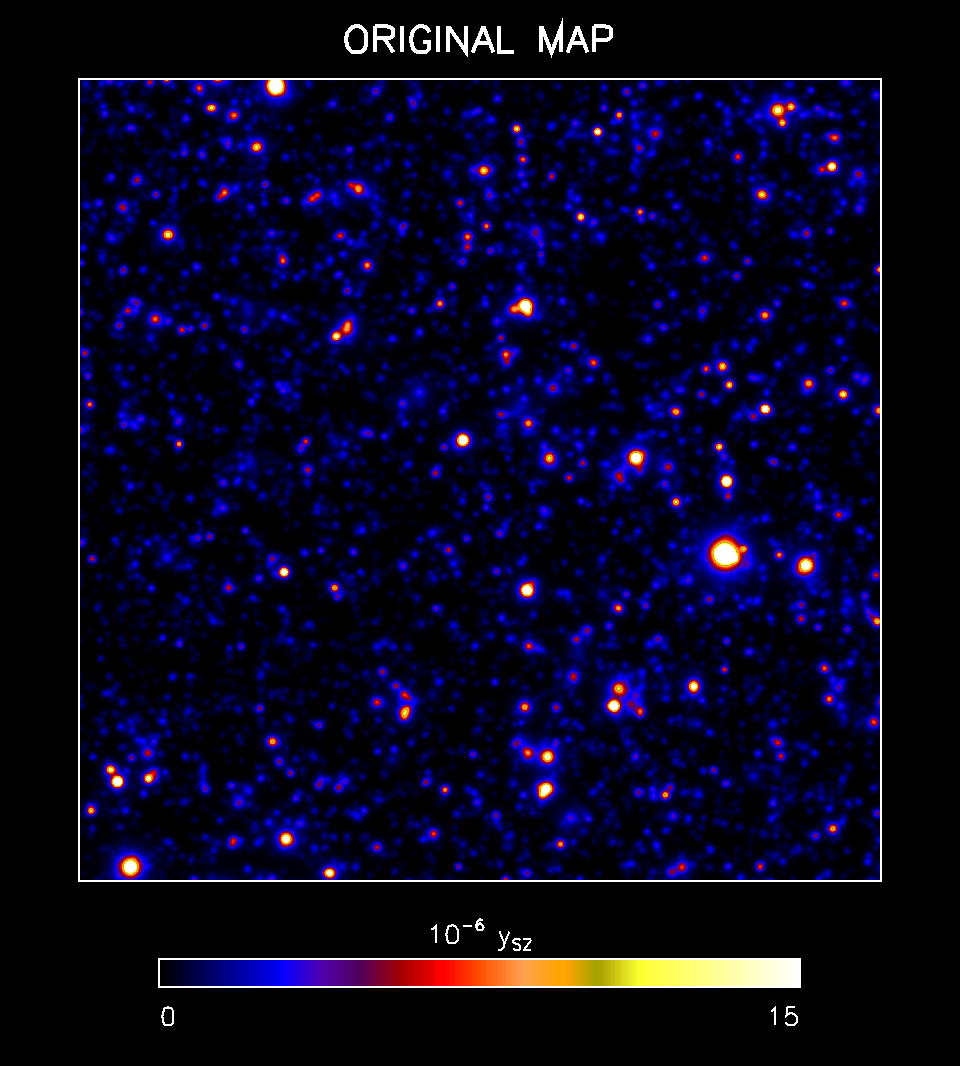}
\includegraphics[width=0.495\textwidth]{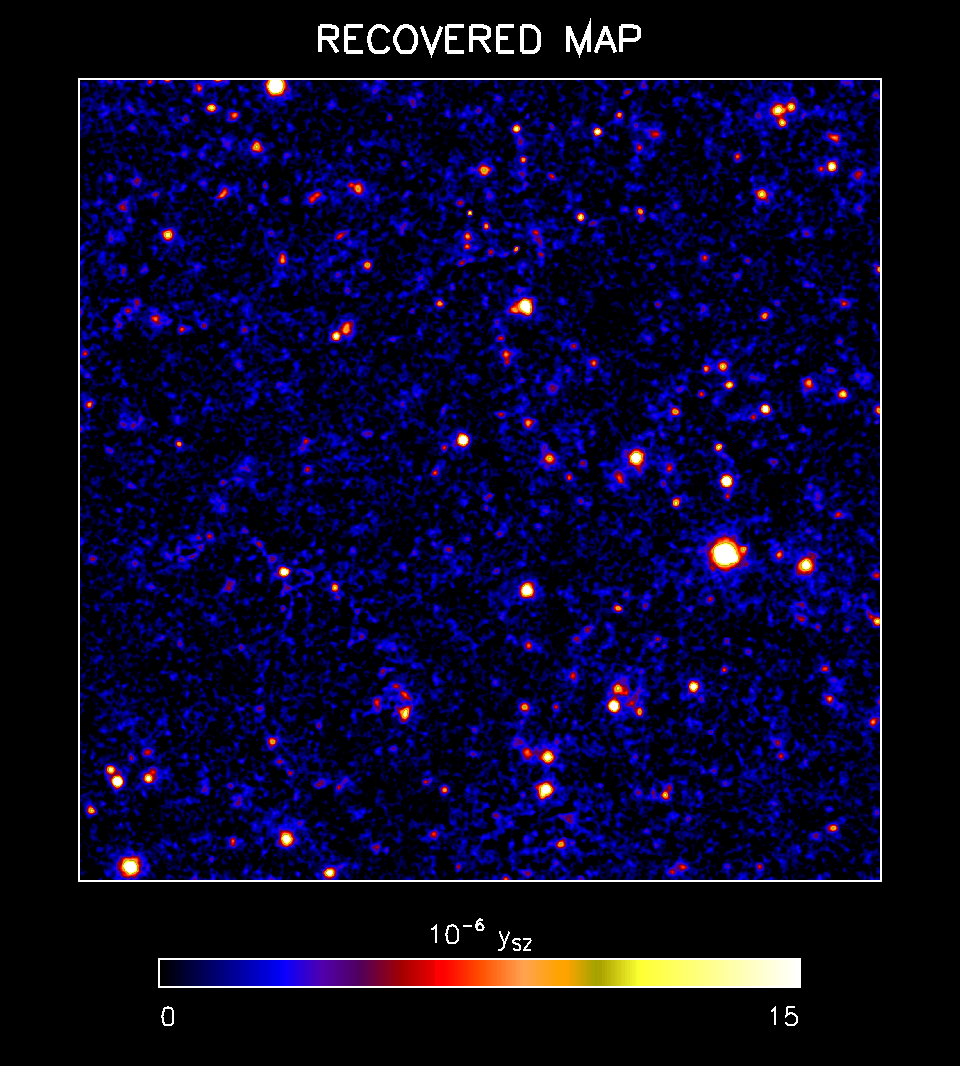}
\caption{{\small Extraction of thermal SZ emission from a set of simulated \mission\ maps. Most of the clusters 
in the original map are clearly visible after separation of the tSZ component with a straightforward multiscale ILC. 
Hundreds of SZ clusters are detectable on this very small patch (0.025\% of sky).}}
\label{fig:tSZ-recovery}
\end{figure}

Figure \ref{fig:tSZ-recovery} illustrates the outcome of this 
processing. Several hundred clusters matching the input simulation are visible in the recovered tSZ maps, 
demonstrating that a straightforward component separation procedure makes it possible to observe most of the 
original clusters. The number of clusters extracted from the recovered tSZ map in this simulation 
confirms the predicted cluster number counts obtained with the MMF method, supporting the estimated number 
of $10^6$ clusters or more observed by \mission. In addition, the recovered map can also be exploited by 
stacking regions centered on various targets: very faint candidate optical clusters not individually 
detected, expected filaments of the cosmic web between pairs of detected clusters, very distant candidate 
massive proto-clusters detected by the infrared red emission of their dust emission, distant quasars, HI 
galaxies, etc.

\begin{figure}[bht]
\begin{center}
\centerline{
\begin{minipage}{18 cm}
\includegraphics[width=0.30\textwidth,trim=0cm 0cm 3cm 0cm]{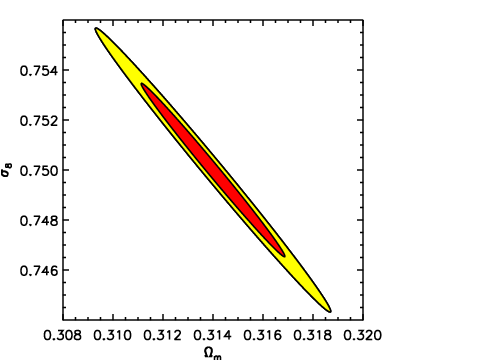}\\
\includegraphics[width=0.30\textwidth,trim=0cm 0cm 3cm 0cm]{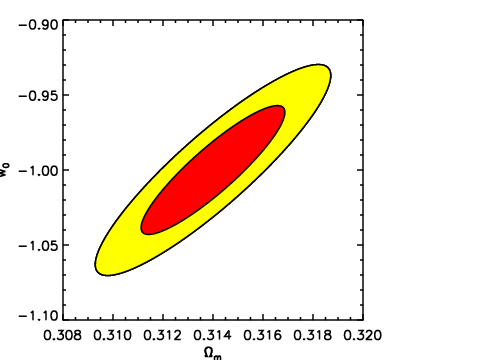}
\includegraphics[width=0.30\textwidth,trim=0cm 0cm 3cm 0cm]{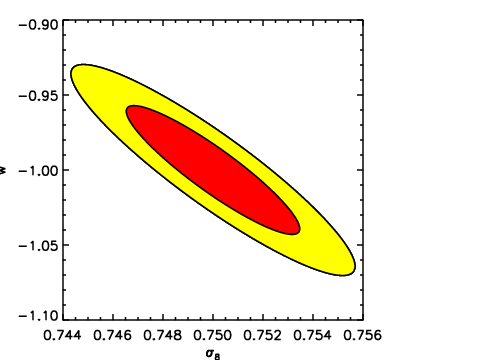}\\
\includegraphics[width=0.30\textwidth,trim=0cm 0cm 3cm 0cm]{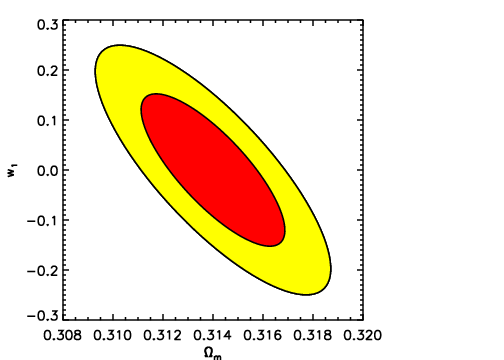}
\includegraphics[width=0.30\textwidth,trim=0cm 0cm 3cm 0cm]{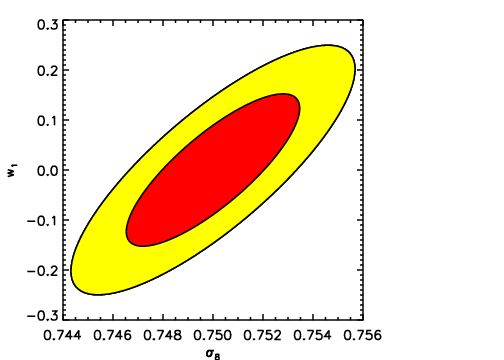}
\includegraphics[width=0.30\textwidth,trim=0cm 0cm 3cm 0cm]{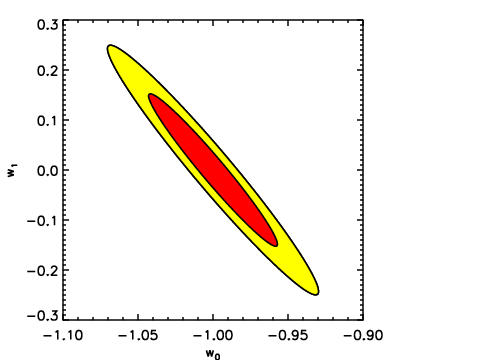}
\end{minipage}}
\caption{{\small Cluster constraints on a four parameter fit with $\Omega_{\rm m}$, $\sigma_8$, 
and dark energy equation-of-state parameters $w_0$ and $w_1$.  Note in the top left panel 
the tight constraints in the $\Omega_{\rm m}$--$\sigma_8$ plane, marginalized over $w_0$ and $w_1$, 
and in the bottom right panel the constraints in the $w_0$--$w_1$ plane, marginalized 
over $\Omega_{\rm m}$ and $\sigma_8$, corresponding to a dark energy FoM of 992.}}
\label{fig:cluster-constraints}
\end{center}
\end{figure}

\subsection{Cluster counts as a cosmological probe} 

As an example of the cosmological constraints that can be obtained from the expected cluster catalog, 
we performed a standard Fisher analysis with the results shown in Fig.~\ref{fig:cluster-constraints}.  The bottom
right panel, in particular, shows constraints on the dark energy equation-of-state parameters $w_0$ and $w_1$
for a four-parameter run, marginalized over $\Omega_{\rm  m}$ and $\sigma_8$ in a standard flat $\Lambda$CDM 
model.  The one-dimensional constraints on the dark energy parameters are $w_0=-1\pm 0.003$ and 
$w_1=0\pm 0.1$.  The Figure-of-Merit (FoM) quantifying the constraining power on the dark energy equation-of-state 
parameters is $({\rm Det} ({{\rm Cov}[w_0,w_1]}))^{-1/2} = 992$ from the cluster counts alone.

Structure evolution is sensitive to neutrino mass, which dampens growth at late times due to 
free-streaming and reduces the amplitude of matter perturbations as measured by $\sigma_8$.  Cluster counts are therefore a powerful probe of neutrino mass \citep{rozo2013}, given constraints from the primary CMB anisotropies.  For instance, changing the total neutrino mass from 0.057 to 0.096 eV (lower limits for normal and inverted hierarchy respectively) changes the value of $\sigma_8$ by 0.01, three to four times the standard deviation of the contour ellipses in Fig. \ref{fig:cluster-constraints}. Despite their simplicity, these examples illustrate the power of the \mission\ cluster catalog as a cosmological probe.

The above forecasts rely only on the cluster abundance data. However, \mission\ will also provide extensive data on the large-scale
distribution of detected clusters that can be exploited for correlation analysis
on large scales, or a large scale structure (LSS) density distribution reconstruction that has an
equivalent and complementary constraining power
for cosmological parameters, and hence will improve the overall result.
Here there may be a very unique application of the data: in some models, Dark Energy can have
LSS fluctuations on very large scales (i.e., on GpC scales, where they are 
not damped) \citep{1999ApJ...521L...1M}; \mission\ is likely to be the first experiment to provide the observations that can
probe the LSS with clusters on these very large scales.

\subsection{Cosmic velocity field} 

\mission\ will map the peculiar velocity field through the kinetic SZ effect \citep[kSZ,][]{sz1972,rephaeli1991,birkinshaw1999}, 
providing a new, independent probe of dark matter and large-scale structure evolution.  The mass limit to which we expect to measure a velocity of 300 km/s to $5\sigma$ on individual clusters is  about $2\times14\, \Msolar$ at $z<2$ (see blue line in Fig.~\ref{fig:clusterdetmasses}).  This mass limit implies that \mission\ will obtain velocity measurements for $\sim 2\times 10^5$ clusters out to the highest redshifts, and can further constrain velocity flows on large scales by stacking individual measurements below that detection threshold in cells of solid angle and redshift. 
Peculiar velocities directly trace the matter distribution.  Measurements of the evolution of cosmic velocities therefore
complement lensing measurements of the mass distribution; in fact, velocities provide much more local information 
on the matter distribution than lensing, which measures the mass projected along the line-of-sight over the broad 
redshift range of the lensing efficiency kernel.

While the relation between velocity and mass distributions is fixed in standard gravitational theory, the relation 
is often violated in modified gravity theories.  By comparing measured velocities to mass concentrations, for example 
from Euclid lensing, galaxy surveys (including dusty galaxies and CIB fluctuations detected by 
\mission---see section \ref{sec:extragalactic}), and CMB lensing correlated with mass tracers, 
we can test the theory of gravity on cosmic scales and to high redshift.  This science is unattainable by any other means, and \mission\  will fully exploit this untapped avenue \citep[e.g.,][]{hernandez2006}.

\subsection{Patchy reionization}

The kSZ effect also generates non-Gaussian anisotropy on the sky, with contributions from the epoch of reionization because of variations in the ionization fraction related to the patchiness of the process \citep{mcquinn2005}.  At $\ell \simeq 1000-2000$ the CMB power spectrum has a substantial contribution due to this signal.  \mission\ has the sensitivity to constrain the duration of reionization and the characteristic bubble size of the ionization regions, informing us about the nature of the first sources \citep{mcquinn2007}.  The small change in the small-scale CMB power spectrum caused by the patchy optical depth screen \citep{dvorkin2009, natarajan2012} should also be detectable.

\subsection{Relativistic and non-thermal effects}
 
Relativistic corrections to the thermal SZ spectrum depend on cluster temperature and optical depth 
\citep{rephaeli1995,challinor1998,itoh1998,nozawa1998,sazonov1998,birkinshaw1999,nozawa2006}.  They are small for low temperature clusters ($T < 4$\,keV) but become relevant for hot clusters. Wide spectral coverage combined with arcminute 
angular resolution enable \mission\ to map the temperature of clusters in an independent way from X-ray 
measurements.  As shown in Fig.~\ref{fig:clusterdetmasses}, we will determine the temperature of clusters 
down to a mass limit just above $10^{14}\Msolar$.  Since X-ray measurements depend on the square of the 
electron density and the SZ effect depends linearly on this density, differences in the X-ray and SZ determined
temperatures reveal interesting physics.  Furthermore, cluster temperature, like the integrated SZ signal, is 
a good cluster mass proxy.  Comparing the two with each other and also with lensing masses will deepen our 
understanding of cluster scaling relations, critical to cosmological studies with clusters.

Combined with the thermal SZ signal, the temperature measurement gives us a determination of the cluster 
\emph{gas} mass.  The is needed to calibrate the optical depth as a function of SZ signal, required to extract 
velocities from the kSZ signal.  Moreover, evolution of cluster gas mass is central to understanding the 
baryon cycle of cooling and feedback at the heart of galaxy formation  With \mission\ we can study gas mass in 
individual cluster systems to the highest redshifts where we detect clusters, thanks to the unique {\it distance 
independence} of SZ signals.

In addition to the thermal relativistic effect, non-thermal processes leave their imprint on the spectrum 
of the SZ effect.  These could signal the presence of highly energetic particles, dark matter annihilation 
products, AGN outflows, intracluster cavities and shocks \citep{cola2003,cola2009, 
cola2011,cola2004,cola2005,cola2008}.  We may even study the temperature structure of the most massive 
systems \citep{cola2010,debernardis2012} and derive the electron distribution function \citep{prokhorov2011}.

\subsection{Diffuse SZ and the cosmic web} 

The diffuse, unresolved SZ effect probes a different mass and redshift range than observations of 
individually detected objects.  PRISM will study this diffuse effect through the power spectrum and higher order 
moments of an SZ map of the sky, such as shown in Fig.~\ref{fig:tSZ-recovery}.  \Planck\ recently extracted the 
first all-sky Compton parameter ($y$-fluctuation) map \citep{planck2013-XXI}, but the results are limited by 
foregrounds and noise. With many more spectral bands and much better sensitivity and resolution, \mission\ will 
significantly improve the results.  With its ability to isolate and subtract detected clusters over a vast range of 
mass and redshift, \mission\ can study the diffuse gas in the remaining, smaller halos, and even 
within the cosmic web itself.

Apart from this statistical measure, we will also be able to directly map the gas content of the cosmic web 
in the larger filaments.  A significant fraction of the baryons of the Universe are expected to reside in 
the warm-hot intergalactic medium (WHIM) and remain undetected at low redshifts. The WHIM phase, 
with over-densities between 5 to 200 times the critical density and temperatures around $10^5$--$10^7$\,K, 
is expected to reside mostly in filaments, but also around and between massive clusters 
\cite{cen1999,cen2006}. The SZ thermal pressure contribution of this cosmic web of gas 
\cite{dolag2006}
is potentially detectable with \mission, either by direct detection of the bridge 
of inter-cluster matter between galaxy clusters \citep[e.g.,][]{pip8}, or even the individual filaments 
around the brightest clusters.  This provides important information on the physical state of the baryons 
in this little understood, moderately dense region of the structure hierarchy.

We will explore the gas content of dark matter halos down to very low masses, a research area pioneered by 
\Planck\ by stacking SZ measurements based on known objects to detect the signal down below $10^{13}$ solar masses 
\citep{planck2011-pepXII,planck2012-pipXI}.  This measurement over such a vast mass range is unique to the SZ 
effect and a highly valuable constraint on feedback mechanisms at the heart of galaxy 
formation.  \mission\ greatly expands this important science area by pushing to the lowest possible masses and
accessing a broad redshift range.  We will probe the gas content and its evolution in halos hosting different types 
of objects and as a function these object properties.  Combined with \mission's lensing measurements, we will wield a 
fundamentally new and exceptionally powerful tool to study the relation between luminous and dark matter. 

\subsection{Polarized SZ effect} 

The polarized SZ signal is generated by any quadrupole
CMB anisotropy seen at the position of a cluster, such as created by cluster peculiar velocity or the primordial 
CMB quadrupole itself \citep{ramos2012}.  Measurements of the polarized SZ signal therefore give access to the transverse component of 
cluster velocities and a way to estimate the CMB quadrupole at distant locations.

By stacking measurements of the polarized SZ signal from \mission, we will attempt to estimate the CMB quadrupole 
at different positions within the Hubble volume.  This will allow us to reduce the cosmic variance on this multipole that
spans the largest cosmic scales.  Averaging the signal of thousands of clusters at high redshift in opposite 
directions of the sky will dampen the effect from their peculiar velocity component, leaving the component that is 
proportional to the intrinsic CMB quadrupole. Such an estimate of the quadrupole as seen by 
objects at high-z would help understand why the low quadrupole seen by \Planck\ and {\it WMAP} seems to be at odds 
(marginally) with the standard model. Kamionkowski \& Loeb (1997), Sazonov \& Sunyaev (1999) and others show 
that the expected polarization signal is at the level of $10^{-2}$ $\mu$K for typical clusters, which could be 
achieved by an experiment with sub-$\muK$ after averaging over thousands of clusters.

X-ray observations from {\it Chandra} reveal the presence of cold fronts associated with 
fast moving (supersonic) bulk motions of the gas in the core region. These bulk motions leave an imprint in the kinetic 
SZ (proportional to the radial component of the velocity) and  also on the polarization pattern of scattered 
CMB photons (proportional to the square of the tangential component of the velocity).  As shown by Diego, 
Mazzotta, \& Silk (2003), in a typical cluster one expects kSZ distortions of order 50 $\mu$K and 
polarization intensities of 0.3 $\mu$K associated with these bulk gas motions.  Although 50 
$\mu$K should be easily detected by \mission, the polarization will prove more challenging; however, in extreme 
cases like the {\it bullet} cluster, where the velocity is significantly higher, the polarization could be 
boosted to the level of 1 $\mu$K (due to the squared dependence of the polarization intensity with the 
velocity) and be detected by \mission. Combining the kSZ (radial component) with the polarized SZ 
(tangential) would allow a determination of the three-dimensional direction of the velocity, opening the door 
to novel dynamical studies of galaxy clusters.


\section{Extragalactic sources and the cosmic infrared background }
\label{sec:extragalactic}
\def\lsim{\,\lower2truept\hbox{${<\atop\hbox{\raise4truept\hbox{$\sim$}}}$}\,}
\def\gsim{\,\lower2truept\hbox{${> \atop\hbox{\raise4truept\hbox{$\sim$}}}$}\,}

\subsection{Early evolution of galaxies} 

Although \Herschel\ and \Spitzer\ produced spectacular advances in our understanding of early, dust enshrouded phases of galaxy evolution, our knowledge of star-formation history in the distant universe is still largely incomplete. The {\mission} mission will yield essential progress thanks to its unique properties: full sky coverage and unparalleled frequency range.

\begin{figure}[thb]
\begin{center}
	\includegraphics[trim=1.5cm 0cm 0.5cm 0cm, clip=true, width=0.49\textwidth]{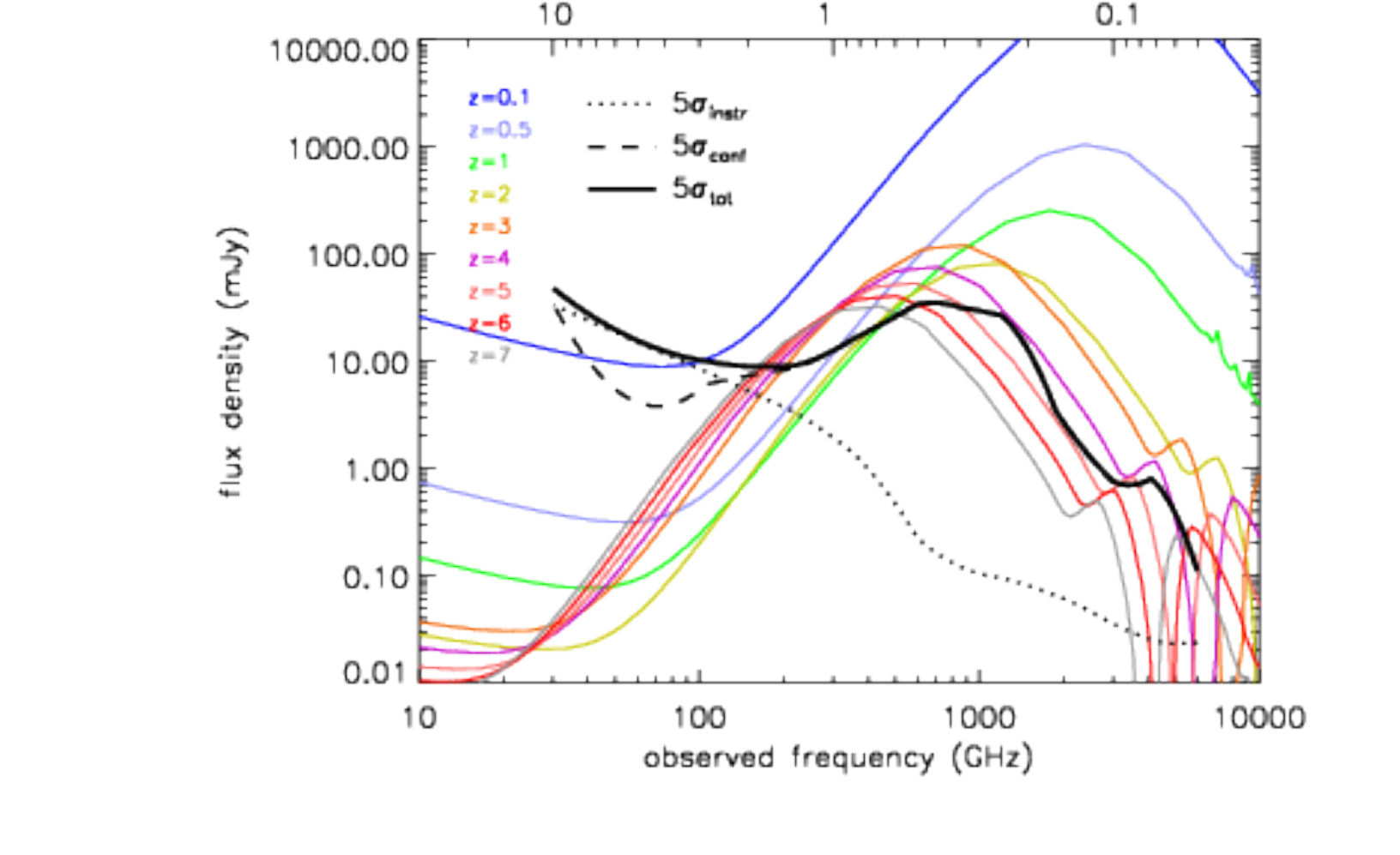}
	\includegraphics[trim=1.5cm 0cm 0.5cm 0cm, clip=true, width=0.49\textwidth]{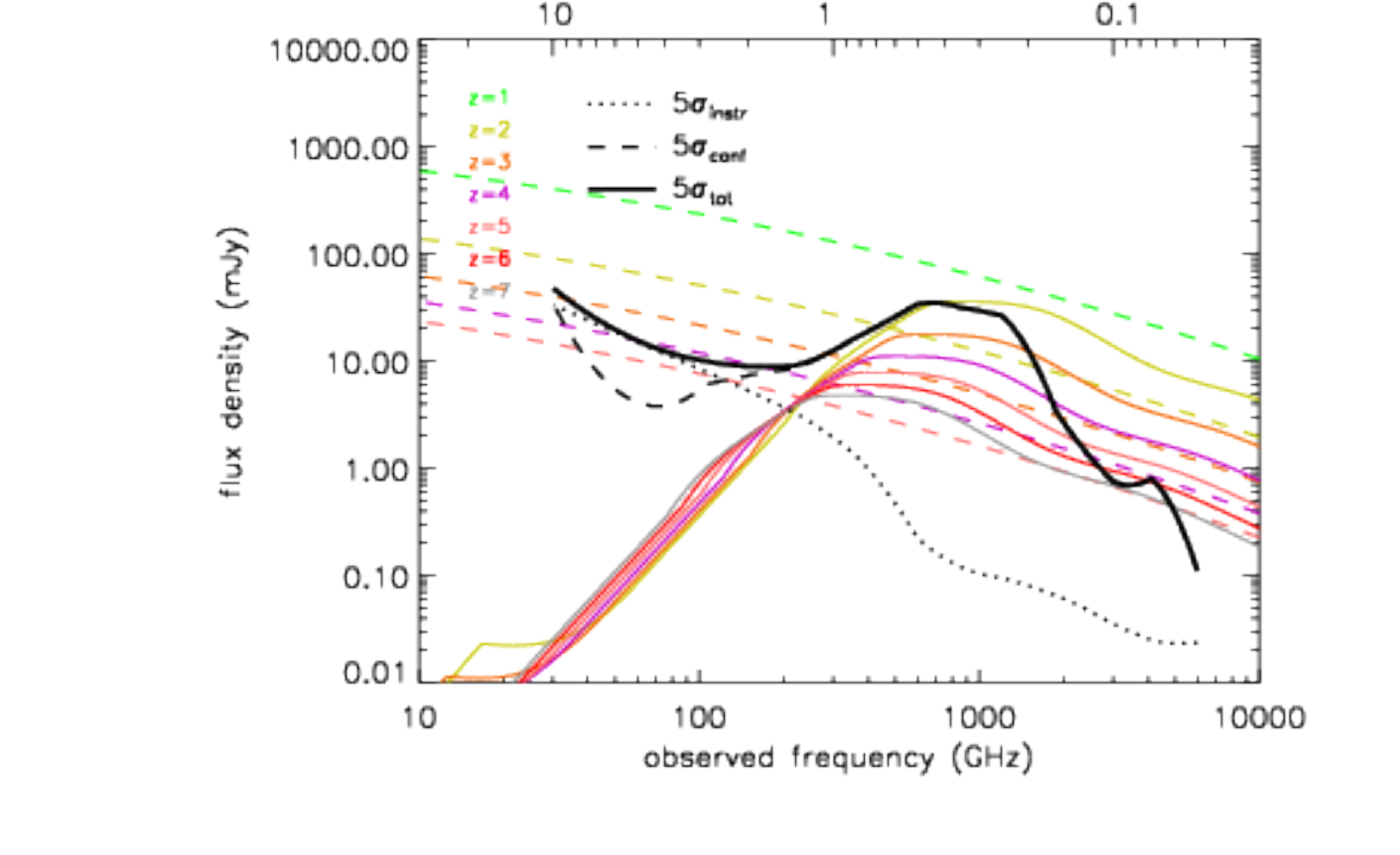}
\caption{\small
SEDs of dusty galaxies (left panel) and of AGNs (right panel) at different redshifts compared with
estimated $5\sigma$ detection limits (solid black line) taking into account instrumental and
confusion noise summed in quadrature. The instrumental noise refers to the full mission, in 1 arcmin. 
pixels. The
$5\sigma$ detection limits allowing for either component are shown by the dotted and the dashed
black lines, showing that {\mission} is confusion limited above $\approx 150\,$GHz. We have assumed
that component separation techniques, extensively validated both on simulations and on real data, can
efficiently remove diffuse emissions such as the CMB (that would otherwise dominate the fluctuation
field for $\nu\lsim 220\,$GHz) and Galactic emissions. In the left panel, at $z=0.1$ and 0.5 we have
plotted the Arp\,220 SED scaled to an IR (8--$1000\,\mu$m) luminosity of $10^{12}\,L_\odot$. At $z\ge
1$ we have used the SED of the $z\approx 2.3$ galaxy SMM~J2135-0102 scaled to $L_{\rm
IR}=10^{13}\,L_\odot$ for $z=1$ and $z=2$, and to $L_{\rm IR}=3\cdot 10^{13}\,L_\odot$ \citep[the
luminosity of the $z=6.34$ galaxy detected by \Herschel/SPIRE,][]{Riechers2013} for $z\ge 3$. In the
right panel, the solid colored lines represent SEDs of a type-2 QSO (contribution of the host-galaxy
subtracted) with $L_{\rm IR}=10^{13}\,L_\odot$ at several redshifts $\ge 2$, while the dashed colored
lines show a schematic representation of the SED of the prototype blazar 3C\,273 shifted to redshifts
from 1 to 5. }\label{fig:SED}
\end{center}
\end{figure}

As illustrated in Fig.~\ref{fig:SED}, {\it PRISM}'s unprecedented frequency coverage provides direct measurements of the bolometric luminosities of star-forming galaxies from $z$ of a few tenths up to $z\approx 6$--7. At $z\gsim 2$, i.e. in the redshift range where both the cosmic star formation and the accretion rate onto supermassive black-holes are maximum, both the IR peak associated to the dusty torus around the AGN ($\lambda_{\rm p,AGN}\approx 30\times(1+z)\,\mu$m) and the peak of dust emission in the host galaxy are within the covered range. 

Moreover, measurements of the complete far-IR to mm-wave SEDs will vastly improve the accuracy of photometric redshift estimates that have a rms error of $\approx 0.2(1+z)$ with SPIRE alone \citep{Lapi2011}. This means that the {\mission} survey will allow us to characterize with high statistical accuracy the evolution with redshift of the bolometric luminosity function, bypassing the need of applying uncertain bolometric corrections.

At $z\gsim 2$ analyses of the full SEDs will return information on the evolution of the relationships between star-formation and nuclear activity: What fraction of the bolometric energy radiated by star-forming galaxies is produced by accretion onto supermassive black holes in active galactic nuclei (AGN)? What are the evolution properties of far-IR selected AGNs? What fraction of them is associated to active star formation? Are the growth of 
central super-massive black hole and the build-up of stellar populations coeval, or does the AGN evolution occur 
earlier as indicated by some recent data \citep{2011ApJ...742..107B}? The substantially higher spatial resolution (thanks to the shorter 
wavelength channels) and the correspondingly higher positional accuracy compared to \Herschel/SPIRE%
\footnote{Note that deep \Herschel/PACS surveys have covered only a few $\hbox{deg}^2$. The large area {\Herschel} surveys have 
taken PACS data in parallel mode and have shallow detection limits.} will greatly improve the identification of reliable 
counterparts in other wavebands (X-ray, UV, optical, near-infrared, radio), necessary for a comprehensive understanding of 
the properties of detected galaxies.

Millimeter-wave observations of low-$z$ dusty galaxies will measure their free-free emission, that remained elusive so far. For example only for three galaxies detected by \Planck\  this emission was clearly identified \citep{Peel2011}. {\mission} will increase the sample to a few hundred galaxies. The free-free emission is a direct measure of the star-formation rate (SFR), unaffected by extinction and not contaminated by the effect of old stellar populations that can contribute to dust heating and, therefore, to the far-IR emission. A comparison of the estimates of the SFR coming from the free-free emission with those from the far-IR will disentangle the contributions to the latter from young stars and from the evolved stellar population, thus providing insight into the intensity of the interstellar radiation field and on the heating of the interstellar dust.

Its all-sky coverage makes {\mission} uniquely suited to study rare phenomena. Examples are the `maximum starburst' galaxy at $z=6.34$ detected by \Herschel/SPIRE \citep{Riechers2013}  or the hyper-luminous IR galaxies (HyLIRGs) such as the binary system, pinpointing a cluster of star-bursting proto-ellipticals at $z=2.41$, discovered by \citep{Ivison2013}. The estimated surface density of the former objects is of $\sim 0.01\,\hbox{deg}^{-2}$, that of binary HyLIRGs of 3--$9\times 10^{-2}\,\hbox{deg}^{-2}$, implying that {\mission} will detect from several hundreds to thousands of such objects. The $z=6.34$ galaxy was found when looking for {\it ultra-red} sources with flux densities $S_{250\mu\rm m} < S_{350\mu\rm m} < S_{500\mu\rm m}$. The {\mission} survey will allow us to look for even redder sources, potentially at even higher redshifts, and will provide a test of our understanding of the interstellar medium and of the star-formation under extreme conditions.

Strongly gravitationally lensed systems have long been very difficult to identify in sufficiently large numbers to be statistically useful. This situation changed drastically with the advent of (sub-)mm surveys. One of the most exciting \textit{Herschel}/SPIRE results was the direct observational confirmation that almost all the galaxies brighter than $\approx 100\,$mJy at $500\,\mu$m are either strongly lensed or easily identifiable low-$z$ spirals plus a small fraction of flat-spectrum radio quasars  \citep{Negrello2010}. The surface density of high-$z$ strongly lensed galaxies above this limit is $\approx 0.3\,\hbox{deg}^{-2}$, implying that an all-sky survey can detect $\sim 10^4$ such systems. The typical magnification is of a factor $\mu \sim 10$. The fact that these sources are very bright makes redshift measurements with CO spectrometers and high resolution imaging with millimeter interferometers (ALMA, NOEMA, SMA, ...) relatively easy. This will allow us to get detailed information on obscured star formation in the early Universe and on processes driving it, in observing times a factor $\approx \mu^2$ shorter that would be possible without the help of gravitational amplification and with effective source-plane resolution $\approx \sqrt{\mu}$ higher than can otherwise be achieved.

Optical spectroscopy of galaxies acting as lenses can be exploited to measure the mass distribution of their dark matter halos as a function of redshift. Note that {\Euclid} will directly provide redshifts for the majority of the lenses out to $z\sim 1$ in its area. Taking into account the large number of newly identified strongly lensed galaxies, this will allow a direct test on the evolution of large-scale structure by measuring the mass distribution of dark matter halos as a function of redshift.

Samples of thousands strongly lensed galaxies are also essential for many other astrophysical and cosmological applications \citep{Treu2010}. Further constraints on the evolution of large scale structure, hence on cosmological parameters that drive it, can be obtained from the cross-correlation between background sub-mm galaxies and foreground optical or IR galaxies induced by weak lensing \citep{2011MNRAS.414..596W,Hildebrandt2013}.

Large numbers of strongly lensed galaxies are also expected from large area optical surveys. It should be noted, however, that sub-mm selection has important distinctive properties. The selected lensed galaxies are very faint in the optical, while most foreground lenses are passive ellipticals, essentially invisible at sub-mm wavelengths so that there is no, or little, contamination between
images of the source and of the lens. This makes possible the imaging of lensing events with small impact parameters. Also, compared to the optical selection, the (sub-)mm selection allows us to probe earlier phases of galaxy evolution.

{\mission} will study the angular correlation function of detected sources with much better statistics than was possible with \Herschel 's extragalactic surveys that, altogether, cover little more than 2\% of the sky. Also, the accurate photometric redshifts will allow us to follow the evolution of clustering with cosmic time. Clustering properties measure the mass of dark matter halos associated to galaxies and are a powerful discriminant for galaxy formation and evolution models \citep{Xia2012}.  In particular this study will allow us to detect high-$z$ {\it proto-clusters} of dusty galaxies. We thus investigate an earlier evolutionary phase of the most massive virialized structures in the Universe. This science is possible only in the wavebands covered by {\mission}.

The {\mission} clustering data will extend to much higher redshift than {\Euclid}, whose wide-area survey will accurately map the galaxy distribution up to $z\sim 1$. The {\mission} data will provide information primarily over the redshift range $2<z<3$, corresponding to the peak in star formation activity. Moreover, optical and near-IR data severely underestimate the SFR of dust obscured starbursts and may entirely miss these objects, which are the main targets of far-IR/sub-mm surveys such as {\mission}. {Only the combination of {\mission} and {\Euclid} data will provide a complete view of the spatial distribution of galaxies and of how star formation is distributed among dark matter halos.}

\subsection{The cosmic infrared background}

The {\mission} sensitivity and spectral coverage will provide substantially improved measurements of the cosmic infrared background (CIB) spectrum with an accurate removal of all contaminating signals. The CIB mean intensity around its peak at $\lambda_p\sim 160\,\mu$m is still quite uncertain \citep{2005ARA&A..43..727L}. Its determination is obviously crucial to determine the cosmic energy budget and to assess to what extent it has been resolved by existing surveys.

{\mission} will also measure, in a uniform way, the CIB power spectrum over an unprecedented range of frequencies and of angular scales (from $\sim 10\,$arcsec to tens of degrees). {\Herschel} and {\Planck} surveys have provided measurements at $\lambda\ge 250\,\mu$m \citep{PlanckCollaborationXVIII2011,Viero2012} which, while generally consistent, show significant differences. Estimates of the power spectrum at 100 and $160\,\mu$m, from \textit{Spitzer\/}/MIPS data \citep{Penin2012} are far more uncertain. The much better statistics due to the much larger area covered by {\mission} compared to \Herschel\ and to the much better sensitivity and angular resolution compared to \Planck\  will allow us to obtain a highly accurate assessment of the CIB power spectrum and of its frequency dependence.

On large angular scales the CIB power spectrum measures the linear clustering bias which is a strong function of halo mass. On small scales the clustering properties are determined by the non-linear evolution and probe the processes that drive the formation of multiple galaxies within large dark matter halos. Specifically, the fluctuation level at multipoles $\ell\sim 1000$ at far-IR/sub-mm wavelengths mostly probes the $z<1$ Universe and is very sensitive to the 1-halo term, thus to the effect of environment on star formation \citep{2013A&A...557A..66B}.

The composition and the redshift distribution of faint sources (below the detection limit) contributing to the CIB varies with frequency. The characteristic redshift of sources of the CIB systematically increases with increasing wavelength, from far-IR to mm-waves. This change is related to the origin of the early/late-type galaxy dichotomy, since present-day spheroidal galaxies have been forming most of their stars at $z\gsim 1.5$ while stellar populations of late type galaxies are substantially younger and indeed these galaxies have been continuously forming stars up to the present time. The observed \textit{downsizing} in galaxy evolution implies that the change with frequency of the dominant galaxy population translates in a variation of the clustering amplitude.  Thus the frequency dependence of the CIB power spectrum also contains information on the origin of the early/late-type galaxy dichotomy and on its relationship with halo mass.

\subsection{Radio sources}

{\mission} will extend the counts of radio sources, both in total and in polarized intensity, by at least one order of magnitude downwards in flux density compared to \Planck. Above 217 GHz, the counts will be determined for the first time over a substantial flux density range with good statistics\footnote{Sources whose sub-mm emission is synchrotron dominated are rare. The sky coverage of \Herschel\ surveys was insufficient to provide samples large enough to determine their counts.}. This will make possible the first investigation of the evolutionary properties of radio sources at (sub-)mm wavelengths.

{\mission} will provide measurements of the spectral energy distribution (SED) of many thousands of radio sources and of multifrequency polarization properties for hundreds of them. The vast majority of these sources are expected to be blazars, and the accurate determination of their spectra will allow us to understand how physical processes occurring along relativistic jets shape the SED \citep{PlanckCollaborationXV2011}. We will also get numerous samples of `extreme' radio sources \citep{PlanckCollaborationXIV2011} allowing us to investigate the rich phenomenology of radio sources at (sub-)mm wavelengths. 

A few hundred steep-spectrum radio sources will also be detected.  The multi-frequency measurements will provide the distribution of break frequencies due to electron aging, allowing an unbiased estimate of the distribution of radio source ages. 

These observations will also shed light on the relationship between nuclear radio emission and star formation activity in the host galaxies. In particular, the present notion is that the blazar hosts are giant passive ellipticals, but already \Planck\  has provided examples of catalogued blazars that are hosted by galaxies with strong star-formation activity. This mission will allow us to quantify the abundance of hosts of this kind and to investigate relationships between blazar and host properties.

\subsection{Extragalactic sources and the integrated Sachs-Wolfe effect}

The cross-correlation of radiosource data (NVSS) and CMB maps has bean already probed as a fruitful way to detect the integrated Sachs-Wolfe (ISW) signal. Since the redshift distribution of sub-mm selected galaxies peaks at $z=2$--3, it will be possible, combining the extragalactic IR source catalogue from the survey data with \mission\ CMB maps, to measure the Integrated Sachs-Wolfe (ISW) effect out to substantially higher redshift than can be done with \Planck\  $+$ large area optical surveys (limited to $z\lsim 1$). At these high redshifts the ISW is insensitive to a cosmological constant and will probe alternative explanations of the cosmic acceleration. Since the ISW effect shows up on large angular scales, the all-sky coverage of {\mission} will allow substantially better statistics than any optical survey. This will be complemented by a similar correlation of \mission\ CMB maps with a large catalogue of HI galaxies detected with SKA. In about 1-yr, a SKA survey will observe more than $10^{9}(f_{sky}/0.5)$ HI galaxies in a redshift range $0 < z < 1.5$, offering yet another view on the impact of dark energy on the dynamics of the large scale structures in our Universe.


\section{Inflation and CMB primordial B-modes}
\label{sec:CMB-B-modes}

\def\planck{\textit{Planck}}

At the heart of our current Standard Model of the Universe is a set of 
initial conditions laid down at very early times by what is known as 
{\it cosmic inflation}. During inflation, the Universe undergoes a 
period of ultra-rapid accelerated expansion. The simplest model involves 
a slowly evolving fundamental scalar $\phi $, much like the Higgs field, with a 
potential energy $V(\phi)$ that dominates over its kinetic energy. 
The energy density of such a scalar field is well approximated by 
$\rho=V$ while its pressure is $p=-V$. As such, it drives the scale 
factor of the Universe to evolve approximately as 
$a(t)\propto \exp(\sqrt{8\pi G V/3}t)$. 

During this period of inflation, quantum fluctuations of spacetime 
and the scalar field $\phi$ are amplified and stretched to 
macroscopic scales. The result is a quasi-Gaussian stochastic distribution of 
density perturbations with an amplitude $A_S,$ which depends on scale, 
or wavenumber $k$. It is convenient to write the dependence of the 
{\it scalar spectral index} as $n_S=1+d\ln A^2_S(k)/d\ln k$. In fact, 
the theory predicts that $A_S$ and $n_S$ will depend on the details 
of $V$ and how itself depends on $\phi$. Furthermore, interactions of $\phi$ with itself and with 
other fields induce cross-correlations between perturbation modes, leading to non-Gaussianity 
which can be detected in higher order statistics (bispectrum, trispectrum).
Inflation will also generate 
a bath of primordial gravitational waves (tensor perturbations of the 
spacetime), 
which can also be characterized by an amplitude $A_T$ and the 
{\it tensor spectral index} $n_T=d\ln A^2_T(k)/d\ln k$ . Remarkably, 
in the simplest models of inflation, the ratio between the tensor and 
scalar perturbations $r$ is a direct probe of $V$ in the early universe: 
$r\equiv (T/S)=16(A_T/A_S)^2\approx M^2_{Pl}(V'/V)^2$.  
From present observations we have that $V^{1/4}=3.3\times 10^{16} r^{1/4}$ 
GeV, so that measuring $r$ effectively translates into a measurement 
of the energy scale of inflation.

Clearly, a measurement of $r$, $n_S,$ and possibly even $n_T,$ 
can directly probe the physics of the early universe for which 
there is a very rich phenomenology. Single field inflation models 
can relate $r$ directly with the evolution of $\phi$ at early times. 
Indeed, for an inflationary expansion lasting long enough to 
provide the observed level of homogeneity and isotropy, we have that $\Delta \phi/M_{\rm Pl}\approx (r/0.01)^{1/2}$. 
Multiple field inflation models can arise in string theory and 
other proposals for unification at high energies and can lead to 
even higher values of $r$. In addition, particle and string 
production during the inflationary period can be an additional source 
of copious gravitational waves  which in some cases will dominate over 
the usual mechanism. Moreover, very recent models propose that the 
Electroweak Higgs could also be the inflaton if we allow for nonminimal 
couplings to gravity. In this case, there is a direct connection between 
early universe phenomena like inflation with late universe dynamics 
like dark energy, and future surveys like Euclid could help us discriminate 
among alternatives.

Over the past few years, the techniques of Effective Field Theory have
been applied to the inflationary era allowing a systematic analysis of
a broad swathe of theory space. With these techniques it is possible to
use constraints on $A_S$, $A_T$, $n_S$, $n_T$ as well as
non-Gaussianity parameters such as $f_{NL}$ and $g_{NL}$ to place
stringent constraints on the types of couplings that are allowed at
ultra-high energies. By sifting through these couplings it is then possible
to make profound statements about the type of unified theory which
should be valid at Planck scales (such as for example Supergravity)
and what ingredients should be considered in a quest for ultra-violet
completion of our current fundamental theory of particles and
fields. Hence, beyond just constraining individual theories, 
measurements of these quantities can play a crucial role
in helping us in the age old quest for a theory of unification.
 
Primordial gravitational waves imprint a unique, as yet undetected, signature in the CMB 
polarization. CMB polarization is a spin-two field on the sky, and is decomposed into the 
equivalent of a gradient---the E-mode---and a curl---the B-mode. Gravitational wave fluctuations 
are visible as the B-mode polarization of the CMB and are the only primordial contribution to 
B relevant at the time of recombination. Hence a detection of B-modes is a direct probe of 
$r,$ and thus the energy scale of inflation and other primordial energetic processes. 
Furthermore, in the simple case of slow-roll inflation we have that $r\approx-8n_T$. 
Additional detailed measurements of the shape of the temperature and polarization spectra 
will measure higher derivatives of the inflationary potential.

The 2013 \planck\ data release has significantly improved previous constraints on 
inflationary models.  In the simplest $\Lambda$CDM scenario, 
where no tensor perturbations are included, \planck\  
results provide $n_s=0.9603\pm0.0073$. 
When tensor modes are included, the same analysis yields $n_s=0.9624\pm0.0075$ 
and $r<0.12$ using a pivot scale of 0.002\,Mpc$^{-1}$.These results are notable 
because
exact scale invariance (i.e., $n_S=1$) of primordial perturbations is ruled out 
at more than $5\,\sigma$, a result which still holds when allowing for a variation in primordial 
Helium abundance and effective neutrino number.
 
\begin{figure}
\begin{center}
\includegraphics[trim=0cm 0cm 0cm 0cm, clip=true,width=15cm]{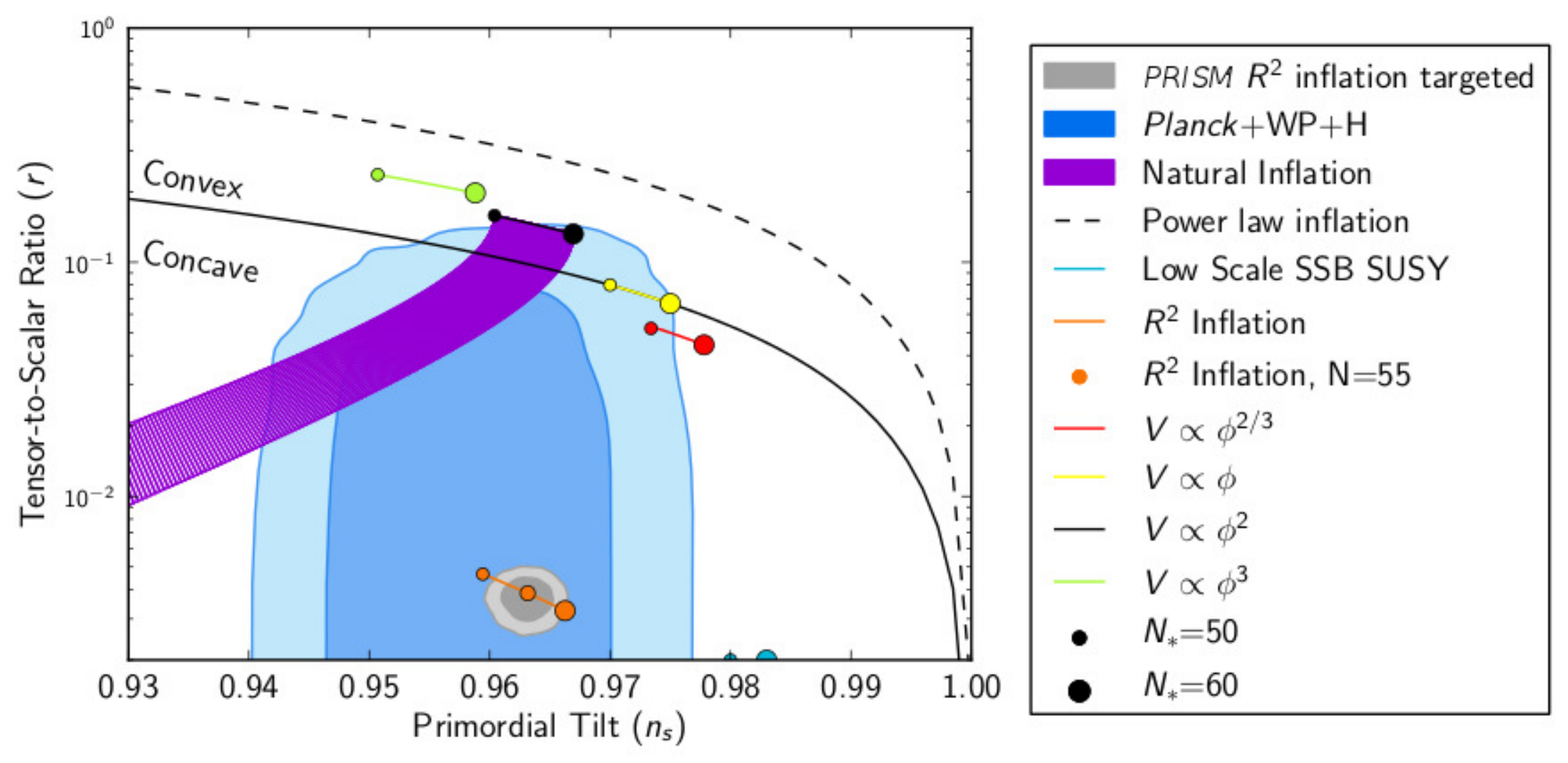}
\caption{\small Constraints on inflationary potentials from \planck\ and the predicted constraints from {\mission} (not assuming de-lensing) for a fiducial value of $r=5\times 10^{-2}$ (adapted from \citep{2013arXiv1303.5082P}).
}
\label{fig:Bmodes}
\end{center}
\end{figure}

 When specific inflationary models are considered, \planck\ imposes 
significant constraints on the potential (Fig.~\ref{fig:Bmodes}), as discussed in 
Ref.~\citep{2013arXiv1303.5082P}. Convex potentials are disfavored by 
\planck\  data. Chaotic inflation with $V\propto\phi^2$ 
is excluded at 95\% C.L. for $N_*=50$ and only marginally compatible 
for $N_*=60$, $N_*$ being the number of inflation e-foldings. In general, 
power-law potentials $V\propto\phi^n$ for $n>2$ are ruled out. 
The spontaneously broken SUSY model
is disfavored, unless some assumptions on reheating are relaxed. 
\planck\  does not have, in general, the power to discriminate 
among different concave potential shapes. 
Inflationary models based on extended theories of gravity are 
also constrained. Remarkably, the original $R^2$ inflation model 
proposed by Starobinsky 
predicts $n_s=0.963$ and $r\approx4\times 10^{-3}$ for $N_*=55$, 
and is thus fully compatible with \planck\  observations. 
Non-minimally coupled Higgs inflation has predictions similar 
to the $R^2$ scenario, and is also well accommodated by \planck .

The \planck\ results have also led to stringent constrains, via the
Effective Field Theory methods, on the broad type of theories which can be
at play in the very early universe. In particular couplings between
the inflaton and any hidden sector are supressed by of order $10^{-2}$
relative to the energy scale of inflation. With trilinear couplings,
the supression can be as large as $10^{-5},$ which for most models
of inflation is at the Planck energy scale. This places extremely tight
constraint models of string compactification in the supergravity
regime, a prime model for unification. Furthermore these constraints
begin to enter the quantum gravity regime, opening up a completely new
phenomenological window on the Universe. A detection of tensor modes
would greatly sharpen these constraints.

 In summary \planck\ has shown that it is possible to test many 
inflation models using the CMB temperature data, yet even a forecast \planck\ limit $r < 
0.05$ would leave many interesting models unprobed. Given that the stochastic background of 
gravity waves is the smoking gun of inflation, it is crucial to map as accurately as 
possible the CMB polarization and in particular characterize the B-mode angular power 
spectrum.

\begin{figure}
\begin{center}
\includegraphics[trim=0cm 0cm 5cm 0cm, clip=true,width=10cm]{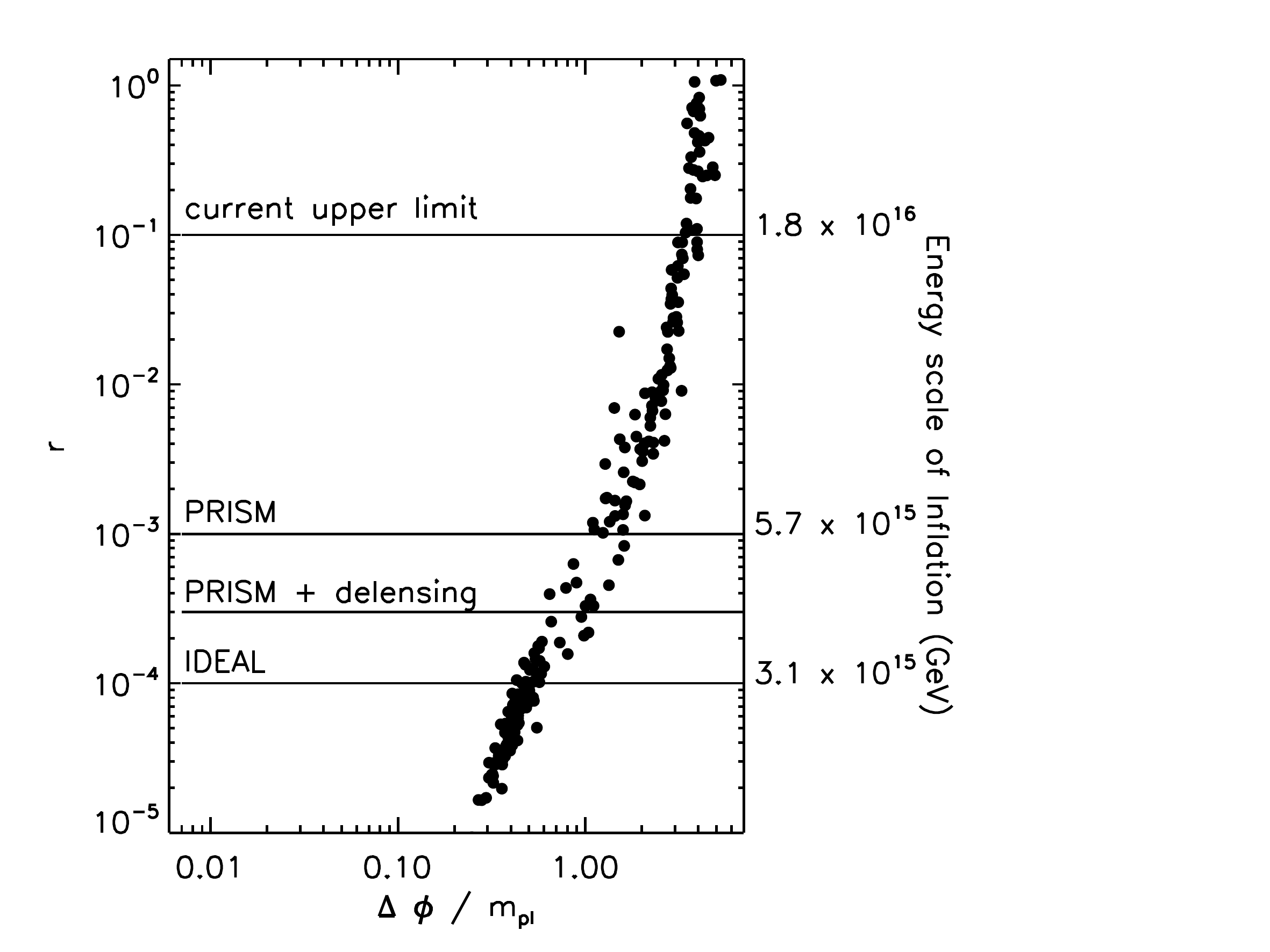}
\caption{\small Distribution of inflationary model parameters generated using a model independent approach that Monte-Carlo samples the  inflationary flow equations. 
While these simulations cannot be interpreted in a statistical way (e.g., \citep{19, 26, 27}), they show that models cluster around attractor regions (adapted from \citep{ultimatepol}).
}
\label{fig:Bmodes-2}
\end{center}
\end{figure}

To forecast how well we would be able to measure the power spectrum of the B-modes, it is 
important to recognize that the foreground signal is likely to dominate the cosmological 
signal at low $\ell$, where the most constraining information on $r$ is situated.
If we propagate the uncertainties connected to foreground contamination into the parameter error forecasts \citep{ultimatepol,baumanncmbpol,coreWP}, we find that
the proposed experimental set-up will enable us to  explore most large field (single field) inflation models (i.e., where the field moves for $\ge $M$_P$) and to  rule in or out all large-field models, as illustrated in Fig.~\ref{fig:Bmodes-2}. 

As the work by \citep{smith_etal_2008} indicates, the instrumental sensitivity, 
angular resolution and, as a result, foreground control and subtraction will enable us to 
achieve a detailed mapping of the lensing signal, and in particular to implement de-lensing 
techniques for the measurement of $r$, improving by a factor of three our constraint on $r$. 
This implies that {\mission} will detect $r\sim 3\times 10^{-4}$ at more than 3$\sigma$. This 
performance is very close, within factors ${\cal O}(1)$, to what an ideal experiment (i.e., 
with no noise and no foregrounds) could achieve, allowing {\mission} to {\it directly} probe 
physics at an energy scale a staggering twelve orders of magnitude higher than the 
center-of-mass energy at the Large Hadron Collider (LHC).

\begin{figure}
\begin{center}
\includegraphics[width=10cm]{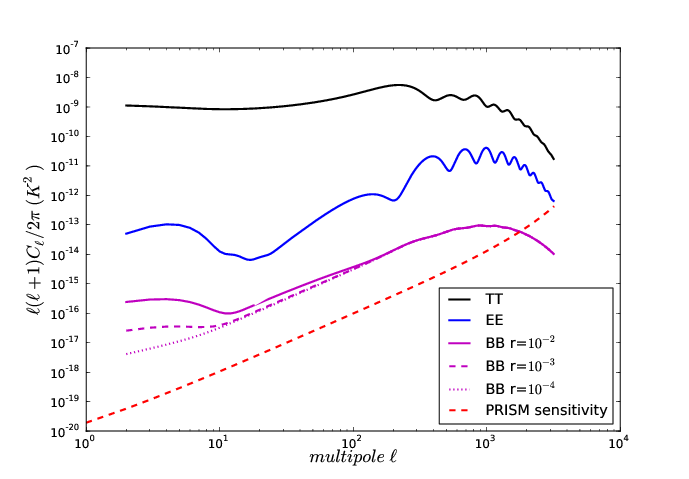}
\caption{\small Expected noise level for PRISM observations in individual CMB multipoles.}
\label{fig:CMB-Cl-sensitivity}
\end{center}
\end{figure}

Figure \ref{fig:CMB-Cl-sensitivity} illustrates the expected noise level of the PRISM CMB observations, in comparison with the 
CMB temperature and polarization power spectra.


\section{CMB at high resolution}
\label{sec:CMB-highres}

The temperature anisotropies of the CMB have proved to be a remarkably clean probe of the 
high-redshift universe and have allowed cosmological models to be tested to high precision. 
However, the accuracy of the recent results from \Planck~\cite{2013arXiv1303.5076P}, based on the 
temperature anisotropies, are now close to being limited by errors in modelling 
extragalactic foregrounds. Fortunately, further progress can be made with the polarization 
anisotropies on small angular scales since the degree of polarization of the anisotropies 
is relatively larger there (around 4\% by $\ell =2000$) than the foreground emission. This, 
combined with the improved resolution and sensitivity of \mission\ over \Planck\ and the 
exquisite frequency coverage for removal of foregrounds, will allow us to extend the 
current temperature analyses to smaller angular scales in polarization. We discuss the main 
science enabled by such observations -- full exploitation of weak gravitational lensing 
effects in the CMB, improved constraints on early-universe physics from CMB 
non-Gaussianity, precision tests of extensions to the standard cosmological model from the 
polarization power spectra in the damping tail, and powerful searches for cosmic strings 
via the small-scale $B$-mode power spectrum -- in the following subsections.

The science case for high-resolution imaging of the CMB is well recognised, and indeed is a 
major focus of the polarization cameras now operating on the SPT and ACT. While future 
upgrades to these experiments may cover several $1000\,\mathrm{deg}^2$ of sky with 
sensitivity approaching that of \mission, only a space mission will be able to cover the 
full sky in many frequency bands. For the science described in this section, the main 
benefits of a space mission are that it provides uniform calibration over the full sky 
(important, for example, for local-model non-Gaussianity searches where one is looking for 
large-scale modulation of small-scale power), excellent foreground rejection, long-term 
stability and control of systematic effects, and improved sample variance due to near 
full-sky coverage.

\subsection{Probing the dark universe with CMB lensing}

\noindent Gravitational lensing of the CMB temperature and polarization anisotropies 
provides a clean probe of the clustering of matter integrated to high redshift (see 
Ref.~\cite{Lewis:2006fu} for a review). The power spectrum of this lensing signal can be 
used to probe the late-time growth of structure and so access information, such as spatial 
curvature, dark energy and sub-eV neutrino masses, that are otherwise degenerate in the 
primary CMB spectra imprinted at recombination~\cite{Hu:2001fb}. Moreover, by 
cross-correlating CMB lensing with other tracers of large-scale structure, one can 
calibrate the astrophysical and instrumental bias relations between the tracers and the 
underlying density field, which is critical to maximise the returns from future surveys 
(see e.g.\ Ref.~\cite{Vallinotto:2013eva}).

\begin{figure}
\begin{center}
\includegraphics[width=8cm,angle=-90]{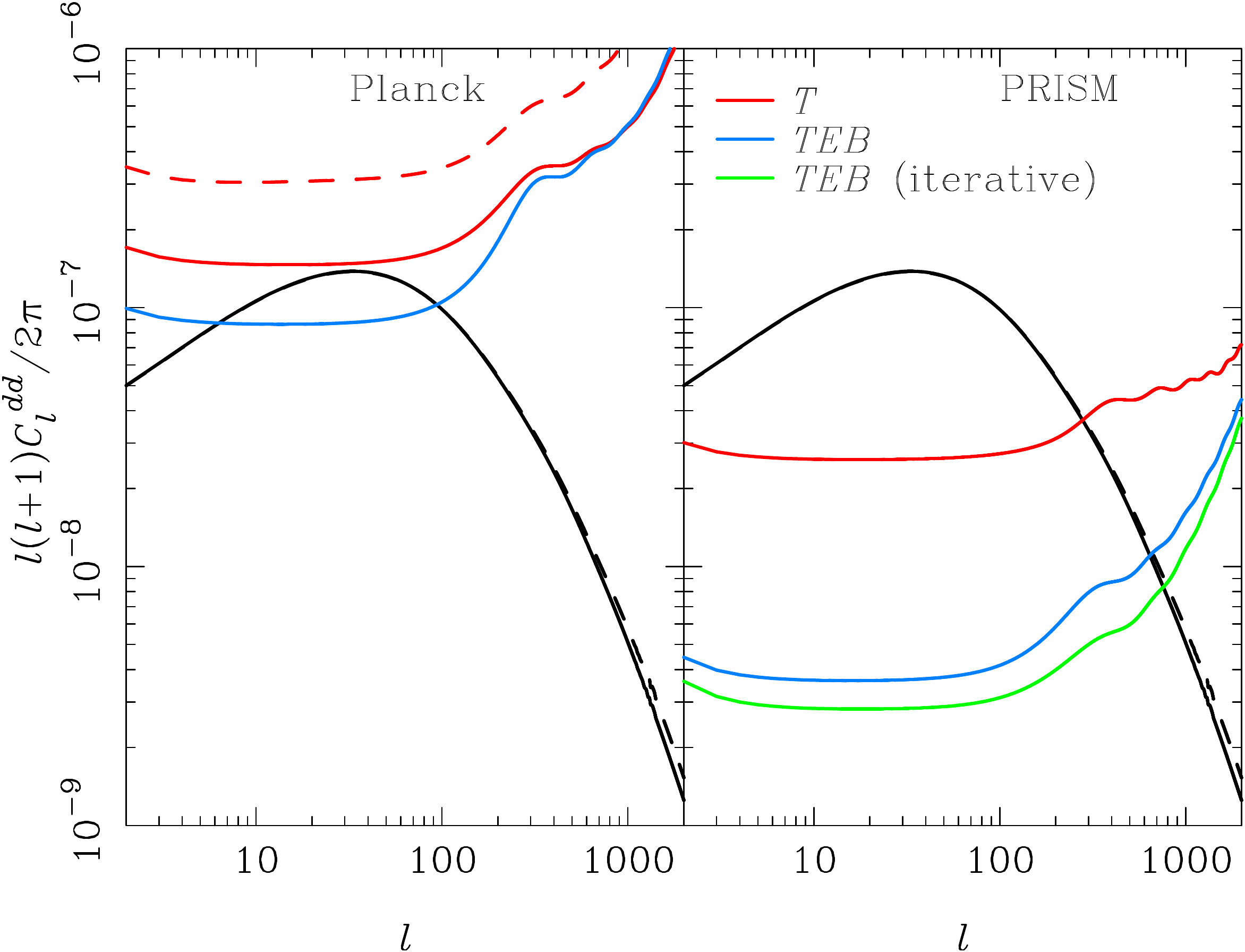}
\end{center}
\caption{\small 
Reconstruction noise on the lensing deflection power spectrum forecast for the full 
\textit{Planck} mission (four surveys; left) and \mission\ (right) using temperature alone 
(red) and temperature and polarization (blue). For \textit{Planck} we also show the 
approximate noise level for the temperature analysis of the nominal-mission data (red 
dashed)~\cite{2013arXiv1303.5077P}, and for \mission, we also show the approximate noise level 
(green) for an improved iterative version of the reconstruction estimator. The deflection 
power spectrum is plotted based on the linear matter power spectrum (black solid) and with 
non-linear corrections (black dashed).
}
\label{fig:CMBlensing}
\end{figure}

Gravitational lensing can be reconstructed from the CMB anisotropies via subtle distortions 
imprinted on the statistics of the CMB by the lenses~\cite{Zaldarriaga:1998te,Hu:2001tn}. 
This approach has been demonstrated recently for the temperature anisotropies by 
ACT~\cite{Das:2011ak,Das:2013zf}, SPT~\cite{vanEngelen:2012va} and 
\Planck~\cite{2013arXiv1303.5077P}, with the latter detecting the lensing power spectrum at the 
$25\sigma$ level. Lensing reconstruction is statistical, with cosmic variance of the 
primary anisotropies giving rise to a statistical noise. With only the temperature 
anisotropies, the lensing $S/N$ cannot exceed unity for multipoles $\ell >300$. However, CMB 
polarization measurements can greatly improve the $S/N$ since polarization has more 
small-scale structure than the temperature, and the $B$-modes of polarization are much less 
limited by cosmic variance of the primary CMB fluctuations~\cite{Hu:2001kj}. With CMB 
polarization measurements from \mission, at a sensitivity of a 
few\,$\mu\mathrm{K}\,\mathrm{arcmin}$ and resolution of a few arcmin, it is possible to 
reconstruct with $S/N>1$ up to multipoles $\ell =600$--$800$, corresponding to scales of about 
15--20\,arcmin (see Fig.~\ref{fig:CMBlensing}), over nearly the full sky. Significantly, 
\mission\ can extract all of the information in the deflection power spectrum on scales 
where linear theory is reliable. Polarization-based lensing reconstructions have recently 
been demonstrated for the first time by the SPTpol team~\cite{2013PhRvL.111n1301H}; 
cross-correlating an $E$-$B$ reconstruction with CIB measurements at $500\,\mu\mathrm{m}$ 
from \emph{Herschel}-SPIRE, they report a $7.7\sigma$ detection. Significantly, their 
analysis represents the first detection of the $B$-modes induced by gravitational lensing 
of the CMB.

\begin{figure}
\begin{center}
\includegraphics[width=8cm,angle=-90]{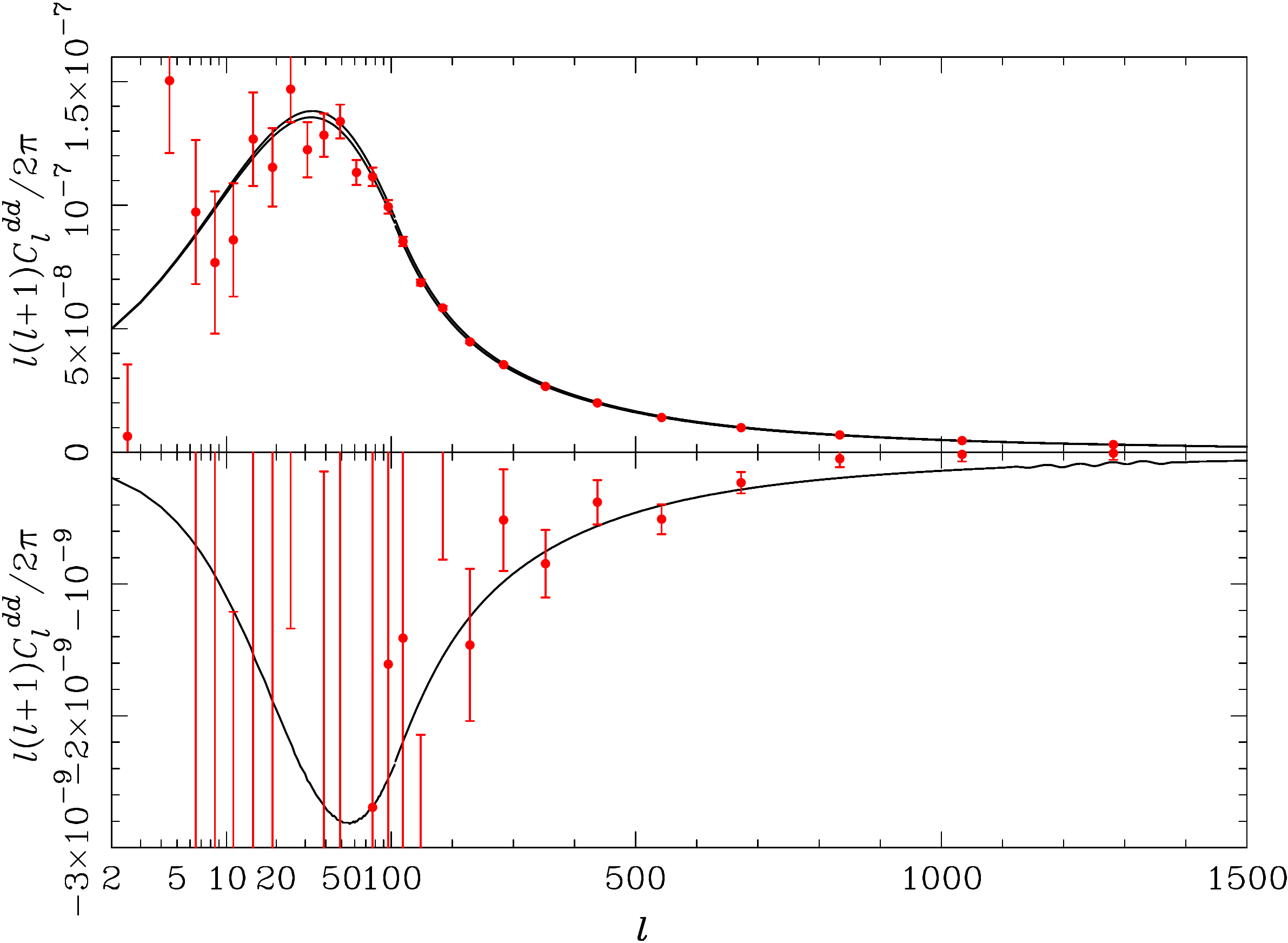}
\end{center}
\caption{\small 
Simulated deflection power spectrum from \mission\ assuming a minimal-mass inverted 
hierarchy ($\sum m_\nu \approx 0.1\,\mathrm{eV}$). In the upper panel, the solid lines are 
the theory power spectrum for this scenario (lower) and for three massless neutrinos 
(upper). The difference between these spectra is plotted in the lower panel illustrating 
how \mission\ can distinguish these scenarios with its lensing power spectrum in the range 
$\ell > 200$.
}
\label{fig:CMBlensing_nu}
\end{figure}

The recent ACT, SPT and \Planck\ lensing analyses have further demonstrated the ability to 
constrain dark parameters from the CMB alone via lensing. For example, in models with 
spatial curvature, the combination of the \Planck\ temperature and lensing power spectra 
measure $\Omega_\Lambda$ at 4\% precision, and $\Omega_K$ to around 1\%~\cite{2013arXiv1303.5076P}. 
As an illustration of the constraining power of the lensing measurements from \mission\ 
alone, we consider constraints on the mass of neutrinos. Oscillation data imply that 
neutrinos must be massive, but the data are insensitive to the absolute mass scale. For a 
normal hierarchy of masses ($m_1, m_2 \ll m_3$), the mass summed over all eigenstates is at 
least $0.06\,\mathrm{eV}$, while for an inverted hierarchy ($m_3 \ll m_1, m_2$) the minimal 
summed mass is $0.1\,\mathrm{eV}$. The individual masses in these hierarchical limits are 
well below the detection limit of current and future laboratory $\beta$-decay experiments, 
but can be probed by cosmology. Massive neutrinos suppress gravitational clustering on 
scales below the horizon size at the non-relativistic transition, reducing the lensing 
power spectrum (as well as galaxy clusters number counts---see Sec.~\ref{sec:clusters}). 
We illustrate the capabilities of \mission\ to distinguish the minimal-mass 
inverted hierarchy from a model with three massless neutrinos via its lensing power 
spectrum in Fig.~\ref{fig:CMBlensing_nu}. Combining the anisotropy and lensing power 
spectra of \mission, we forecast a 1$\sigma$ error of $0.04\,\mathrm{eV}$ for the summed 
mass, comparable to the hierarchical target masses. This constraint can be improved further 
by combining with near-future BAO measurements, for example by a factor of two using BOSS, 
at which point it becomes possible to distinguish between the normal and inverted 
hierarchies (in the hierarchical limits)~\cite{Hall:2012kg}.

The cosmological constraints from lensing with \mission\ would be highly complementary to 
those from upcoming cosmic shear surveys in the optical, such as \Euclid. For example, the 
systematic effects are quite different with non-linearities being much less of an issue for 
CMB lensing with its peak sensitivity to large-scale structures around $z=2$. Moreover, 
there are no intrinsic alignments of galaxy ellipticities to worry about. The combination 
of the two probes of mass is particularly promising, since it allows calibration of 
multiplicative bias effects such as due to PSF corrections in the optical (e.g.\ 
Ref.~\cite{Vallinotto:2013eva}).

\begin{figure}[bht]
\includegraphics[height=0.33\textwidth,angle=-90]{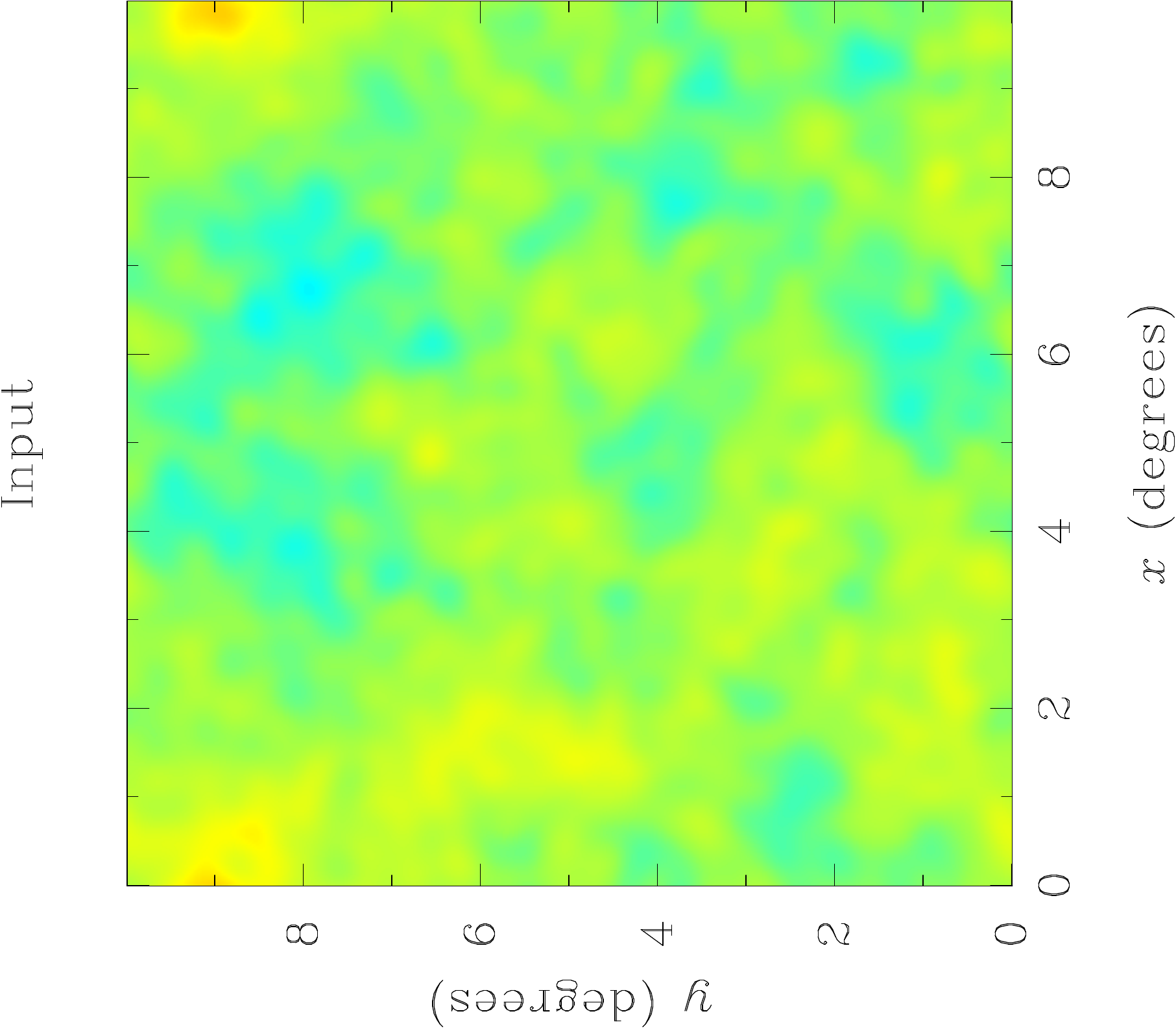}
\includegraphics[height=0.33\textwidth,angle=-90]{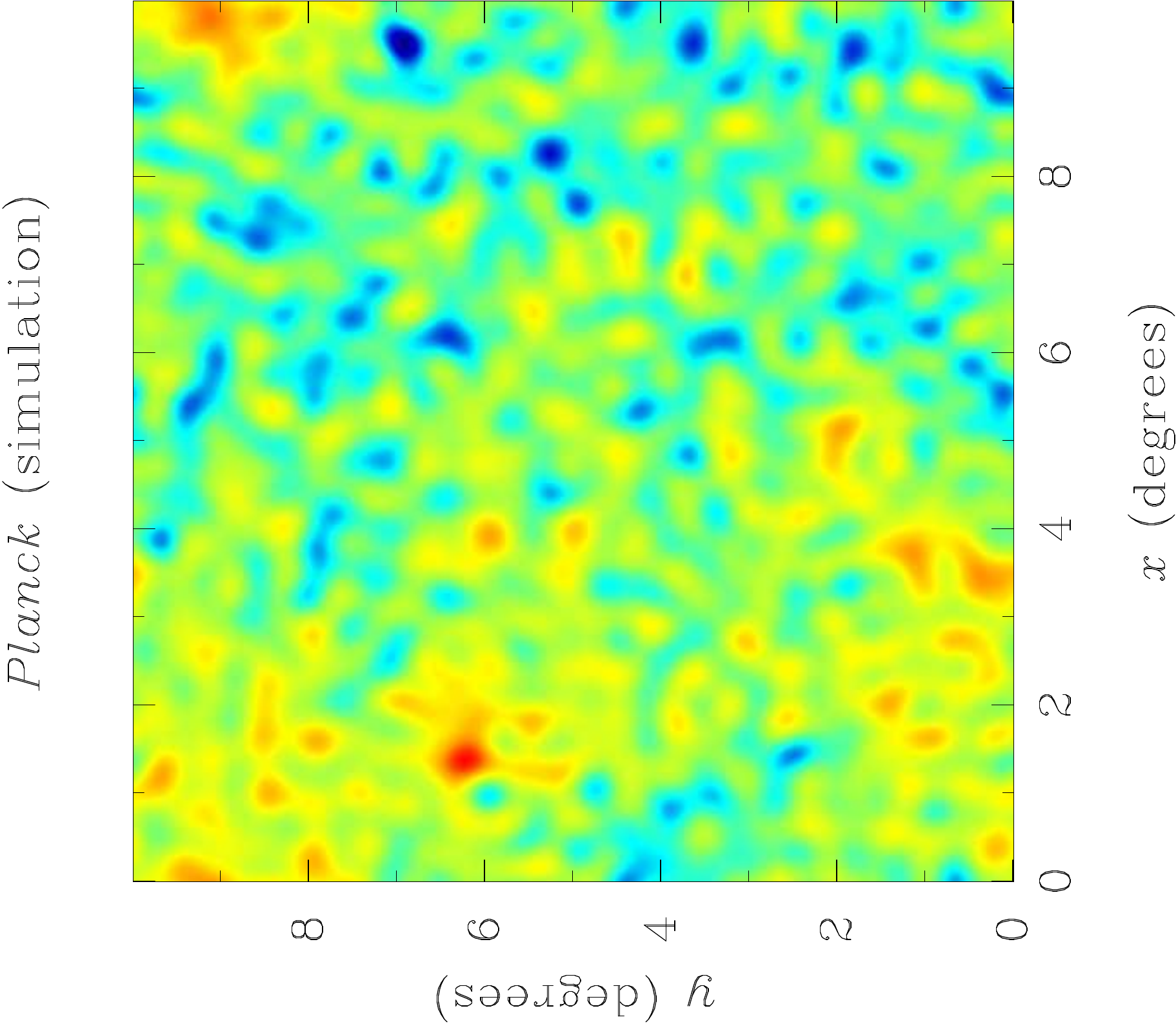}
\includegraphics[height=0.33\textwidth,angle=-90]{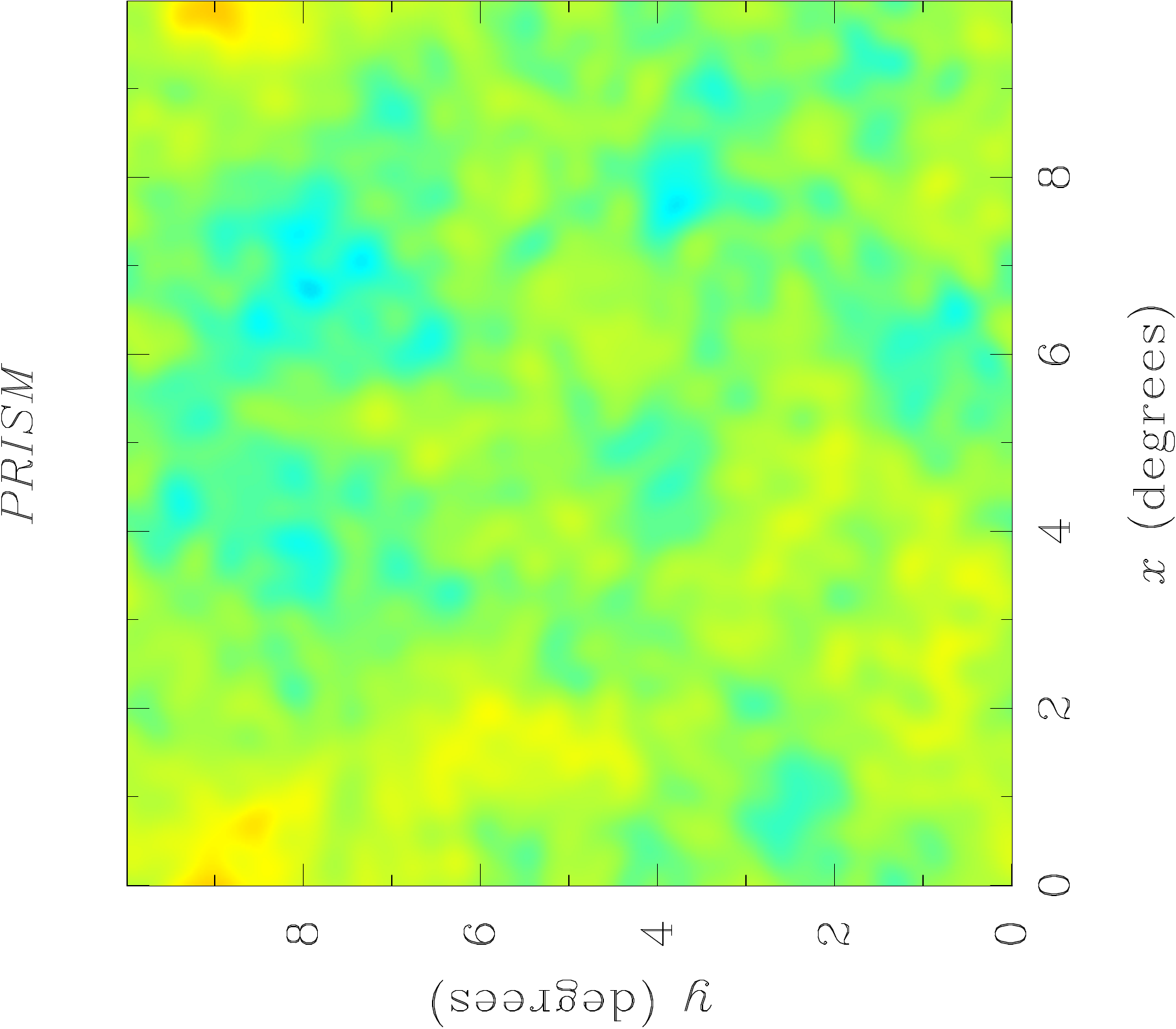}
\caption{{\small 
Reconstruction of the CMB lensing potential. Right: $10^\circ \times 10^\circ$ simulated 
map of the (filtered) lensing potential for a standard $\Lambda$CDM cosmology matching 
\planck\ cosmological parameter constraints. Middle: simulated reconstruction of the 
lensing potential on the same field with the sensitivity of the full \planck\ mission. 
Right: simulated reconstruction with the \mission\ sensitivity. The same color scale is 
used for all three maps.}}
\label{fig:lensing-maps}
\end{figure}

The high $S/N$ reconstruction of the lensing potential by \mission\ is illustrated in 
Fig.~\ref{fig:lensing-maps}. The left panel is a simulated map of the filtered lensing 
potential (in Fourier space, the image is $\ell \phi(\mbox{\boldmath $\ell $})$ with 
$\phi(\mbox{\boldmath $\ell $})$ the Fourier transform of the lensing potential, so has the 
same dimensions as the deflection field) over a small patch of the sky. The middle and 
right panels illustrate the fidelity of the lensing reconstructions expected from the full 
\planck\ mission and \mission, respectively. A similar quality reconstruction will be 
obtained with \mission\ over nearly the full sky.

Cross-correlating CMB lensing with other probes of large-scale structure, such as 
galaxies~\cite{Smith:2007rg,Hirata:2008cb,Sherwin:2012mr,2012ApJ...753L...9B,2013arXiv1303.5077P,2013ApJ...776L..41G}, 
the Ly$\alpha$ forest~\cite{Vallinotto:2009wa} or CIB 
clustering~\cite{2013arXiv1303.5078P,Holder:2013hqu,2013PhRvL.111n1301H} (see also Sec.~\ref{sec:extragalactic}), also 
has exceptional promise, allowing self-calibration of the tracer's bias relation at the 
sub-percent level. Such cross-correlations with tracers selected as a function of redshift 
will make possible the three-dimensional characterization of the lensing potential, and 
hence the reconstruction of the distribution of matter inhomogeneities in three dimensions 
over most of our Hubble volume. A combination of this with the bulk-velocity constraints 
obtained from kSZ observations will give an unprecedented view of cosmic structure 
formation in the entire observable Universe.

\subsection{Primordial non-Gaussianity}

\begin{figure}
\centerline{\includegraphics[width=8cm]{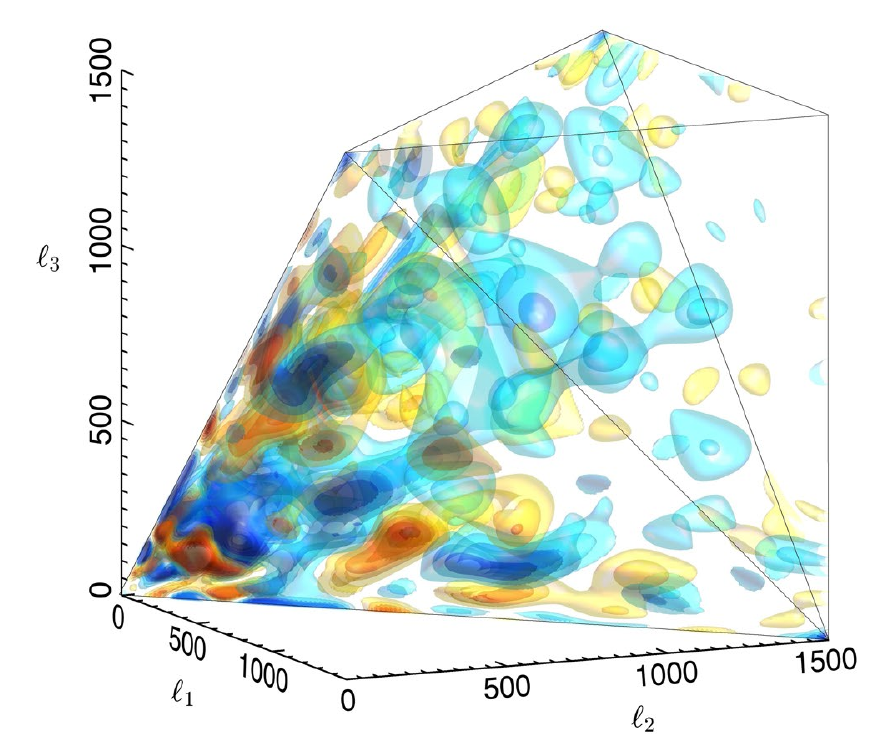}}
\caption{\small 
\Planck\ CMB temperature ($TTT$) bispectrum~\cite{2013arXiv1303.5084P}.  The bispectrum depends on three multipoles 
$\ell _1, \ell _2, \ell _3$ subject to a triangle constraint, so it contains rich 3D shape information in a tetrahedral 
domain.   Isocontours are plotted with red positive and blue negative.  
Note the periodic CMB ISW-lensing signal in the squeezed limit along the edges (seen at about the 2.5$\sigma$ level).  
\mission\ would detect this signal at 9$\sigma$ -- a unique window on dark energy. Scale-invariant signals predicted 
by many inflationary models are strongly constrained by this data though `oscillatory' and `flattened' features 
hint at new physics. \mission\ will probe these hints with an order of magnitude more resolved triangle 
configurations as well as polarization cross-terms.}
\label{fig:recon}
\end{figure}

\noindent With the first \Planck\ cosmology data release, non-Gaussianity (NG) crossed a 
significant threshold to become a robust quantitative probe of cosmological 
physics~\cite{2013arXiv1303.5084P}.  \Planck\ results dramatically improved previous NG analyses, 
offering the most stringent test to date of inflationary theory (with $f^{\rm loc}_{\rm NL} 
= 2.7\pm 5.8$) while also detecting for the first time ISW-lensing and diffuse point source 
bispectra.  Already \Planck\ offers enticing clues about the nontrivial `shape' of the CMB 
bispectrum of our universe (as illustrated in Fig.~\ref{fig:recon}), the origin of which is 
yet to be explained.  \mission\ would offer the highest precision reconstructions of the 
CMB temperature and polarization bispectra and trispectra which, in conjunction with the 
CMB power spectra, will provide a decisive and unambiguous probe of primordial cosmology 
back to the \Planck\ era.  At the same time, \mission\ NG data will open new windows for 
investigating dark energy, gravitational physics, and cosmological parameters, as well as 
astrophysical sources, large-scale structure and galactic history.

\begin{figure}[t]
\centering
\includegraphics[width=5.0cm]{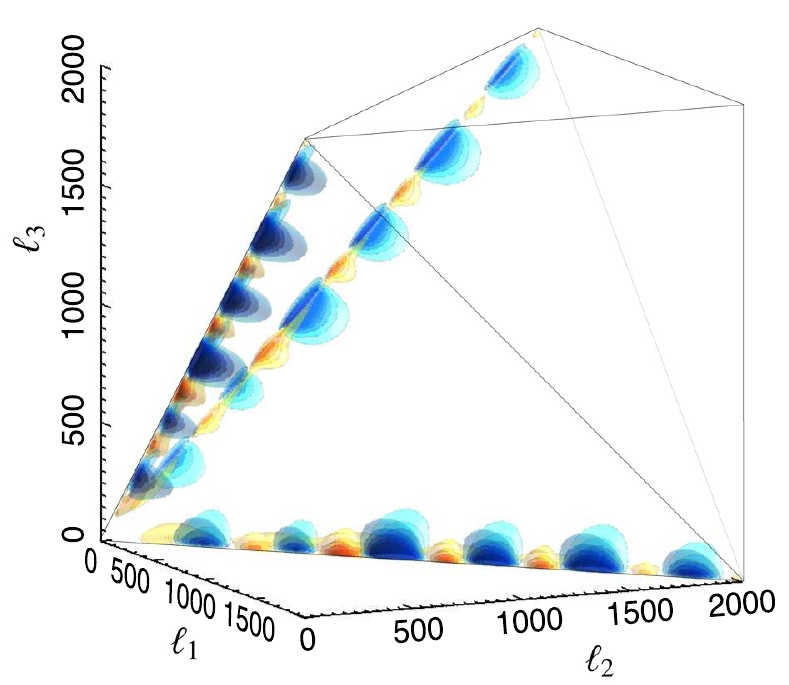}
\includegraphics[width=5.6cm]{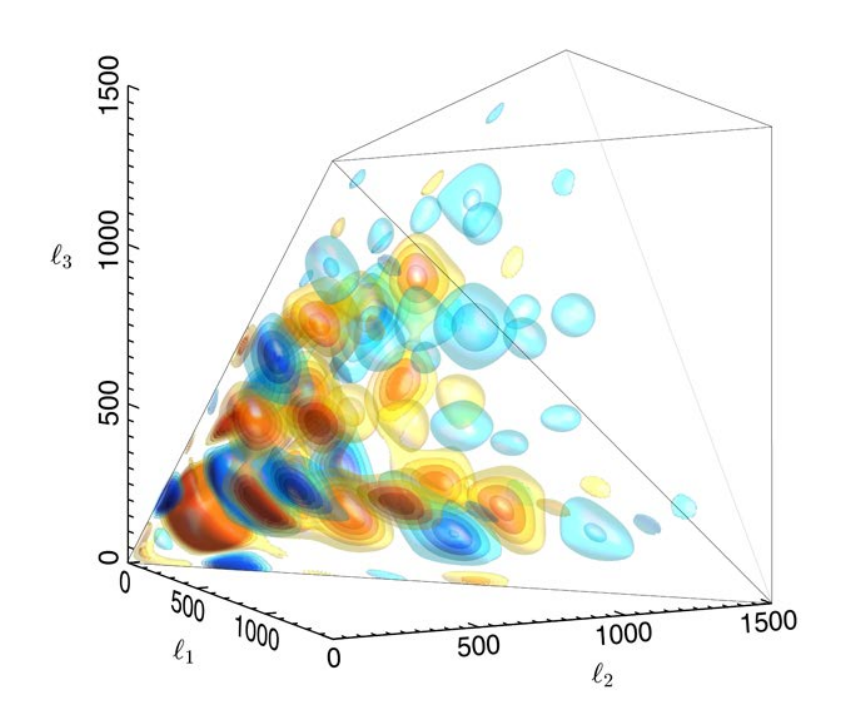}
\includegraphics[width=5.0cm]{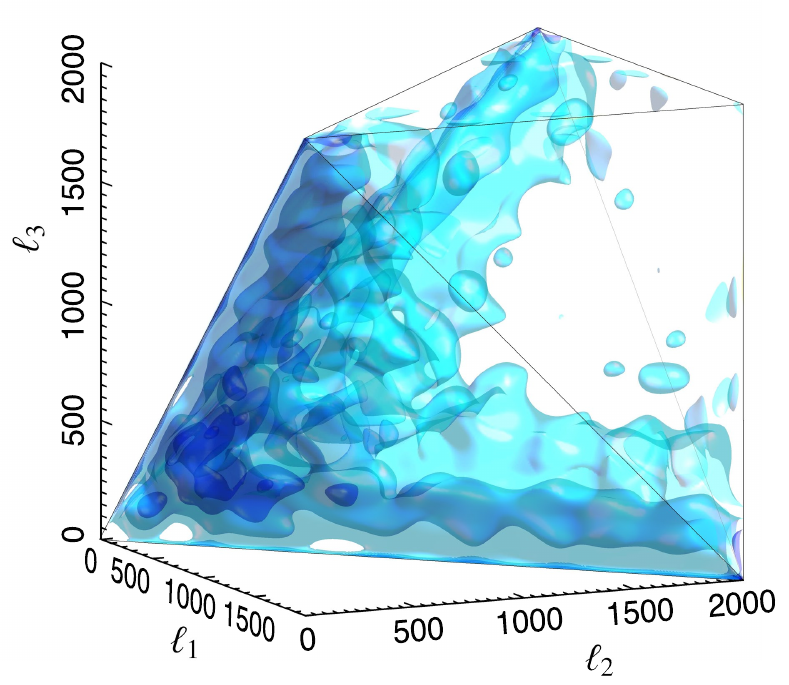} 
\small
\caption{\small Primordial and late-time non-Gaussian shapes~\cite{2013arXiv1303.5084P,2013arXiv1303.5085P}.
Here the distinct CMB bispectra $B_{l_1 l_2 l_3}$ from three theoretical models are plotted 
(from left to right): the ISW-lensing signal (observed in Fig.~\ref{fig:recon});
an inflationary `feature' model; and the late-time bispectrum generated by cosmic strings.}
\label{fig:NG}
\end{figure}

An important and unique advantage of \mission\ over non-CMB probes of NG results from its 
ability to recognize the distinct patterns that physical mechanisms leave in the 
\textit{shape} of higher-order correlators, as illustrated in Fig.~\ref{fig:NG}. The 
competitive strength of \mission\ will therefore be a vastly enhanced exploration of 
physically-predicted NG shapes compared to any other projected probe of NG. While other 
probes claim competitive power for detecting `local' bispectral non-Gaussianity, these 
claims do not yet address the full suite of non-linear systematic and biasing effects.

Only a high-resolution polarization CMB mission such as \mission\ can provide the high 
sensitivity for detecting and distinguishing differing NG bispectral shapes. In fact, 
polarization maps contain more information than temperature maps and the best constraints 
arise from combining the two.  Polarization is the more powerful probe because the ratio of 
the expected contaminant signal (mainly Galactic dust polarization) to the primordial 
signal is smaller for polarization than for temperature for the majority of resolved modes.  
To construct a figure of merit comparing the predicted impact of \mission\ on physical 
forms of NG, we compare the predicted constraint volume in bispectrum space spanned by the 
local, equilateral and flattened bispectra. Compared to \textit{WMAP} data, the first 
\Planck\ release reduced the constraint volume by a factor of 20 and full-mission \Planck\ 
data including polarization is expected to improve this by an additional factor of 2.5; 
\mission\ would go beyond this by another factor of 30.  Considering the constraint volume 
based only on the polarization maps (which provide information independent of the 
temperature maps and hence provide an important consistency check), we find a volume 
reduction factor from \Planck\ to \mission\ of order $110$. \mission\ would therefore 
greatly advance unambiguous precision tests of the standard inflationary paradigm. 
will be distinct from other experiments because of its full-sky coverage, polarization 
sensitivity, high resolution, strong rejection of systematics, and the benign environment 
in space at L2 making \mission\ the ideal CMB NG probe.

The improvement compared with \Planck\ will be even more dramatic when considering combined 
bispectrum and trispectrum constraints on a larger range of NG shapes. The forecast 
precision with which local trispectrum parameters could be measured with \mission\ are 
$\Delta g_{\rm NL}=3\times 10^4$ and $\Delta \tau_{\rm 
NL}=1\times10^2$~\cite{smidtetal2010}. These could investigate consistency conditions 
between trispectra and bispectra, which can be used to test large classes of multi-field 
inflation models, in addition to single-field inflation.  There are other alternative 
inflationary scenarios for which an observable non-Gaussian signal is quite natural, 
including models with non-canonical kinetic energy, additional vector fields, features or 
periodicity in the inflationary potential, or remnants of a pre-inflationary phase. There 
are also more exotic paradigms which can create NG such as cosmic 
(super-)strings~\cite{2013arXiv1303.5085P} or a contracting phase with a subsequent 
bounce~\cite{Lehners:2010fy}.  Each of these models has a distinct fingerprint, many 
uncorrelated with the standard three primordial shapes and, in all cases, \mission\ would 
significantly improve over present \Planck\ constraints, offering genuine discovery 
potential.  Beyond searches for primordial NG, \mission\ is guaranteed to make important 
observations of late-time NG.  For example, \mission\ will decisively detect and 
characterize the lensing-ISW correlation, driven by dark energy, achieving a $9\sigma$ 
detection, resulting in a new probe of dark energy physics from the CMB alone.  
Second-order recombination effects can also be probed; present calculations (temperature 
only) suggest \mission\ could detect these at about 3$\sigma$ 
\cite{2009PhRvD..79b3501K,Su:2012gt}.  Furthermore, ancillary signatures will also 
constrain modified gravity alternatives to the standard cosmological model.

CMB spectral distortions (further discussed in Sec.~\ref{sec:CMB-spectrum}) can also open up a new window through which to search for primordial NG~\cite{Pajer:2012vz}.
     We know almost nothing about NG on the small scales that can
     be probed via
     these observations. In particular the 2-point
     cross-correlation
     between $\mu$-type distortions and CMB anisotropies turns out
     to be naturally sensitive to the very squeezed limit of the
     primordial
     bispectrum (probing scales as  small as $50 \leq k\,\mathrm{Mpc} \leq
     10^4$).
     Also, the power spectrum of $\mu$-distortions can probe the
     trispectrum of primordial fluctuations. Such measurements can
     be particularly constraining for models where the power
     spectrum grows on small scales~(see,
     e.g. Ref.~\cite{Chluba:2012we}). Values as low as
     $f^{\rm loc}_{\rm NL} <1$ can be achieved. Also, $\mu$-type
     distortions
     can shed light into modifications of the initial state of
     quantum fluctuations. For a large class of inflationary
     models
     characterized by a non-Bunch-Davies vacuum (whose bispectrum
     is enhanced in the squeezed limit with respect to the local form) a
     high $S/N$ can be
     achieved~\cite{Ganc2012}. Being sensitive to 
     small scales, the cross-correlation between $\mu$-type
     distortions
     and CMB anisotropies can also be employed to probe primordial
     NG that runs with scales~\cite{2013PhRvD..87f3521B}.

\subsection{Parameters from high-resolution polarization spectra}

\mission\ will measure the CMB temperature and polarization angular power spectra with 
outstanding precision to small angular scales. Combining the five channels in the 
105--200\,GHz frequency range, the $TT$ spectrum is cosmic-variance limited to $\ell =3850$ (Fig.~\ref{fig:CMB-Cl-sensitivity}). 
Although uncertainties in foreground modelling, in particular the background of radio and 
infrared sources, might limit the maximum useful multipole to lower values, the large 
frequency coverage of \mission\ will help correct for such contamination in the best 
possible way. As noted above however, high-$\ell $ polarization is expected to be cleaner, and 
with \mission\ the $EE$ spectrum is cosmic-variance limited to $\ell =2500$ (and the $B$-mode 
power from lensing to $\ell =1100$)\footnote{The current \textit{Planck} $TT$ power spectrum is 
cosmic-variance limited for $\ell  \ltorder 1500$, while forthcoming \textit{Planck} 
polarization data is expected to be noise-dominated for nearly all multipoles.}. Such a 
remarkable measurement of the polarization of the CMB damping tail will be an invaluable 
source of information for several reasons. Ongoing experiments, such as \textit{Planck} 
have confirmed that the tail of the $TT$ power spectrum can not only constrain inflationary 
models by providing a long lever-arm for measuring the shape of the power spectrum of the 
primordial curvature perturbation, but also provide detailed information about the 
fundamental matter content of the Universe. This includes the effective number of 
relativistic species $N_{\mathrm{eff}}$, for which a non-standard value (as hinted by 
pre-\textit{Planck} anisotropy measurements) could be due to sterile neutrinos, as 
advocated in particle physics to explain certain anomalies in the neutrino sector, the 
helium abundance $Y_{\mathrm{P}}$, which provides a clean test of standard big-bang 
nucleosynthesis, and the neutrino mass. Measuring the high-$\ell $ damping tail in polarization 
will allow us to place strong constraints on a number of different models, free from the 
foreground issues that limit inferences made from the temperature measurements.

\begin{figure}[t] 
\includegraphics[width=0.33\textwidth,angle=0]{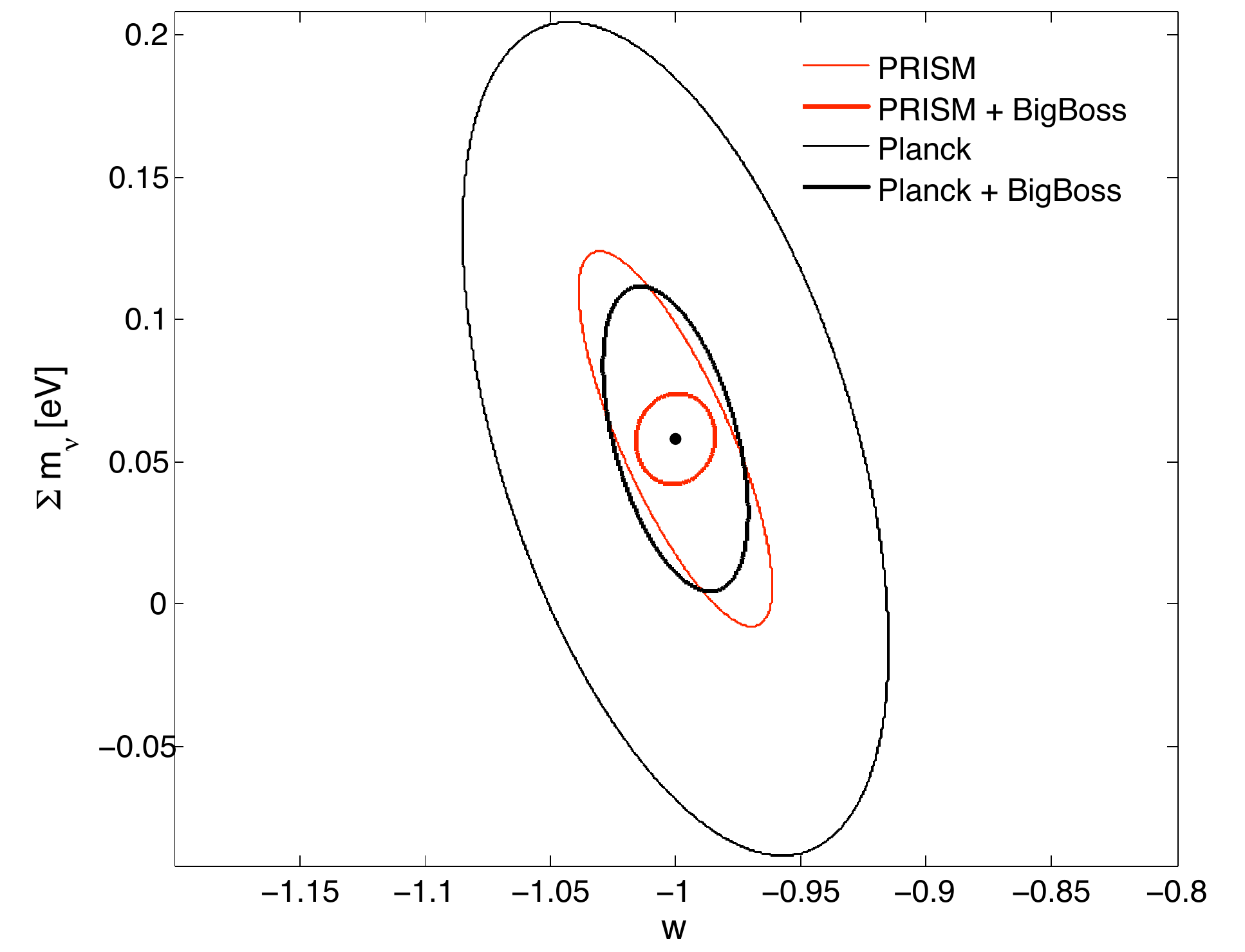}
\includegraphics[width=0.32\textwidth,angle=0]{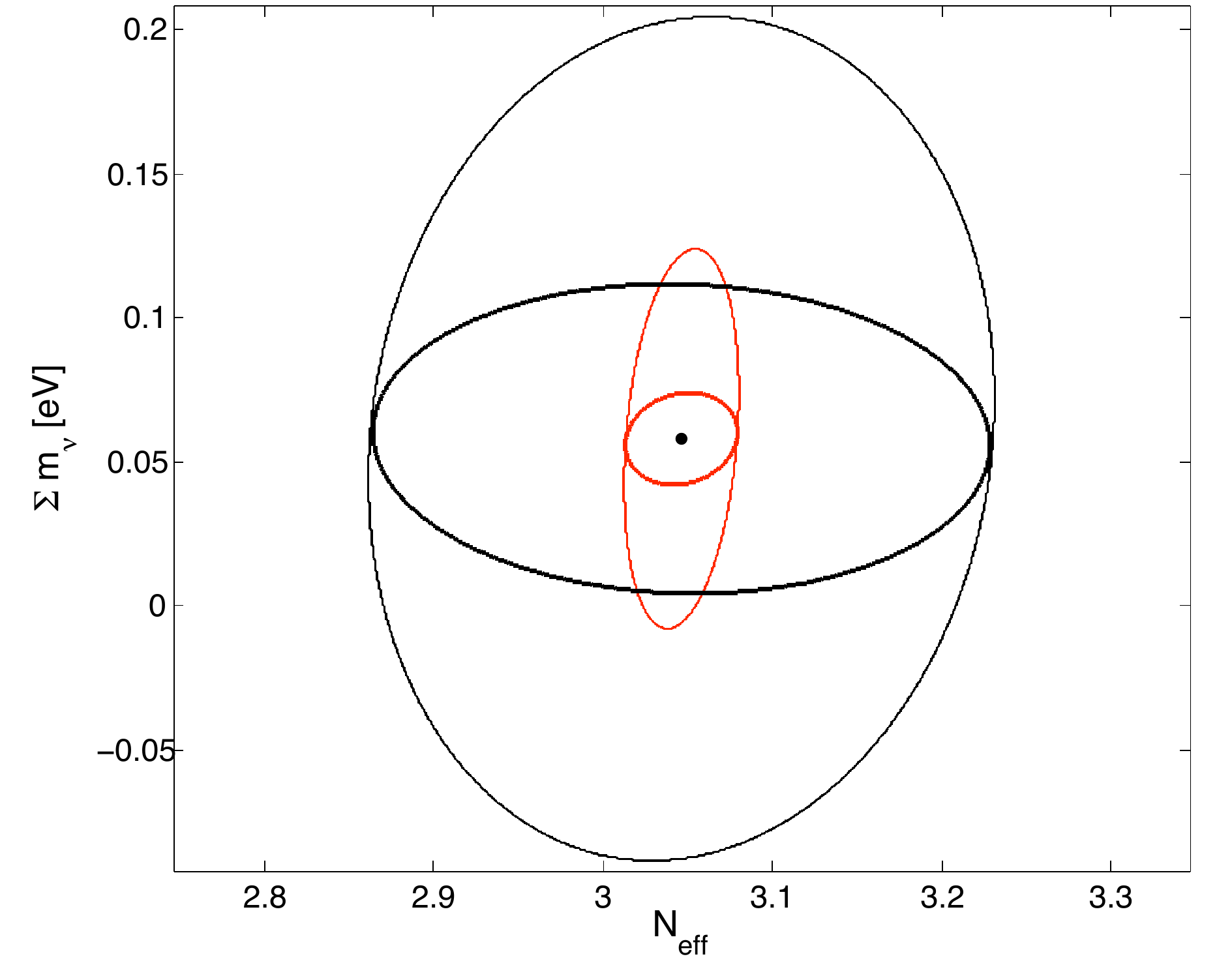}
\includegraphics[width=0.33\textwidth,angle=0]{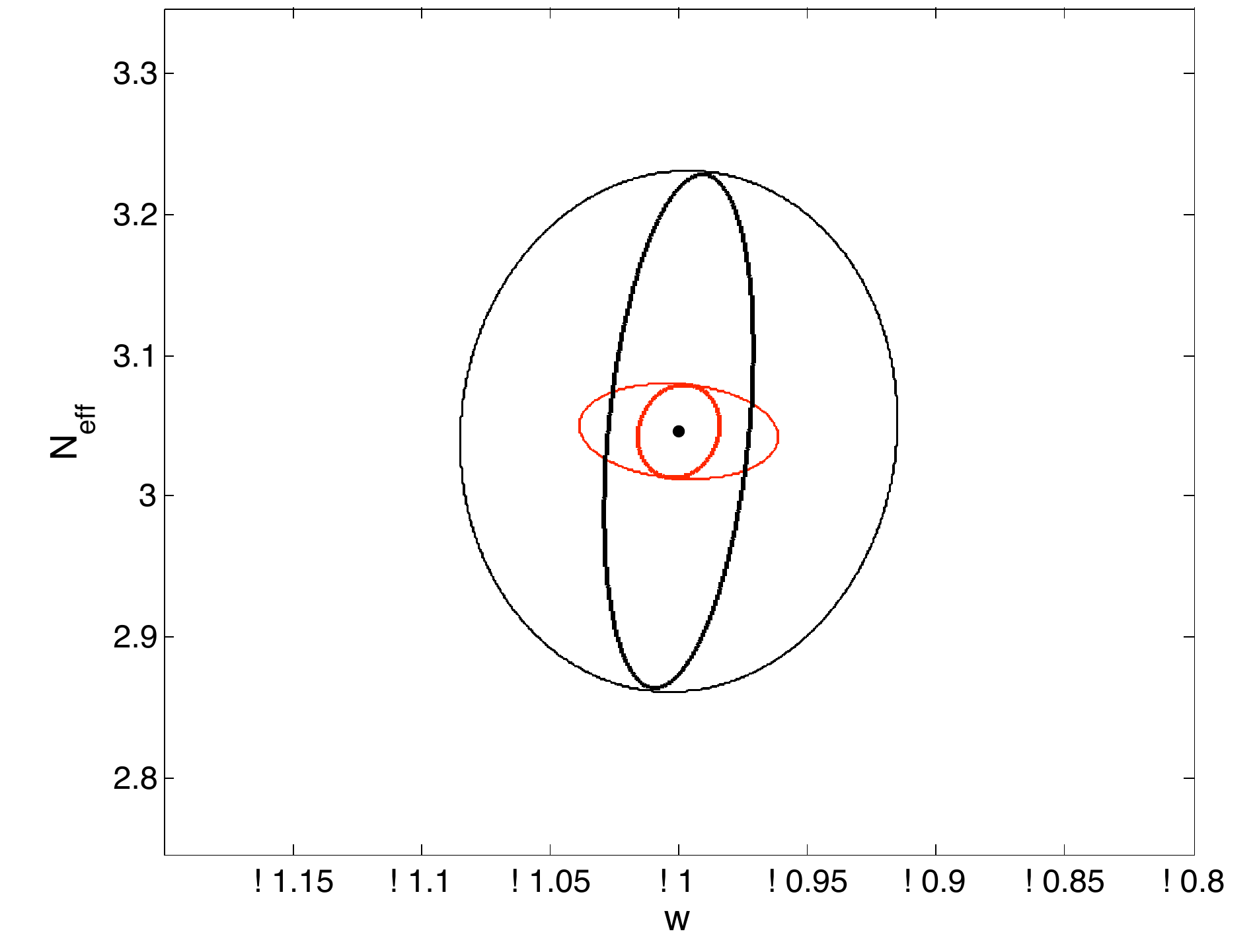}
\caption{{\small 
Fisher forecasts for the 68\% confidence regions in 10-parameter models (the standard six 
parameters plus the summed neutrino mass $m_\nu$, effective number of relativistic degrees 
of freedom $N_{\rm eff}$, equation of state of dark energy $w$, and spatial curvature). 2D 
marginalised constraints are shown in the $\sum m_\nu$--$w$ plane (left), $\sum 
m_\nu$--$N_{\rm eff}$ (middle), and $N_{\rm eff}$--$w$ (right) for the full-mission 
\Planck\ data including lens reconstruction (thin black lines) and further combining with 
BAO data from BigBoss/DESI (thick black lines). Results for \mission\ are shown as thin red 
contours, and for \mission+BigBoss in thick red.}}
\label{fig:highell_params}
\end{figure}

In order to illustrate the potential improvements that \mission\ will bring, we have made 
MCMC parameter forecasts, following Ref.~\cite{Galli:2010it}. In $\Lambda$CDM models, the 
spectral index and its running will be measured more precisely than with current \planck\ 
data by factors of five and three, respectively. The Hubble constant (a point of tension 
between \Planck\ data and direct astrophysical measurements) will be measured a factor of 
10 better than currently (and $2.5$ times better than expected from the full \planck\ 
data). In one-parameter extensions to the matter content beyond $\Lambda$CDM, we forecast 
that \mission\ will measure $N_{\mathrm{eff}}$ to 2\% precision and $Y_{\mathrm{P}}$ to 
1\%. These values indicate that a $2\sigma$ anomaly hinted at by \Planck\ could be 
confirmed decisively with \mission. Moreover, from its measurement of the $B$-mode power 
spectrum, \mission\ should extend the range of sensitivity to cosmic strings by an order of 
magnitude over the recent \planck\ constraints~\cite{2013arXiv1303.5085P,Avgoustidis:2011ax}.

The combination of temperature and polarization power spectrum measurements to small 
angular scales \emph{and} lensing reconstruction with \mission\ will allow precise 
exploration of more complex models. For example, models in which an extended neutrino 
sector, e.g.\ including the mass of the active neutrinos as well as additional massless 
species (parametrised by $N_{\rm eff}$), is combined with evolving dark energy (with 
equation-of-state parameter $w$) and spatial curvature. In such complex models, parameter 
degeneracies are significant for current CMB data and adding external data, such as BAO, 
leads to large improvements in statistical errors. Forecasts based on the Fisher 
information\footnote{In models with near degeneracies, Fisher forecasts are unreliable.
The relative improvements in the constraints plotted in 
Fig.~\ref{fig:highell_params} are more accurate than the absolute forecasted errors.} for 
such a 10-parameter model are shown in Fig.~\ref{fig:highell_params} for the full \Planck\ 
mission (including lens reconstruction and rather less conservative sky fractions than 
adopted for the 2013 temperature analysis), full \Planck\ combined with next-generation BAO 
measurements from BigBOSS/DESI, \mission\ (including lens reconstruction), and \mission\ 
combined with BigBOSS/DESI.  Significantly, \mission\ \emph{alone} is almost as powerful 
for the late-time parameters $\sum m_\nu$ and $w$ as the combination of the full \Planck\ 
mission and next-generation BAO measurements, and is much more powerful for early-time 
parameters such as $N_{\rm eff}$. The combination of \mission\ and next-generation BAO 
measurements is particularly powerful, improving over full \Planck\ plus next-generation 
BAO by a factor of three on the (summed) neutrino mass and factors of two for $w$, the 
curvature parameter and $N_{\rm eff}$. With this data combination, constraints on the 
absolute mass scale of neutrinos at the $0.02\,\mathrm{eV}$ are robust to extensions to the 
dark energy model and the spatial geometry.


\section{CMB spectral distortions}
\label{sec:CMB-spectrum}
\newcommand{\zmu}{{z_{\mu}}}
\newcommand{\pot}[2]{#1 \times 10^{#2}}
\newcommand{\ion}[2]{{\text{{\sc #1}\,{\sc #2}}}}
\newcommand{\JC}[1]{{#1}}
\newcommand{\Supercore}{{\mission}}

\def\aap{A\&A}
\def\apj{ApJ}
\def\apjs{ApJS}
\def\apjl{ApJL}
\def\mnras{MNRAS}
\def\aj{AJ}
\def\nat{Nature}
\def\aaps{A\&A Supp.}
\def\pra{Phys.Rev.A}         
\def\prb{Phys.Rev.B}         
\def\prc{Phys.Rev.C}         
\def\prd{Phys.Rev.D}         
\def\prl{Phys.Rev.Lett}      
\def\araa{ARA\&A}       
\def\gca{GeCoA}         
\def\pasp{PASP}              
\def\pasj{PASJ}              
\def\apss{Astrophysics and Space Science}
\def\jcap{JCAP}
\def\plb{Phys. Lett. B.}
\def\jhep{JHEP}


 
\begin{figure*}
\centering
\includegraphics[width=0.8\columnwidth]{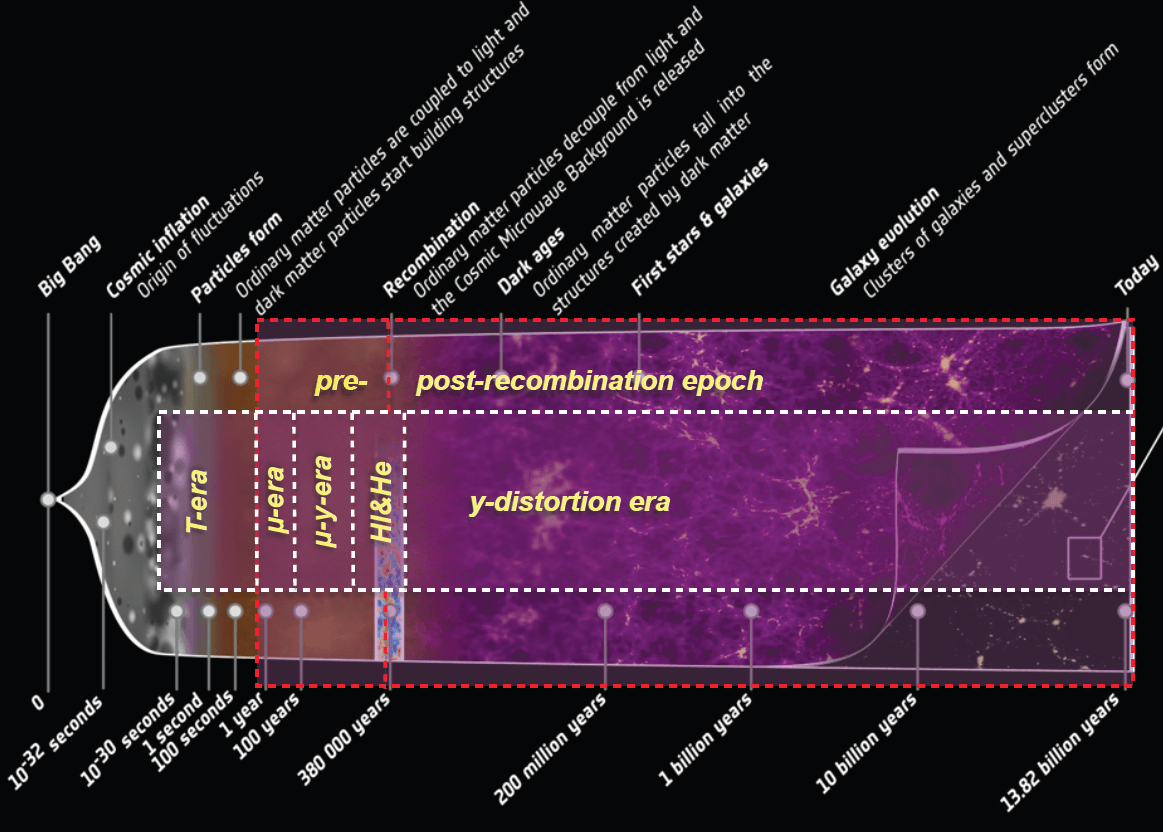}
\caption{\small CMB spectral distortions provide a sensitive probe of energy release throughout the cosmic 
history. At very early stages ($z\gtrsim {\rm few}\times 10^6$), thermalization is extremely efficient 
and only a shift in the average CMB temperature is produced, leaving an unmeasurable distortion signature. At 
later stages ($3\times 10^5 \lesssim z \lesssim {\rm few}\times 10^6$), when the Universe was only a few months 
old, energy release causes a pure $\mu$-distortion. This type of distortion is unique to the early Universe and 
thus tells a story about processes occurring deep in the pre-recombination epoch. At redshift $z\lesssim 10^4$, 
on the other hand, a pure Compton $y$-distortion is created, providing the means to learn about reionization and 
structure formation. At intermediate stages ($ 10^4 \lesssim z \lesssim 3\times 10^5$), the distortion 
interpolates between $\mu$- and $y$-distortion, providing additional information about the time dependence 
of the energy release scenario encoded in the smaller residual (non-$\mu$/non-$y$) distortion. 
\Supercore\ will target this signal, which will inform us
about decaying particles and the small-scale power spectrum, 
observations that would otherwise be impossible. The hydrogen and helium recombination lines, which are created at 
redshifts $z\approx 10^3$, might furthermore allow \Supercore\ to distinguish pre- and post-recombination 
$y$-distortions.}
\label{spectral_targets}
\end{figure*}
 
\begin{figure*}
\centering
\includegraphics[width=0.56\columnwidth]{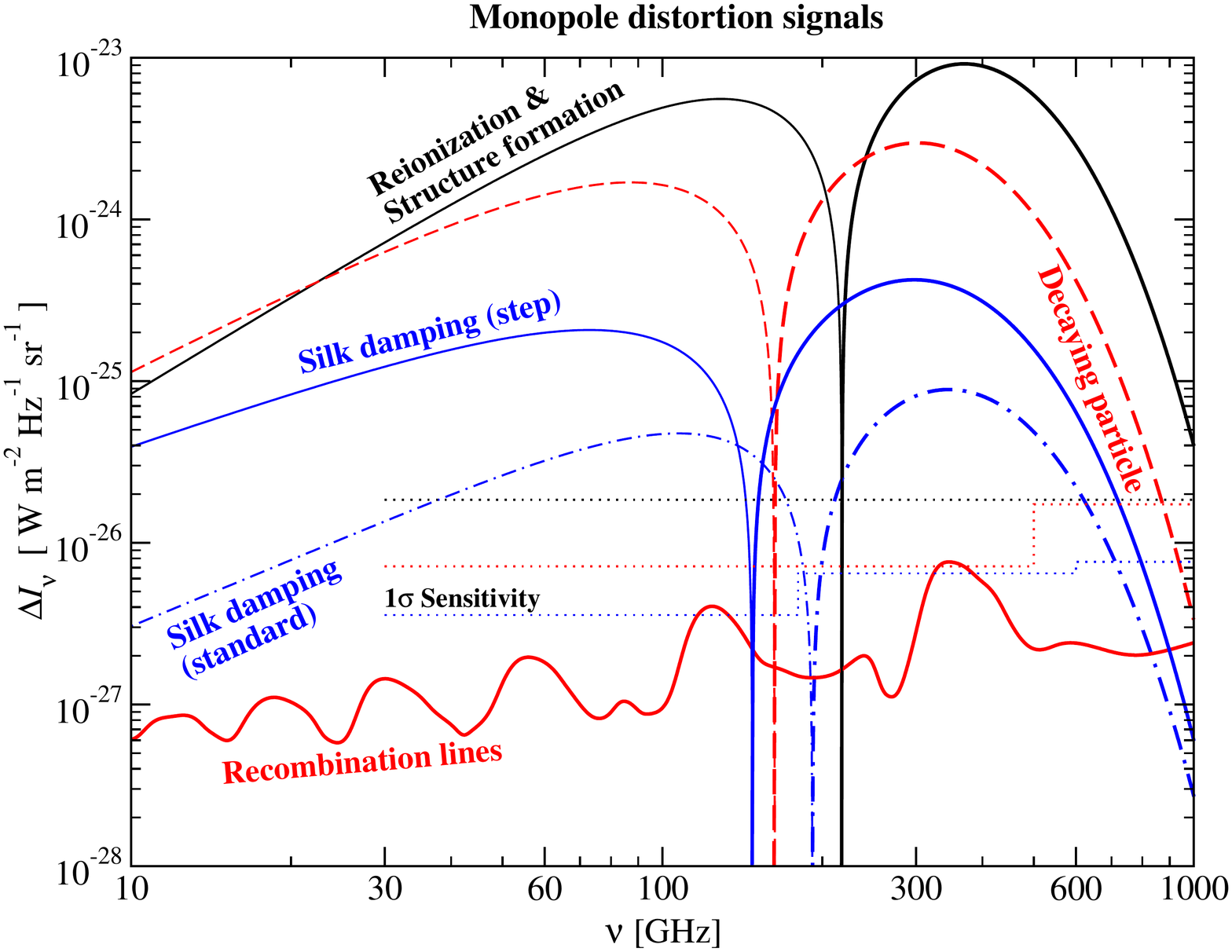}
\includegraphics[width=0.43\columnwidth]{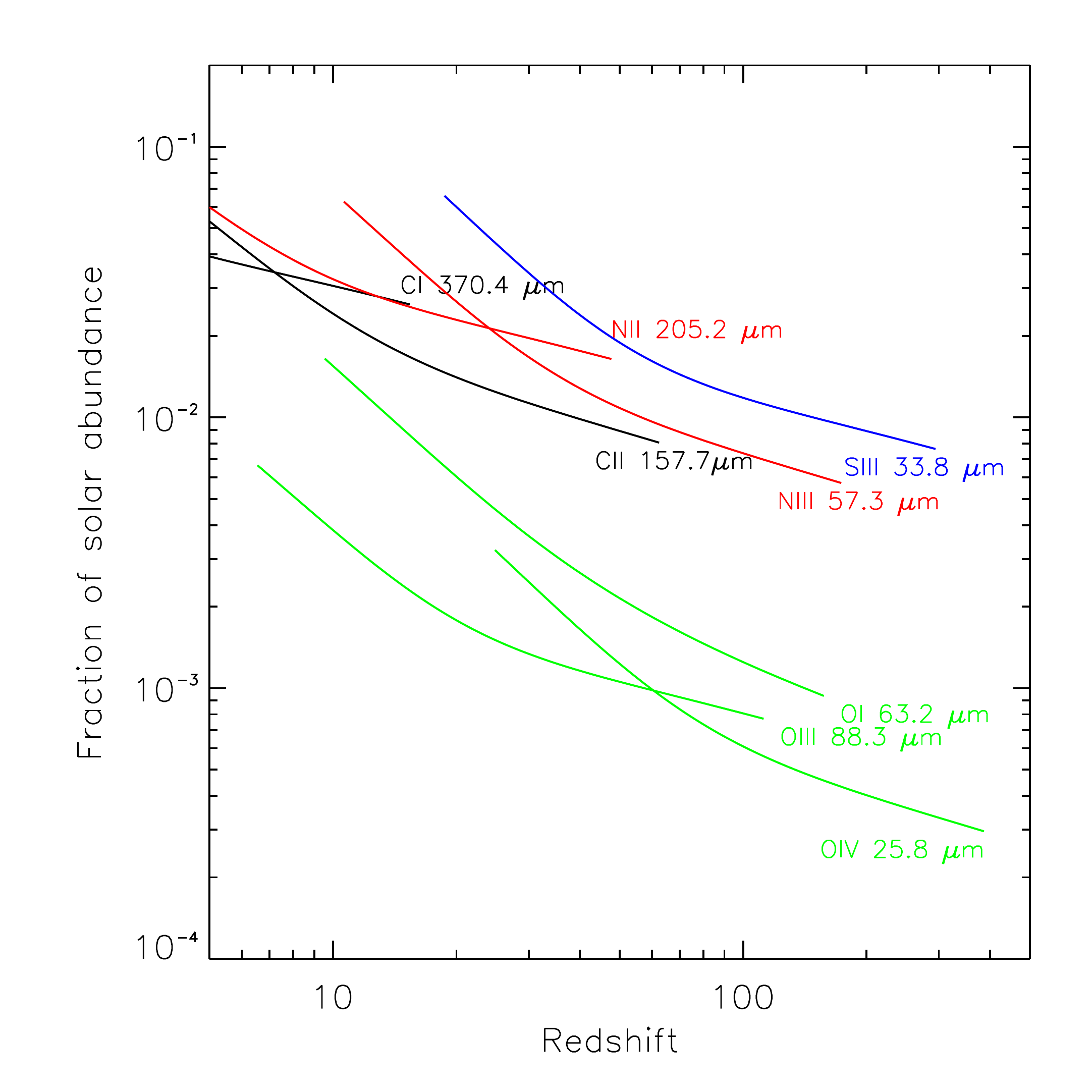}
\caption{\small
Left: spectral distortions for different scenarios. Thick lines denote positive, and thinner lines negative 
signal. The $1\sigma$ sensitivities of \Supercore\ for different designs are also indicated. The largest 
distortion is from the heating of the medium during reionization and structure formation, here for $y\approx 
\pot{5}{-7}$. The decaying particle scenario is for lifetime $t_{\rm X}\approx \pot{3.6}{9}\,{\rm sec}$ and total 
energy release $\Delta \rho_\gamma/\rho_\gamma \approx \pot{6.3}{-7}$. Two cases for the distortion caused by 
dissipation of small-scale curvature perturbations are shown, one for the standard power spectrum, extrapolated 
from large-scale CMB measurements all the way to $k\approx \pot{\rm few}{4}\,{\rm Mpc^{-1}}$, the other assuming 
an additional step $\Delta A_\zeta=10^{-8}$ at $k=30\,{\rm Mpc^{-1}}$ in the power spectrum. The signal caused by 
recombinations of hydrogen and helium is also, in principle, directly detectable 
(combining adjacent frequencies). Right: Projected constraints on the 
abundance of different metal ions versus redshift assuming inter-channel calibration accuracy of $\approx 
10^{-5}$.}
\label{spectralfigure}
\end{figure*}

The {\it COBE} {\it FIRAS} instrument established that the average CMB spectrum is extremely 
close to a perfect blackbody, with possible departures limited to $\Delta I_\nu/I_\nu \lesssim 
\pot{\rm few}{-5}$ \citep{Mather1994, Fixsen1996}. Similarly, the tiny directional variations 
of the CMB intensity and polarization, imprinted by perturbations of the photon-baryon 
fluid at the last scattering surface, all have a thermal spectrum \citep{Smoot1992, 
WMAP_params, Planck2013params}.
This observation places very tight constraints on the thermal history of our 
universe, ruling out cosmologies with extended periods of significant energy release at 
redshifts $z \lesssim \pot{\rm few}{6}$ \citep{Zeldovich1969, Sunyaev1970diss, Illarionov1974, 
Danese1977, Burigana1991, Hu1993, Chluba2005, Chluba2011therm, Khatri2012b}.
However a large number of astrophysical and cosmological processes cause 
inevitable spectral distortions of the CMB at a level detectable using 
present day technology. The measure of the absolute spectrum by PRISM
will open an unexplored window to the early universe, allowing detailed studies of many expected or possible
processes of energy injection throughout cosmic history (Fig.~\ref{spectral_targets}), targeting several sources of early energy release 
(Fig.~\ref{spectralfigure}).

\subsection{Reionization and structure formation} 

Radiation from the first stars and galaxies 
\citep{Hu1994pert, Barkana2001}, feedback by supernovae \citep{Oh2003} and structure formation 
shocks \citep{Sunyaev1972b,cen1999,Miniati2000} heat the IGM at redshifts $z\lesssim 10-20$, 
producing hot electrons that up-scatter CMB photons, giving rise to a Compton $y$-distortion 
with average amplitude $\Delta I_\nu/I_\nu \approx 10^{-7}-10^{-6}$.
This signal will be detected at more than a $100\,\sigma$ with \Supercore, providing a 
sensitive probe of reionization physics and delivering a census of the missing baryons in the 
local Universe. \Supercore\ furthermore has the potential to separate the spatially varying 
signature caused by the WHIM and proto-clusters \citep{Refregier2000, Zhang2004}.
It also offers a unique opportunity to observe the free-free distortion associated with 
reionization \citep{ponenteetal2011}. The amplitude of the signal depends on the clumping of 
matter and thus provides complementary way to study the late evolution of inhomogeneities.

\begin{figure*}
\centering
\includegraphics[width=0.8\columnwidth]{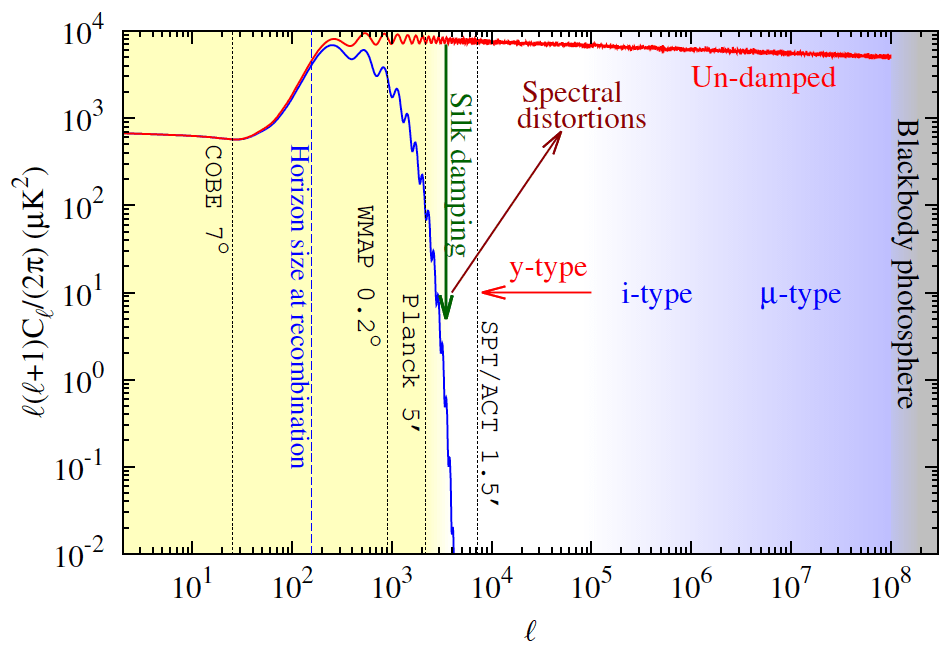}
\caption{\small Silk damping erases primordial anisotropies on small scales from the CMB, and current experiments 
have reached this resolution limit set by damping. CMB spectral distortions allow 
us to bypass this limit and extend our knowledge of initial conditions by many orders of magnitude in scale, 
complementing CMB anisotropy experiments. CMB spectral distortion measurements provides information 
that could be otherwise obtained (in the absence of damping) with a conventional anisotropy experiment 
with the extreme angular resolution of 6 milli-arcsec. Figure adapted from \cite{Khatri2013forecast}.
}
\label{powerspectrum_probe}
\end{figure*}

\begin{figure*}
\centering
\includegraphics[width=0.49\columnwidth]{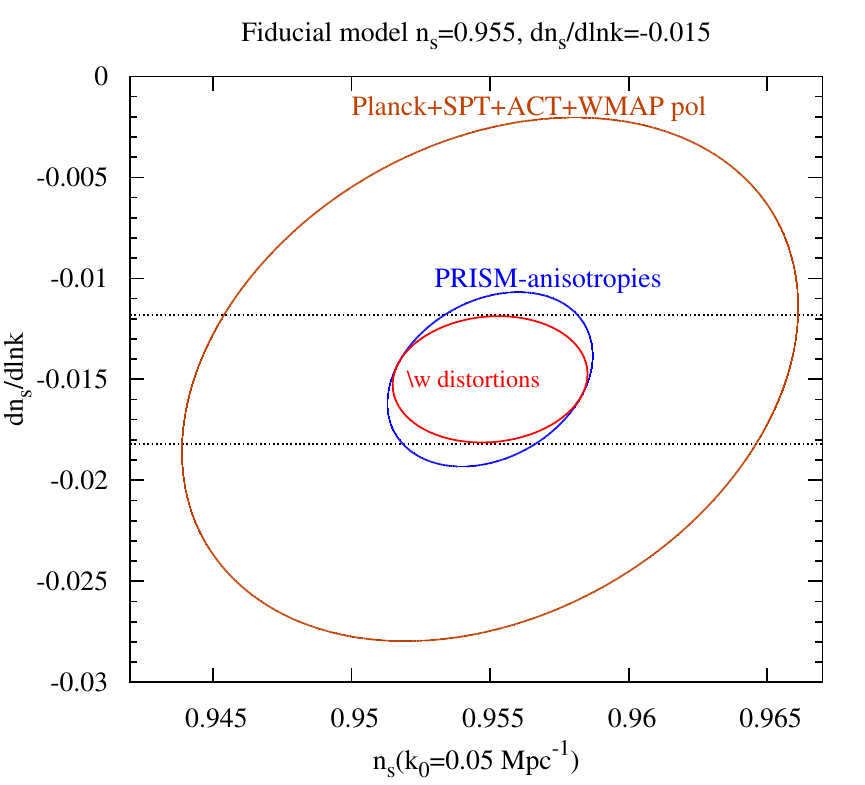}
\includegraphics[width=0.49\columnwidth]{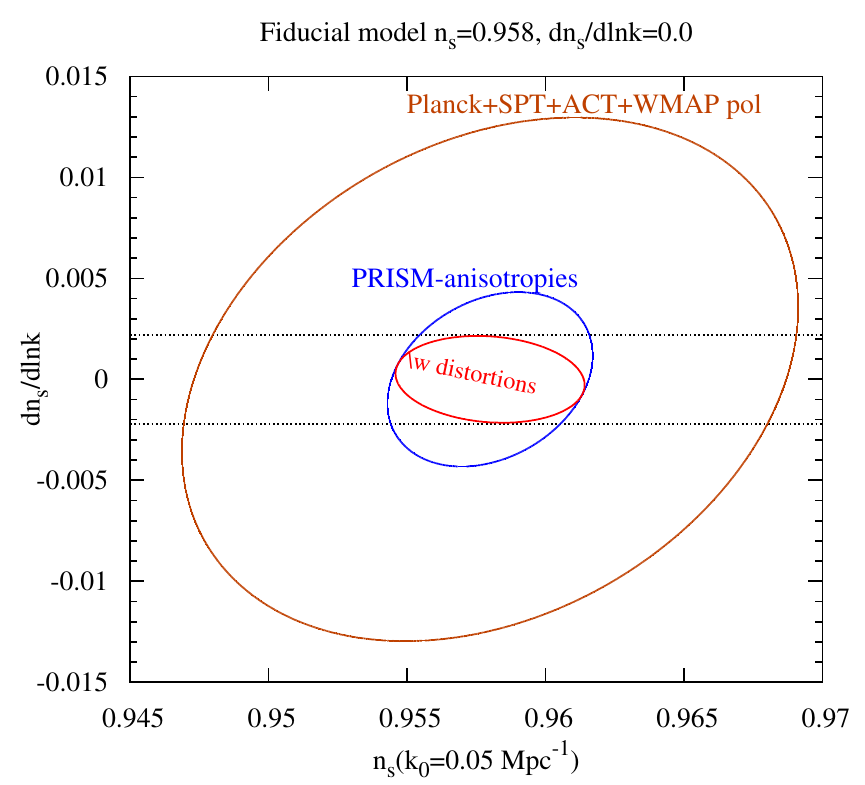}
\caption{\small Fisher forecasts for constraints on the primordial power spectrum from \Supercore\ for a power 
spectrum with (left panel) and without (right panel) running \citep[][was used for the 
computation]{Khatri2013forecast}. Combined power of anisotropies and CMB spectral distortions will nail down the 
running of power spectrum to about $\pm 0.002$ providing stringent tests of single-field inflation 
\cite{Lidsey1997RvMP...69..373L, Easther2006JCAP...09..010E}, significantly improving the constraints from 
current CMB experiments.}
\label{powerspectrum_constraints}
\end{figure*}


\subsection{Constraining the inflaton} 

\begin{figure*}
\centering
\includegraphics[width=0.4\columnwidth]{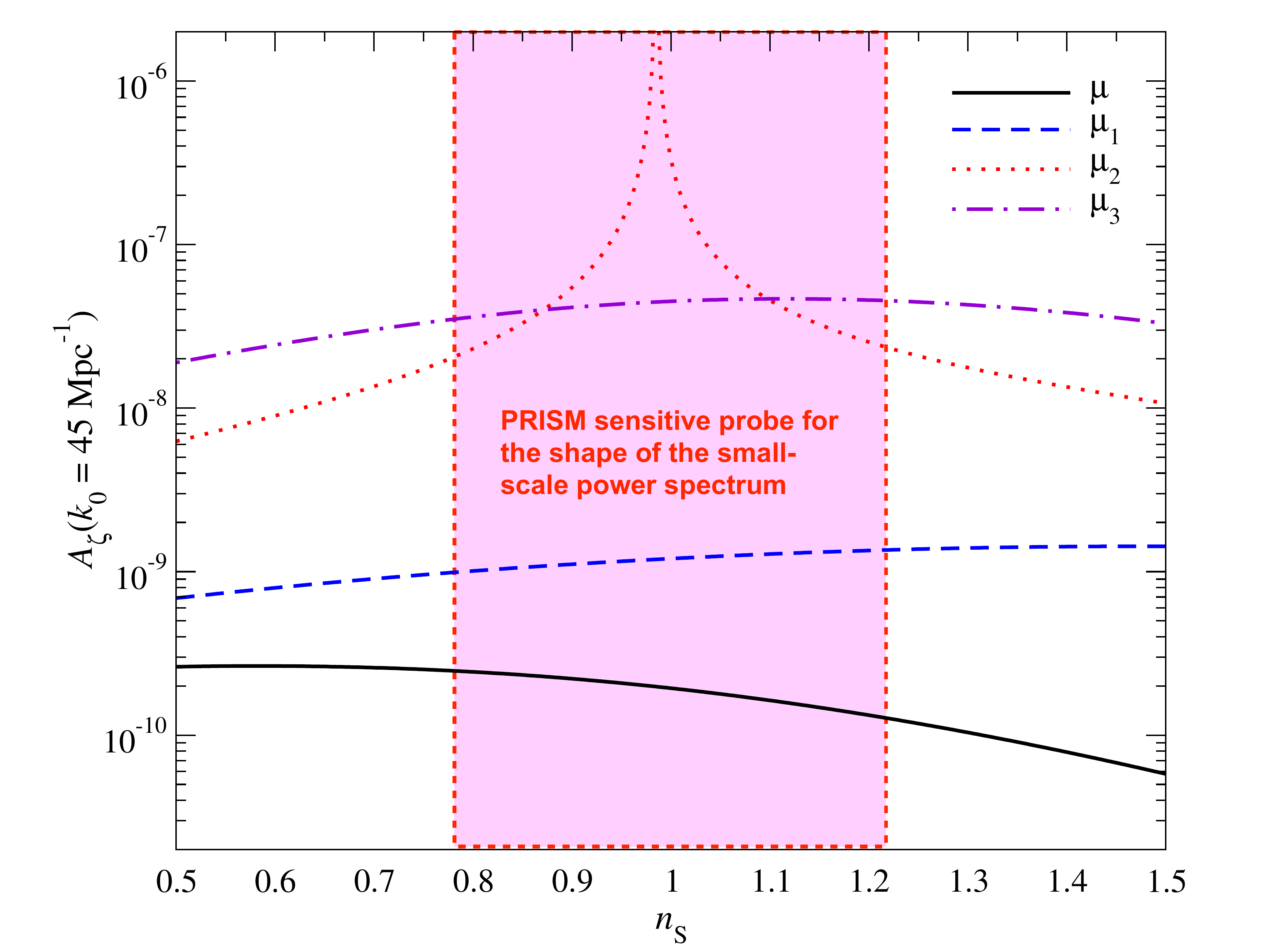}
\hspace{6.5mm}
\includegraphics[width=0.42\columnwidth]{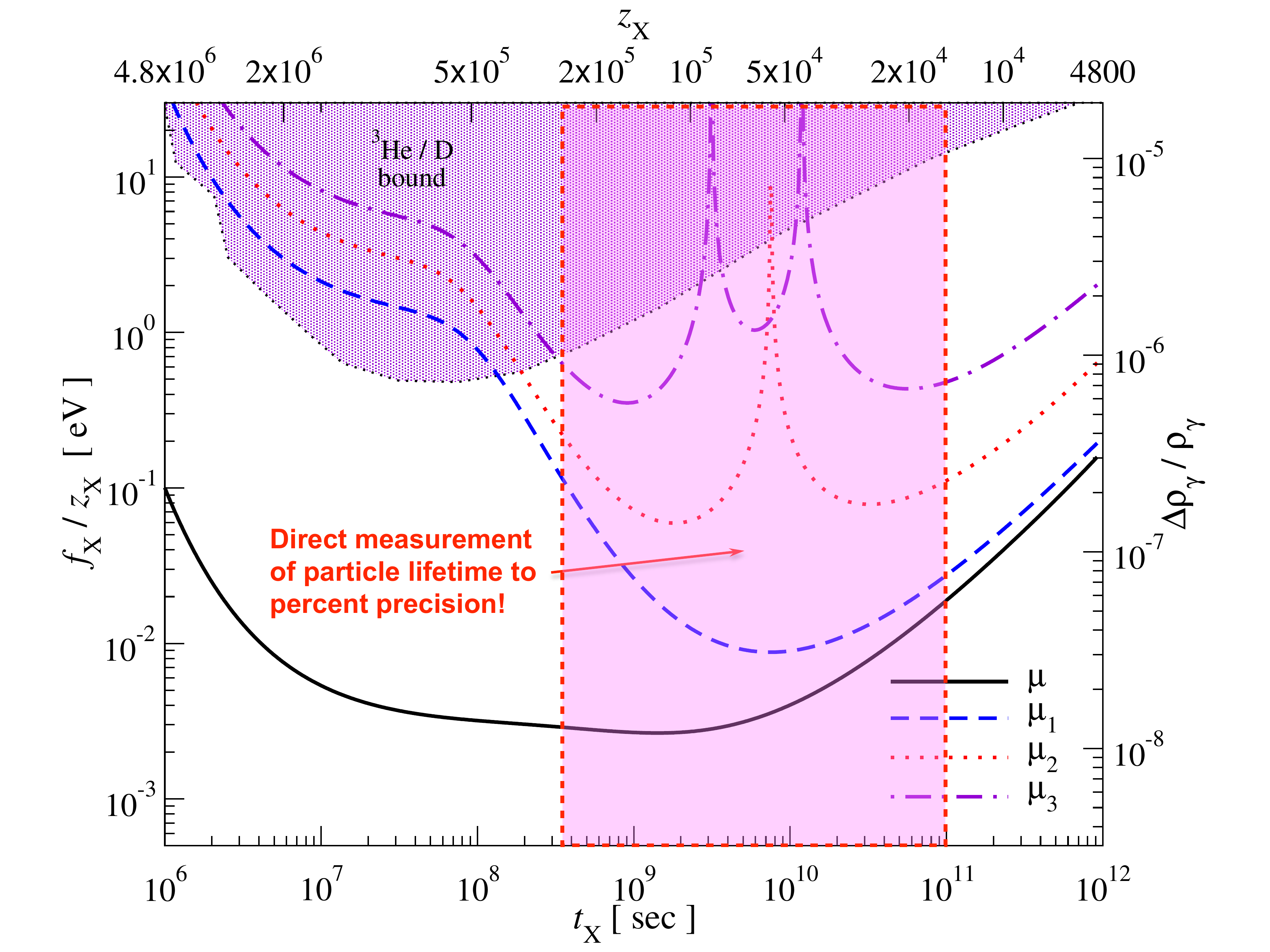}
\caption{\small 
\Supercore\ will provide a unique probe of the small-scale power spectrum (top panel) and decaying particles 
(bottom panel). Here we show the $2\sigma$-detection limits deduced from the $\mu$-parameter and the first few 
distortion principle components, $\mu_i$, \citep[see][for details]{Chluba2013PCA}. The extra information 
contained in the residual (non-$\mu$/non-$y$) distortion will independently constrain the amplitude and 
spectral index of the primordial power spectrum around pivot scale $k\approx 45\,{\rm Mpc}^{-1}$, superseding 
the constraint deduced from ultracompact minihalos or black hole abundances \citep{BSA11} by at least 2 orders of 
magnitude. The signatures of decaying particles will furthermore independently constrain their abundance 
(parameterized by $f_{\rm X}$) and lifetime, $t_{\rm X}$, with highest sensitivity to particles with $t_{\rm 
X}\approx {\rm few}\times 10^8\,{\rm sec}-10^{11}\,{\rm sec}$. \Supercore\ will improve existing limits 
from light elements \citep[we show those adapted from Fig. 42 of][]{Kawasaki2005} by some 2 orders of magnitude. 
These figures are adapted from \citep{Chluba2013PCA}.}
\label{decay_constraints}
\end{figure*}

Silk damping of small-scale perturbations of the photon fluid gives rise to CMB distortions 
\citep{Sunyaev1970diss,Daly1991,Barrow1991,Hu1994} that depend directly on the shape and amplitude of the 
primordial power spectrum at scales $0.6\,{\rm kpc}\lesssim \lambda \lesssim 1\,{\rm Mpc}$ (or multipoles 
$10^5\lesssim \ell \lesssim 10^8$) \citep{Chluba2012,Khatri2012short2x2} \JC{(see Fig.~\ref{powerspectrum_probe} 
and \ref{powerspectrum_constraints} for illustration)}. This allows constraining the trajectory of the inflaton 
at stages unexplored by ongoing or planned experiments \citep{Chluba2012inflaton,Powell2012, Khatri2013forecast, 
Chluba2013PCA}, extending our reach from 7 e-folds of inflation probed with the CMB anisotropies to a total of 17 
e-folds \JC{(see Fig.~\ref{decay_constraints} for detection limits)}. The signal is also sensitive to the 
difference between adiabatic and isocurvature perturbations \citep{Barrow1991,Hu1994isocurv, Dent2012, 
Chluba2013iso}, as well as primordial non-Gaussianity in the ultra squeezed-limit, leading to a spatially varying 
spectral signal that correlates with CMB temperature anisotropies as large angular scales \citep{Pajer2012, 
Ganc2012}. This effect therefore provides a unique way to study the scale-dependence of $f_{\rm NL}$ 
\citep{2013PhRvD..87f3521B}. CMB spectral distortions thus provide a complementary and independent probe of 
early-Universe physics, with \Supercore\ capitalizing on the synergies with large-scale B-mode polarization 
measurements.

\subsection{Decaying and annihilating relics}

The CMB spectrum allows placing tight limits on decaying and annihilating particles during the pre-recombination 
epoch \citep{Hu1993b, McDonald2001, Chluba2010a, Chluba2011therm}. This is especially interesting for decaying 
particles with lifetimes $t_{\rm X} \approx \JC{{\rm few}\times }10^{8}\,{\rm sec}-10^{11}\,{\rm sec}$ 
\citep{Chluba2013fore, Chluba2013PCA}, as the exact shape of the distortion encodes when the decay occurred 
\citep{Chluba2011therm, Khatri2012mix, Chluba2013Green}. Decays or annihilations associated with significant 
low-energy photon production furthermore create a unique spectral signature that can be distinguished from simple 
energy release \citep{Hu1993, daneseburigana94} \JC{(see Fig.~\ref{decay_constraints})}. \Supercore\ therefore 
provides an unprecedented probe of early-universe particle physics (e.g., dark matter in excited states 
\citep{Pospelov2007, Finkbeiner2007} and Sommerfeld-enhanced annihilations close to resonance 
\citep{Hannestad2011}), with many natural particle candidates found in supersymmetric models \citep{Feng2003, 
Feng2010}.

\subsection{Metals during the dark ages} Any scattering of CMB photons after recombination blurs CMB anisotropies 
at small scales, while producing new anisotropies at large scales. Electrons from the reionization epoch are the 
dominant source of optical depth, causing a frequency-independent signature already detected by WMAP and Planck 
\citep{WMAP_params, Planck2013params}. The resonant scattering of CMB photon by fine structure lines of metals 
and heavy ions produced by the first stars adds to this optical depth, making it frequency-dependent 
\citep{Loeb2001, Basuetal2004}. By comparing CMB temperature and polarization anisotropies at different 
frequencies one can thus determine the abundances of ions such as OI, OIII, NII, NIII, CI, CII at different 
redshifts \citep{Hernandezetal2006,Hernandezetal2007a}. Furthermore, UV radiation emitted by the first stars can 
push the OI 63.2\,$\mu$m and CII 157.7\,$\mu$m transitions out of equilibrium with the CMB, producing a 
distortion $\Delta I_\nu/I_\nu \approx 10^{-8}-10^{-9}$ due to fine structure emission 
\citep{Gongetal2012,Hernandezetal2007b}, providing yet another window within the reach of PRISM 
for probing reionization.

\subsection{Cosmological recombination radiation} The 
recombination 
of H and He
introduces distortions \citep{Zeldovich68, Peebles68, Dubrovich1975} at 
redshifts $z\approx 10^3-10^4$, corresponding to $\approx 260\,{\rm kyr}$ (\ion{H}{i}), $\approx 
130\,{\rm kyr}$ (\ion{He}{i}), and $\approx 18\,{\rm kyr}$ (\ion{He}{ii}) after the big bang 
\citep{Jose2006, Chluba2006b, Jose2008}. The signal is  
small ($\Delta I_\nu/I_\nu 
\approx 10^{-9}$) but its unique spectral features promise an independent determination 
of cosmological parameters (such as the baryon density and {\it pre-stellar} helium abundance)
and direct measurements of recombination dynamics, probing the Universe at stages well 
before the last scattering surface \citep{Sunyaev2009}. 

\begin{figure}
\centering
\includegraphics[height=10cm]{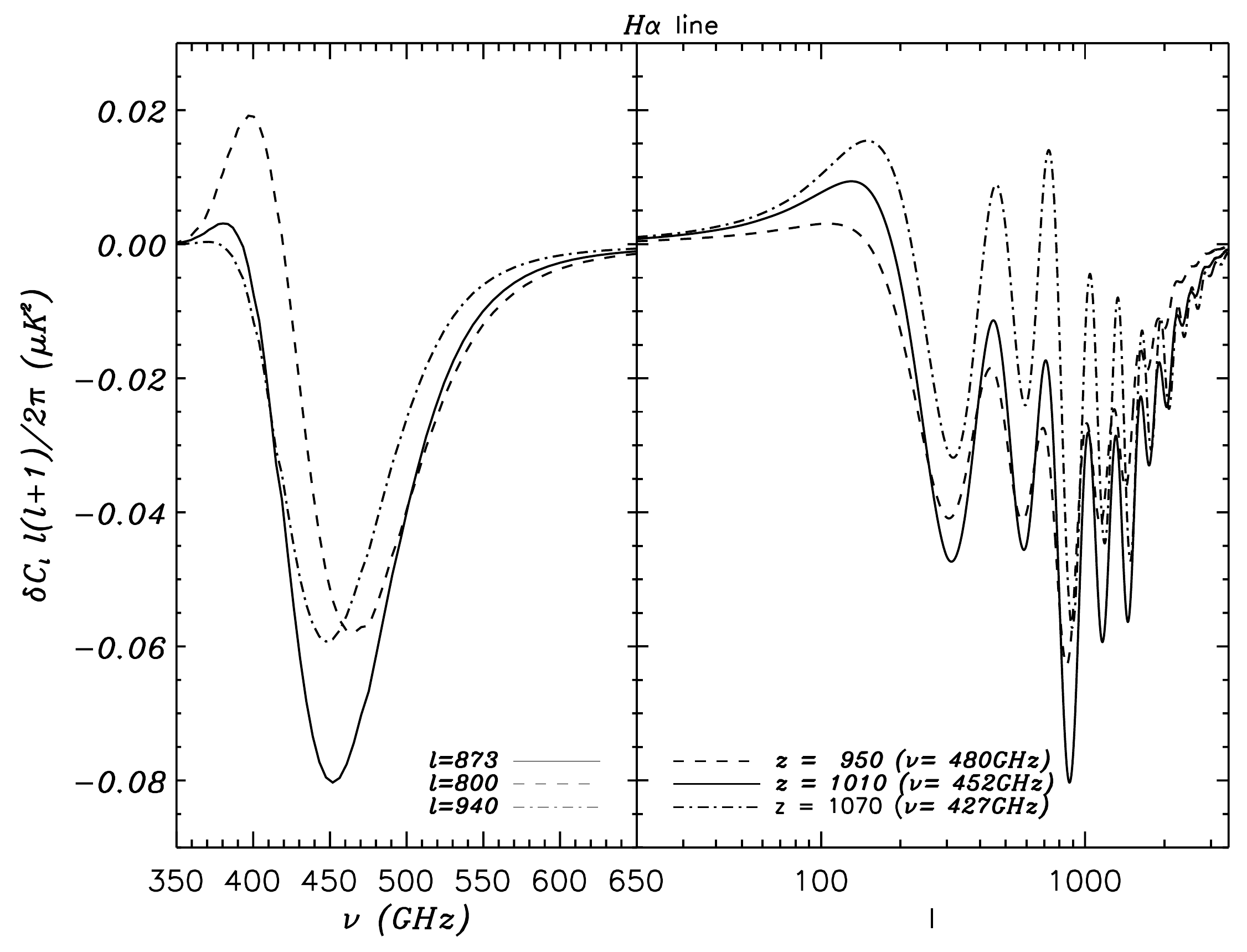}
\caption{\small
Change in the CMB angular power spectrum (TT) arising from the scattering of photons in the H$\alpha$
line during recombination. The left panel illustrates the signal as a function of the redshifted frequency, while 
the right panel shows the angular multipole. Solid lines in both panels refer to the cases in which the signal is 
largest. }
\label{fig:Halpha}
\end{figure}

At the highest sensitivity observing mode, \Supercore\ can directly detect
the H$\alpha$ and P$\alpha$ lines. Furthermore, if something unexpected happened during different stages of the 
recombination epoch, atomic species will respond \citep{Liubarskii83, Chluba2008c} by producing additional 
distortion signals that can exceed those of the normal recombination process. This behavior provides a unique way to 
distinguish pre- from post-recombination energy release \citep{Chluba2008c, Chluba2010a}.

\subsection{Hydrogen line scattering} 

Another interesting signature arises from resonant scattering of CMB 
photons by the hydrogen lines during recombination, introducing a frequency-dependent optical depth contribution 
and hence a CMB anisotropy signal \citep{RHS05, HS05,Hernandezetal2007a}. The effect on the TT power spectrum 
from the first lines of the Balmer and Paschen series is detectable with \Supercore\ and provides an additional 
opportunity to directly constrain the recombination history and obtain independent determinations of cosmological 
parameters (e.g. $\Omega_{\rm b}$ or $\Omega_{\rm m}$).  Figure~\ref{fig:Halpha} shows the predicted signal for 
the H$\alpha$ line, with a maximum amplitude of $\approx 0.3~\mu$K at frequencies of $\approx 450$~GHz and at small 
angular scales corresponding to $\ell \approx 870$. For the P$\alpha$ line, the intensity reaches $\approx 
0.02~\mu$K at $\approx 150$~GHz and the same angular scales.

\vspace{2mm}\noindent

\subsection{Rayleigh scattering} 

Neutral hydrogen produced as the Universe recombines is not completely transparent to the CMB. Rayleigh 
scattering by atomic hydrogen modifies the primary CMB anisotropies in both temperature and polarization by 
providing additional damping on small scales and also producing new temperature and polarization signals 
\citep{Yu2001,Lewis2013}.  \Supercore\ will measure the amplitude of temperature signal from Rayleigh scattering 
with $0.4\%$ accuracy and EE polarization signal at \JC{$\approx 20\sigma$}. The part of the Rayleigh signal 
correlated with primary CMB will constrain the recombination history. The uncorrelated part will probe 10,000 new 
modes at $\ell < 500$ \cite{Lewis2013}. Because of its $\sim \nu ^4$ characteristic frequency dependence, the signal
can be distinguished from the resonant scattering by the H$\alpha$ line \cite{RHS05}. 
The detection of the frequency dependent signals, even on large scales, is not limited by cosmic 
variance but by sensitivity and foregrounds. These signals are therefore ideal targets for \Supercore.

\subsection{Cooling of matter} 

The adiabatic cooling of ordinary matter continuously extracts energy from the CMB photon bath, leading to a small 
but certain distortion characterized by a {\it negative} chemical potential and $y$-parameter at the 
level of $\approx \pot{\rm few}{-9}$ \citep{Chluba2005, Chluba2011therm, Khatri2011BE}. \Supercore\ will be 
sensitive to this signal, delivering an independent way to determine the baryon density.

\subsection{Cosmic strings and primordial black holes} 

Cosmic strings arise naturally in many proposed theories of new physics
beyond the standard model unifying the electroweak and strong interactions as well as in many
superstring inspired inflation models. If the cosmic strings are superconducting cosmic strings, they
would also produce copious electromagnetic radiation, creating CMB spectral distortion of a unique 
shape distinguishable from other sources of distortions if detected at high S/N \citep{Ostriker1987, 
Tashiro2012, Tashiro2012b}. Evaporating primordial black holes provide another source of energy injection, with 
the shape of the resulting distortion depending on the black hole mass function \citep{Carr2010}.
For non-evaporating black holes, CMB spectral distortions could furthermore be used to constrain the black hole 
spin \citep{Pani2013}. \Supercore\ is sensitive to a rich variety of new physics and will in any case provide
constraints on physics beyond the standard model.

\subsection{PRISM's discovery potential using spectral distortion}

The above examples demonstrate that the CMB spectrum provides a rich and unique source of complementary 
information about the early Universe, with several predicted spectral distortions 
within the reach of PRISM. But the science is not limited to this. The CMB spectrum could 
also place interesting constraints on the power spectrum of small-scale magnetic fields 
\citep{Jedamzik2000}, decay of vacuum energy density \citep{BartlettSilk1990, Burigana1993, daneseburigana94}, 
axions \cite{Tashiro2013} and other new physics examples \citep{Lochan2012, Bull2013, Brax2013, deBruck2013, 
Caldwell2013}. Deciphering these signals is a challenge, but holds the potential for important
new discoveries and constraining unexplored processes that cannot be probed by other means.


\section{Structure of the dusty magnetized Galactic ISM}
%
%
%

The data analysis is still on-going but it is already clear that \Herschel\ and \Planck\  will have a profound and lasting impact on  our understanding
of the  interstellar medium and star formation. 
\mission\ holds even greater promise for breakthroughs. 
Dust and synchrotron radiation are the dominant contributions to  the sky emission and polarization to be observed by  \mission . 
Dust emission is an optically thin tracer of the structure of interstellar matter. 
Synchrotron radiation traces the magnetic field over the whole volume of the Galaxy, while 
dust polarization traces the magnetic field within the thin star forming disk, where the interstellar matter is concentrated. 
\mission\ will image these two complementary tracers 
with unprecedented sensitivity and angular resolution. It will also
provide all-sky images of spectral lines, which are key diagnostics of interstellar gas physics. 
No other initiative offers a comparable imaging capability of interstellar components 
over as wide a range of scales. 
In the following subsections we detail 
how \mission\  will address three fundamental questions of Galactic astrophysics: (1) 
What are the processes that structure the interstellar medium? (2)
What role does the magnetic field play in star formation? (3)
What are the processes that determine the composition and evolution of interstellar dust?

\subsection{Structure of interstellar medium} 

\Herschel\ far infrared observations have provided astronomers new insight into 
how turbulence stirs up the interstellar gas, giving rise to a filamentary, web-like structure within the diffuse interstellar medium and  molecular clouds.
\mission\  will extend the \Herschel\ dust observations to the whole sky and provide unique data on emission lines key to quantifying
physical processes.  The spectral range of \mission\  includes
atomic and molecular lines that serve as diagnostics of 
the gas density and temperature, its chemical state, and energy
budget. \Herschel\ has observed these lines along discrete lines of sight with very limited imaging.
By mapping these lines and dust emission over the whole sky at an angular resolution
comparable to that of \Herschel, \mission\  will 
probe the connection between  the structure of matter  and gas cooling across scales. 

The \mission\  sky maps will provide multiple clues to characterize the physical processes that shape
interstellar matter. The CII, CI, and OI fine structure lines and the rotational lines of 
CO and H$_2$O are the main cooling lines of 
the cold neutral interstellar medium and of molecular clouds. These lines
probe local physical conditions and the exchange of energy associated with the formation 
of molecular gas within the diffuse interstellar medium and of stars within molecular clouds. 
The NII lines at 122 and 205$\, \mu$m are spectroscopic tracers 
of the ionized gas. 
These lines are essential for distinguishing the contribution of neutral and ionized gas to the CII emission. 
\mission\  will have the sensitivity to image the CII line emission at sub-arcminute resolution even at the Galactic poles. 
The CII map can be combined with HI and dust observations
to study the formation of cold gas from the warm neutral medium through the thermal instability.
This analysis will probe the expected link, yet to be confirmed observationally, 
between the small-scale structure of the cold interstellar medium and gas cooling.
The CII line emission is also key to studying the formation of molecular gas by tracing the CO-dark H$_2$ gas \citep{2013arXiv1304.7770P}. 
In star forming molecular clouds, the CO, CI, OI, and H$_2$O lines
are the key tracers of the processes creating  
the initial conditions of star formation and of the feedback from newly formed stars on their parent clouds. 

\subsection{Galactic magnetic field and star formation} 

Star formation results from the action of gravity, 
counteracted by thermal, magnetic, and turbulent pressures \citep{2012A&ARv..20...55H}.  
For stars to form, gravity must locally become the dominant force. 
This happens when the turbulent energy has dissipated and matter has condensed 
without increasing the magnetic field by a comparable amount. 
What are the processes that drive and regulate the rate at which matter reaches this stage?
This is a long standing question to which theorists have over the decades offered multiple explanations, 
focusing on either ambipolar diffusion, turbulence, or 
magnetic reconnection to decouple matter from the magnetic field and allow the formation of 
condensations of gas in which stars may form \citep{2012ARA&A..50...29C}.

\mission\  observations of the
polarization in the far-IR and sub-mm will provide unique clues to understand the role of the magnetic field  
in star formation.  
Compared to  synchrotron radiation and  Faraday rotation, 
dust polarization images the structure of the magnetic field through an emission process tracing matter. 
It is  best suited to characterize  the 
interplay between turbulence, gravity, and the Galactic magnetic field. 
\mission\ will provide unique data to  
study magneto-hydrodynamical turbulence because it will drastically increase the spectral range of accurately probed magneto-hydrodynamical modes. 
The data will provide unprecedented statistical information to characterize  
the energy injection and energy transfer down to the dissipation scales.

Polarization data from the  \mission\  survey will have the  sensitivity and angular resolution required 
to map continuously the Galactic magnetic field  over the whole sky down to sub-arcminute resolution even at the Galactic poles. 
The wide frequency range of the mission will make it possible to
measure polarization for separate emission components (with distinct temperatures along
the line of sight). \mission\  will provide a new perspective on the structure of the magnetic field 
in molecular clouds, independent of grain alignment,  by 
imaging the polarization of  CO emission in multiple rotational  lines \citep{1981ApJ...243L..75G}.
No  project offers comparable capabilities.  Planck has provided the first all-sky maps of  dust polarization with 5' resolution but the data 
is sensitivity limited even at the highest Planck frequency (353 GHz).  
Ground based telescopes at sub-mm and millimeter wavelengths of bright compact sources at arcsecond resolution (for example with ALMA)
complement the full-sky survey of extended emission from
the diffuse interstellar medium and molecular clouds that only \mission\   can carry out.

\subsection{Nature of interstellar dust}

The combination of spectral and spatial information provided by  \mission\   will provide new tools for studying
the interstellar dust, in particular its nature and its evolution.
Dust properties (e.g., size, temperature, emissivity) are found to vary from one line of sight to another 
within the diffuse interstellar medium and molecular clouds. These observations indicate that dust grains evolve in a manner
depending on their environment  within 
the interstellar medium. They can grow through the formation of refractory or ice mantles,  or 
by coagulation into aggregates in dense and quiescent regions. They can also be destroyed by 
fragmentation and erosion of their mantles under more violent conditions.  The composition of interstellar 
dust reflects the action of interstellar processes, which contribute to breaking and reconstituting grains over timescales 
much shorter than the timescale of injection by  stellar ejecta. While there is broad consensus on 
this view of interstellar dust, the processes that drive its evolution in space are poorly understood \citep{2009ASPC..414..453D}.   
Understanding interstellar dust evolution is a major challenge  in astrophysics
underlying key physical and chemical processes in interstellar space. In particular, to fully exploit the \mission\ data we will need
to characterize where in the interstellar medium grains are aligned with respect to the Galactic magnetic field and with what efficiency. 

Large dust grains (size $> 10$~nm) dominate the dust mass. 
Within the diffuse interstellar medium, these grains are cold ($\sim10-20$~K) and emit within the \mission\ frequency range. 
Dipole emission from small rapidly spinning dust particles constitutes an additional emission component, known as anomalous microwave emission.
Magnetic dipole radiation from thermal fluctuations in magnetic nano-particles
may also be a significant emission component over the frequency range relevant to CMB studies \citep{2013ApJ...765..159D}.
To achieve the \mission\  objectives on CMB polarization, it is  necessary to characterize the spectral dependence of the polarized signal from 
each of these  dust components with high accuracy across the sky. This is a challenge but also a unique opportunity for dust studies.
The spectral energy distribution of dust emission and the polarization signal can be cross-correlated with the spectral diagnostics of 
the interstellar medium structure to characterize the physical processes that determine the composition and evolution of interstellar dust. 
The same data analysis will also elucidate the physics of grain alignment.


\section{Zodiacal light emission}

Zodiacal light emission (ZLE), or thermal emission from dust particles in our Solar System heated by the Sun, 
is brightest in the mid-infrared~\citep{Leinert1998}. ZLE, however, has been detected even at millimeter
wavelengths~\citep{PlanckCollaborationXIV2013} and constitutes an obstacle to the precise measurement 
of the total emission and large-scale variations of other emission components. The combination of 
very sensitive multi-frequency observations with the \mission\ imager over a frequency range in which 
ZLE and galactic dust have significantly different emission spectra and of \emph{simultaneous} 
absolute sky intensity at about $1.4^\circ$ angular resolution with the \mission\ spectrophotometer 
offers a unique opportunity not only to separate zodiacal light emission from galactic dust and 
other sky emission components, but also to decompose it into main components of different nature.

The bulk of the emission comes from the so-called {\it diffuse cloud}, 
but there are also contributions from a number of IRAS-discovered {\it dust bands} as well as a 
{\it circumsolar ring} and an {\it earth-trailing feature.} Measurements in the infrared and 
far infrared have begun to give us information on the size distribution of particles in the 
diffuse cloud and in the dust bands, but the abundance of millimeter-sized particles is not 
well-constrained in the circumsolar Ring and its associate earth-trailing feature. 

In addition, the Kuiper belt may have detectable
amounts of dust, and there is interest in finding dust associated with comets, Jupiter's fourth
and fifth Lagrange points, and other substructures in our solar system. 
While larger objects have been detected in these systems, only a sensitive survey
such as \mission\ can detect the faint, diffuse emission from larger dust grains which might be at 
these locations. Thus, while mapping the furthest reaches of the cosmos, \mission\
could also help us better understand the diffuse matter in our own neighborhood. 

While Galactic emission and the CMB anisotropies
themselves will always be brighter than zodiacal emission in the wavelength range near 150\,GHz, 
the power spectrum of the emission does reach the $\left(\mu K\right)^2$ level. 
This is comparable to the difference in the power spectra arising from different values of cosmological parameters detectable by \mission, 
particularly at low and intermediate multipoles. As there is enormous interest in CMB anisotropies at the largest angular scales, 
even these small contaminants can mask or mimic interesting effects. In the past, there have been claims of interplay between the large-scale 
CMB anisotropies and emissions or other systematics from our solar system---the so-called {\it anomalies.} 
While this connection seems more tenuous in light of the recent Planck data, the \mission\ combination of high 
precision mapping of temperature and polarization anisotropies with an absolute temperature calibration
will either definitely end such speculation or converge toward a more interesting outcome.

An interesting option would be 
to design the \mission\ orbit and scanning strategy such that it views the distant celestial sphere through different
columns of interplanetary dust, thereby allowing us to best measure the nearby zodiacal signal
and separate it from the background sky emission.



\section{Strawman mission concept}
\label{sec:mission}
\begin{figure}[th]
\centerline{\includegraphics[width=0.75\linewidth]{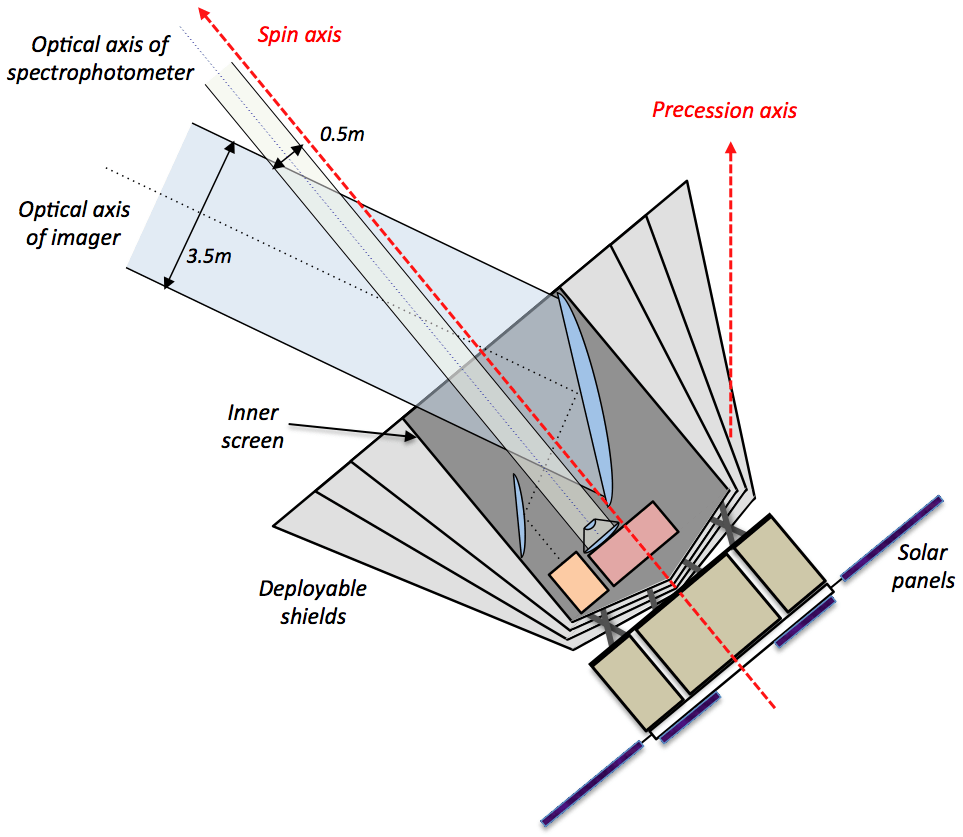}}
\caption{{\small The PRISM spacecraft with its two instruments: (1) the polarimetric imager with a 3.5m effective diameter 
telescope with the field of view at $\approx $30$^\circ$ with respect to the spacecraft spin axis; (2) the absolute spectrophotometer, 
with a $1.4^\circ$ beam aligned with the spin axis. While the general concept is similar to that illustrated in Fig.~7 of the white 
paper proposed to ESA \citep{2013arXiv1306.2259P}, here we propose an option where the spectrophotometer 
is located between the primary and secondary mirror of the imager, rather than behind the primary, allowing extra 
space for the optics of the larger telescope of the imager, as well as for the various screens necessary to provide an effective first
stage of passive cooling for the telescopes and instruments.
}}
\label{fig:compared_instr}
\end{figure}

The science program above requires measuring the sky brightness and polarization at high angular resolution and 
in many frequency bands across a wide spectral range as well as measuring the absolute spectrum of the 
sky background with moderate angular and spectral resolution. As a baseline, we propose to perform the best 
possible spectro-polarimetric sky survey in the 30-6000~GHz frequency range, with two instruments optimized for 
best joint performance sharing a single platform in orbit around the Sun-Earth L2 Lagrange point: (1) a 
\emph{polarimetric imager} (PIM) observing with about 30 broad and 300 narrow spectral bands with a diffraction 
limited angular resolution and a sensitivity limited by the photon noise of the sky emission itself; and (2) an 
\emph{absolute spectro-photometer} (ASP) that will measure sky emission spectra with a spectral resolution 
between 500~MHz and 15~GHz and an angular resolution of about 1.4$^\circ$. These complementary instruments will 
simultaneously map the absolute sky intensity and polarization with high sensitivity and with high spectral or 
spatial resolution. The data from both instruments can be binned (in frequency) and smoothed to obtain matching 
observations with $\delta \nu/\nu \approx 0.25$ and 1.4$^\circ$ resolution, allowing on-sky inter-calibration on 
large scales (and hence absolute calibration of the PIM). This will also enable correction of the ASP spectra 
from foreground contamination using high resolution component maps extracted from PIM data (e.g., $y$-distortion from clusters and galactic emission from regions unresolved in the coarse resolution ASP maps).

Since the scientific outcome of this mission depends on the complementarity of both instruments and on the control 
of systematic errors, a careful optimization of the ASP and the PIM (number and bandwidth of spectral bands vs. 
sensitivity) and of the mission (scanning strategy, joint analysis tools) with comprehensive simulations is an 
essential future phase of the mission study.

\subsection{Instruments}

\begin{table}[h]
\begin{center}
{\footnotesize
\begin{tabular}{|c|c|c|c|c|c|c|c|c|r|}
\hline 
\hline 
$\nu_0$ 		& Range 		& $\Delta \nu/\nu$ 	& $n_{det}$ 	& $\theta _{\rm fwhm}$ 	& \multicolumn{2}{|c|}{$\sigma_I$ per det}            		&  \multicolumn{2}{|c|}{$\sigma_{(Q,U)}$ per det} 	& Main molec. \& atomic lines \\       
 			&          		&           			&         		& 					& \multicolumn{2}{|c|}{\@ 1 arcmin} 	&  \multicolumn{2}{|c|}{\@ 1 arcmin} 	&  \\       
\cline{6-9}
$\rm GHz$ 	& $\rm GHz$ 	& 	&  			& 		   			& $\mu$K$_{\rm RJ}$    	& $\mu$K$_{\rm CMB}$   		& $\mu$K$_{\rm RJ}$    	& $\mu$K$_{\rm CMB}$  			& \\
\hline 
      30		& 26-34		& .25			&  50			& 17'					& 61.9			& 63.4				& 87.6			& 89.7					& \\  
      36		& 31-41		& .25			&  100		& 14'					& 57.8			& 59.7				& 81.7			& 84.5					& \\  
      43		& 38-48		& .25			&  100		& 12'					& 53.9			& 56.5				& 76.2			& 79.9					& \\  
      51		& 45-59		& .25			&  150		& 10'					& 50.2			& 53.7				& 71.0			& 75.9					& \\  
      62		& 54-70		& .25			&  150		& 8.2'				& 46.1			& 50.8				& 65.2			& 71.9					& \\  
      75   		& 65-85 		& .25			&  150     		&  6.8'				& 42.0			& 48.5				& 59.4   			& 68.6	     				& \\   
      90   		& 78-100 		& .25	 		&  200     		&  5.7'				& 38.0 			& 46.7				& 53.8  			& 66.0     					& HCN \& HCO$^+$ at 89 GHz \\   
      105   		& 95-120		& .25	  		&  250     		&  4.8'				& 34.5			& 45.6				& 48.8 			& 64.4     					& CO at 110-115 GHz\\   
      135   		& 120-150	& .25	  		&  300     		&  3.8'				& 28.6 			& 44.9				& 40.4 			& 63.4    					&  \\   
      160   		& 135-175	& .25	  		&  350     		&  3.2'				& 24.4			& 45.5				& 34.5 			& 64.3    					&  \\   
      185   		& 165-210	& .25	  		&  350     		&  2.8'				& 20.8			& 47.1				& 29.4 			& 66.6    					& HCN \& HCO$^+$ at 177 GHz \\   
      200   		& 180-220	& .20	  		&  350     		&  2.5'				& 18.9			& 48.5				& 26.7 			& 68.6    					&  \\   
      220   		& 195-250	& .25	  		&  350     		&  2.3'				& 16.5			& 50.9				& 23.4 			& 71.9    					& CO at 220-230 GHz \\   
      265   		& 235-300	& .25	  		&  350     		&  1.9'				& 12.2			& 58.5				& 17.3 			& 82.8    					& HCN \& HCO$^+$ at 266 GHz \\   
      300  		& 270-330	& .20	  		&  350     		&  1.7'				& 9.6				& 67.1				& 13.6 			&   94.9    					&  \\   
      320   		& 280-360	& .25	  		&  350     		&  1.6'				& 8.4				& 73.2				& 11.8 			&   103    					& CO, HCN \& HCO$^+$ \\   
      395   		& 360-435	& .20	  		&  350     		&  1.3'				&  4.9			& 107				&  7.0 			&   151    					&  \\   
      460   		& 405-520	& .25	  		&  350     		&  1.1'				&  3.1			& 156				&  4.4 			&   221    					& CO, HCN \& HCO$^+$ \\   
      555   		& 485-625	& .25	  		&  300     		&  55"				&  1.6			& 297				&  2.3 			&    420					&  C-I, HCN, HCO$^+$, H$_2$O, CO \\   
      660   		& 580-750	& .25	  		&  300     		&  46"				&  0.85			& 700				&  1.2 			&    990   					&  CO, HCN \& HCO$^+$ \\   
\hline 
\hline 
\cline{6-9}
			& 			& 	&  			& 		   			& nK$_{\rm RJ}$    	& kJy/sr   						& nK$_{RJ}$    	& kJy/sr    					&  
			\\
\hline 
      800   		& 700-900 	& .25			& 200      		&  38"				& 483  		& 9.5  						& 683   		& 13.4      						&  \\   
      960   		& 840-1080	& .25	 		& 200      		&  32"				& 390 		& 11.0						&  552  		&  15.6     						&  \\   
      1150   		& 1000-1300	& .25	  		& 200      		&  27"				& 361 		& 14.6 						&  510  		&   20.7    						&  \\   
      1380   		& 1200-1550	& .25	  		& 200      		&  22"				& 331 		& 19.4						&  468  		&    27.4   						&  N-II at 1461 GHz\\   
      1660   		& 1470-1860	& .25	  		& 200     		&  18"				& 290 		& 24.5 						&  410  		&    34.7   						&  \\   
      1990   		& 1740-2240	& .25	  		& 200      		&  15"				& 241 		&  29.3 						&  341  		&    41.5   						&  C-II at 1900 GHz\\   
      2400   		& 2100-2700	& .25  		& 200      		&  13"				& 188		&  33.3 						&  266  		&    47.1   						&  N-II at 2460 GHz\\   
      2850   		& 2500-3200	& .25	  		& 200      		&  11"				& 146		&  36.4 						&   206 		&    51.4   						&  \\   
      3450   		& 3000-3900	& .25	  		& 200      		&  8.8"				& 113 		&  41.4 						&   160 		&    58.5   						&  O-III at 3393 GHz\\   
      4100  		& 3600-4600	& .25	  		& 200      		&  7.4"				& 98 			&  50.8 						&   139 		&    71.8   						&  \\   
      5000   		& 4350-5550	& .25	  		& 200      		&  6.1"				& 91 			&  70.1 						&   129 		&     99.1  						&  O-I at 4765 GHz\\   
      6000   		& 5200-6800	& .25  		& 200      		&  5.1"				& 87	 		&  96.7 						&    124		&     136  						&  O-III at 5786 GHz\\   
\hline 
\hline 
\end{tabular}
}
\end{center}
\vspace{-4mm}
{
\baselineskip=0pt
\caption{{\small The 32 broad-band channels of the polarized imager
with a total of 7600 detectors. Sensitivities are averages for sky regions 
at galactic latitude and ecliptic latitude both higher than $30^\circ$. 
The mission 
sensitivity per frequency channel is the sensitivity per detector divided by $\sqrt{n_{det}}$. The angular resolution $\theta_{\rm fwhm}$ conservatively assumes under-illumination of the primary mirror by a Gaussian with 30dB edge taper to reduce sidelobe pickup.}}
\label{tab:PIM-bands}
}
\end{table}

\subsubsection{The polarimetric imager}

The optical configuration of the polarimetric imager relies on a dual off-axis mirror telescope with a 3.5$\,$m projected aperture primary mirror (corresponding to a physical size of $3.5 \times 4.2$ m) and a 0.8$\,$m diameter secondary mirror, coupled to a multi-band polarimeter. 
The broad-band PIM comprises 32 main channels of $\delta \nu/\nu \approx .25$ with dual-polarized 
pixel arrays (Table~\ref{tab:PIM-bands}). 
An instrinsic detector noise equal to the photon noise from the sky signal is assumed as well as a conservative 
optical efficieny of $40 \% .$ 
At frequencies below 700 GHz, the emphasis 
is on the sensitivity and control of systematics for CMB and SZ science.
Most of the frequency range will also be covered at higher spectral resolution ($\delta \nu/\nu 
\approx  .025$ or better) to map spectral lines. The exact frequency channels are still to be optimized and are not listed in Table~\ref{tab:PIM-bands}. 

A simple calculation assuming a focal plane unit based on classic feed horn coupled detectors for each spectral band, leads to a focal plane diameter of about 80~cm. While this concept is well understood, new technologies will help to 
reduce the focal plane size and the number of pixels (see discussion on detectors below).

\subsubsection{The absolute spectrophotometer}

A Martin-Puplett Fourier Transform Spectrometer (FTS) will allow for a large throughput and
deliver high sensitivity, differential measurements (in which the sky is compared to an
internal blackbody calibrator as in COBE-FIRAS). This setup also allows for an adjustable
spectral resolution that can be changed in the course of the mission. 
Dichroics at the two output ports can optionally
split the full 30-6000~GHz range into sub-bands with reduced 
photon noise. The instrument is cooled at 2.7K, so that the 
bolometric detector sensitivity is limited by photon noise from the sky.
Two operating modes are available:
high-resolution ($\Delta \nu \sim 0.5\,$GHz) and low-resolution
($\Delta \nu \sim 15\,$GHz). The sensitivity of the high-resolution
mode is 30 times worse than for the low-resolution mode. The
instrument beam is aligned with the spin axis of the satellite, so
that precession has a negligible effect during the
interferogram scan ($\sim$1s/10s long in the low-res/high-res
mode). The main characteristics for three possible configurations
of the instrument are detailed in Table \ref{tab:spectrometer}.

\begin{table}[h]
\begin{center}
{\footnotesize
\begin{tabular}{|c|c|c|c|c|c|}
\hline 
\hline Band & Resolution & $A\Omega$ & Background & NEP$\nu$ & Global 4-yr mission 
\\
(GHz) & (GHz) & (cm$^2$sr) & (pW) &
(W/m$^2$/sr/Hz$\times\sqrt{\rm s}$) & sensitivity (W/m$^2$/sr/Hz)
\\
\hline \hline
30-6000 & 15 & 1 & 150 & $1.8\times 10^{-22}$ & $1.8 \times 10^{-26}$ \\
\hline
30-500 & 15 & 1 & 97 & $7.0\times 10^{-23}$ & $7.2 \times 10^{-27}$ \\
500 - 6000 & 15 & 1 & 70 & $1.7\times 10^{-22}$ & $1.7 \times 10^{-26}$ \\
\hline
30-180 & 15 & 1 & 42 & $3.5\times 10^{-23}$ & $3.6 \times 10^{-27}$ \\
180-600 & 15 & 1  & 57 &  $6.3\times 10^{-23}$ & $6.5 \times 10^{-27}$ \\
600-3000 & 15 & 1 & 20 &  $7.4\times 10^{-23}$ & $7.6 \times 10^{-27}$ \\
3000-6000 & 15 & 1 & 28 &  $1.6\times 10^{-22}$ & $1.6 \times 10^{-26}$ \\
\hline
\hline 
\end{tabular}
}
\end{center}
\vspace{-4mm}
\caption{{\small FTS performance of
three possible configurations for photon noise limited
detectors, operating in a radiative background comprising
CMB, high galactic latitude interstellar dust, CIB, and high ecliptic latitude
interplanetary dust emission. 
With an entrance pupil 50
cm in diameter, the baseline throughput is $\sim 1 \, {\rm cm}^2{\rm sr}$ and
the angular resolution 1.4$^\circ$. The theoretical monopole sensitivity
for each spectral bin is reported in the last column assuming 4 years of observation and 75\% useful sky. The actual
sensitivity taking into account efficiency factors can be 2-3 times worse. 
Line 1 is a configuration with an ultra-wide spectral coverage obtained
with one detector in both output ports. In lines 2-3 the detectors 
at the output ports are sensitive to different bands. 
In lines 4-7 each output port is
split into two sub-bands using dichroics to minimize photon noise in
the low-frequency bins.}}
\label{tab:spectrometer}
\end{table}

Using detectors with $A\Omega \! \sim \! 1 \, {\rm cm}^2{\rm sr}$ and
angular resolution $\sim$1.4$^\circ$, we estimate that the CIB
can be measured with $S/N=10$ in a fraction of a second at 1500 GHz 
and in $\sim 10$ seconds 
at 140 GHz, while a $y$-distortion $\sim 10^{-8}$ can be measured
with $S/N=10$ at 350$\,$GHz in two hours of integration. Recombination
lines could be measured integrating over the whole mission if the
overall stability of the instrument and the quality of the
reference blackbody are sufficient.

The main issue for this instrument is the control of systematic
effects. The instrument design allows for a number of zero tests
and cross-checks on the data. The main problem is to control the
blackness of the reference and calibration blackbodies with the
required accuracy. Reflectivities lower than $R=-50 / \! -\! 60\,$dB
have been obtained in the frequency range of interest in the
Planck and ARCADE references. Building on these experiences,
we plan to achieve $R<-70\,$dB through a combination of
electromagnetic simulations and laboratory emissivity measurements
on improved shapes and space qualified materials.

\subsubsection{Synergy of the two instruments}

There is significant added value to observing with the two instruments
together on the same payload. The observed data cubes from the absolute spectrophotometer comprise CMB and foreground emission, 
mixed and blended by both superposition along the line of sight and averaging across several lines of sights 
smeared by the $1.4^\circ$ beam.
Foreground emissions include galactic dust cirrus, spectral lines in molecular clouds, 
extragalactic sources, clusters of galaxies, and variable radio sources. 

To achieve the ambitious precision goals of the spectrophotometer for CMB spectral distortion measurements, 
it is essential that these sources of foreground emission are characterized and subtracted using the narrow 
beam imager data. 
Without the high angular resolution data from the imager,
precise measurement of the CMB spectral distortions are possible only on a reduced 
fraction of clean sky, resulting in significant degradation of its global sensitivity.
The interpretation of the observations would also rely on how some of the components are modelled. 

Conversely, absolute 
spectrometer data are essential to measure the zero-level of the imager 
maps (and hence to measure absolute column densities for ISM emissions, the total 
emission of the CIB, etc.) as well as for accurate absolute photometric calibration of 
the imager channels across frequencies (with a goal of 0.05\%). 
In this way the absolute calibration is transferred to the imaging polarimeter 
to yield high-resolution absolute maps of sky emission. The precise calibration
of all measurements across frequencies is key to component separation, 
in particular to extracting with minimal contamination faint sources of radiation 
identified using the frequency dependence of their emission.

\subsection{Experimental challenges}

The following subsections address some of the technologically challenging requirements 
of the \mission\ mission. The main features essential to carry out the proposed science program are 
the large telescope, necessary for the high angular resolution needed for SZ, 
extragalactic, and galactic science; telescope cooling to temperature below 10 K, essential 
for the sensitivity at frequencies above 1~THz; a large focal plane of thousands of detectors, 
for sensitivity and spectral coverage; the quality of the calibration, essential for accurate component 
separation; and the combination of the co-observing imager and spectrophotometer, to measure both the absolute level 
of all maps and absolute spectrum of CMB emission, as well as the sky brightness and 
polarization fluctuations down to the smallest possible angular scales.

\subsubsection{Telescope}

The telescope is designed with the largest possible monolithic primary mirror to obtain 
the best angular resolution achievable within the budget of a large ESA mission. 
The surface accuracy must allow for observations down to a wavelength of 50 microns.
An off-axis configuration is preferred because it allows for a larger focal plane area and better polarization 
properties (without, in particular, struts holding the secondary mirror in the optical path).

The science achievable with \mission\ depends on the angular resolution, especially for 
detecting galaxy clusters and measuring their peculiar velocities, and for resolving the distant galaxies that 
constitute the CIB. The dependance of mission performance on telescope size, however, is not very steep, 
so that the appropriate compromise between feasibility, cost, and scientific requirements can be made during
the mission design phase.

\begin{figure}[th]
\centerline{
\includegraphics[height=0.6\linewidth]{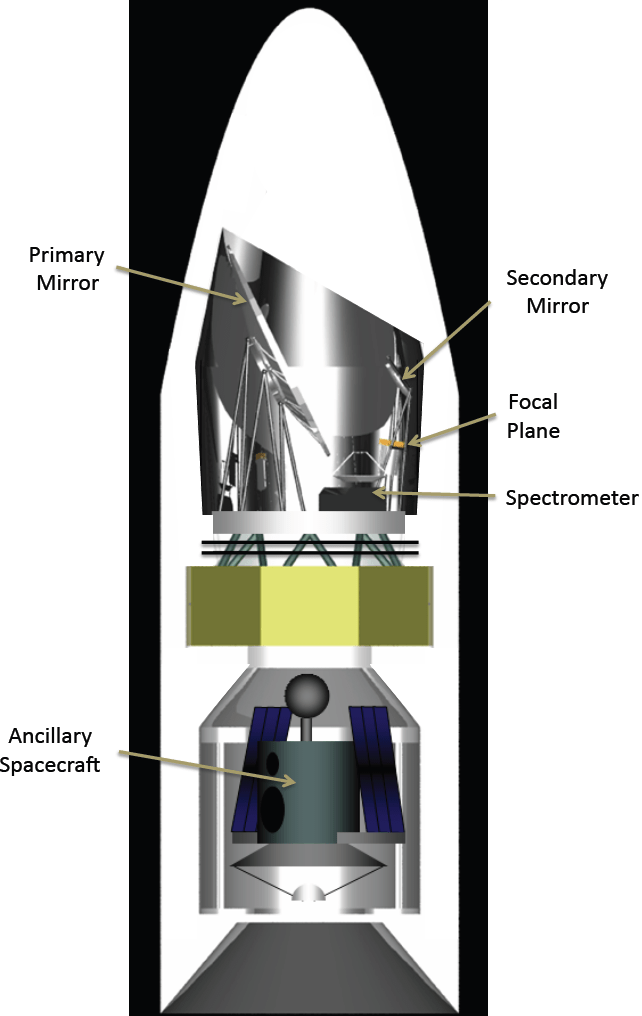}
\hskip 5mm
\includegraphics[height=0.6\linewidth]{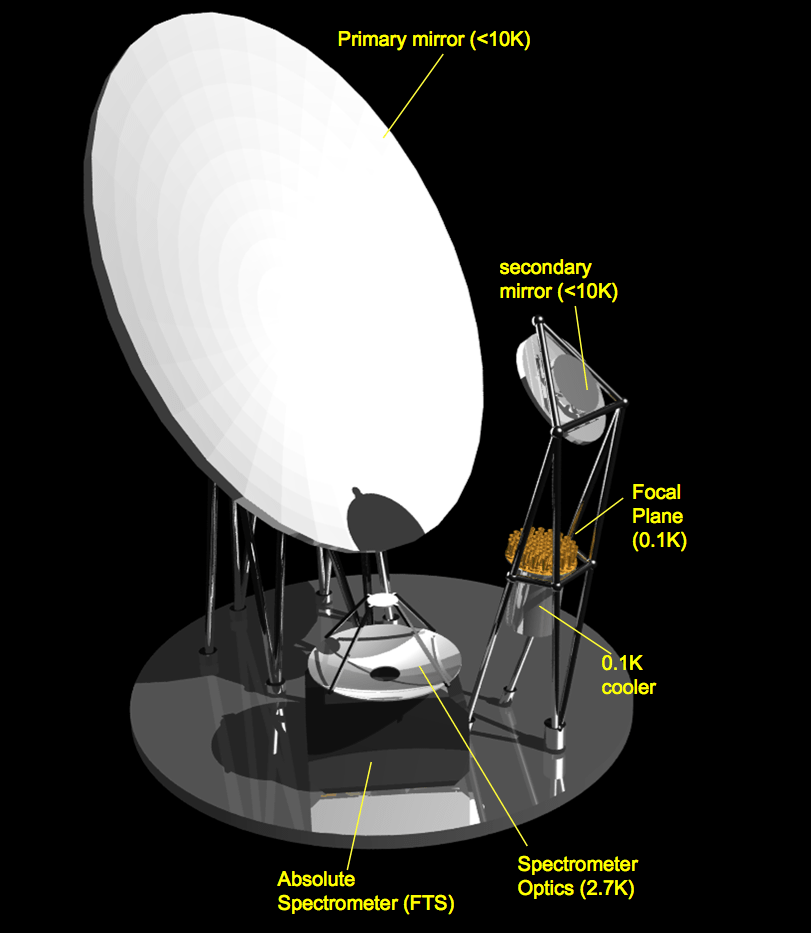}
}
\caption{{\small  
Left panel: The two satellites inside the Ariane-V fairing. 
The main satellite is on top (the Sun shields, in stowed position, are not shown). 
The bottom satellite (displayed in the Ariane-V SYLDA for a possible launch configuration) 
provides a set of calibrators for the observatory and the high gain, high data rate communication system.
Right panel: Detail of main satellite, showing a possible layout for the two PRISM instruments
with the locations of the off-axis telescope with a 3.5 x 4.2 m primary, 
the polarimeter focal plane, and the spectrometer.}}
\label{fig:instruments}
\end{figure}

Figure \ref{fig:instruments} shows the envisaged \mission\ layout for both instruments. The primary 
mirror of the imager is $3.5\textrm{m} \times 4.2\,$m in size. The design and production of the large telescope
will benefit from the \Herschel\ mission heritage 
(3~m class telescope, with instruments operated 
down to 60~microns), and the studies and development activities for the \SPICA\ mission 
($\sim \! 3\,$m diameter primary mirror cooled to about 5~K and instruments operating down to 5~microns).

\begin{figure}[th]
\begin{center}
\includegraphics[width=0.7\textwidth]{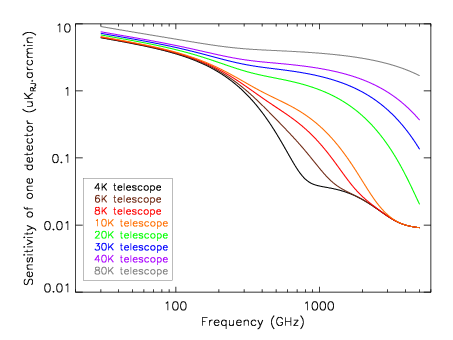}
\caption{\small Sensitivity of one PRISM detector as a function of frequency for various temperatures 
of a 2\% emissive telescope. The sensitivity is plotted as the average of sky regions 
at galactic and ecliptic latitudes higher than $30^\circ$. The background comprises CMB, 
thermal dust emission, CIB, zodiacal light, and telescope emission (assuming 2\% total telescope emissivity). }
\label{fig:telescope_temp}
\end{center}
\end{figure}

Actively cooling the telescope to 4$\,$K (mission objective) instead of 40$\,$K (achievable by passive cooling)
substantially improves the sensitivity, especially for frequencies above 200~GHz (Fig.  \ref{fig:telescope_temp}). 
The goal is to reach a mirror temperature of 4 to 5$\,$K with a requirement of 10$\,$K, necessary to map the 
complete sky at frequencies up to 6 THz (50 microns) with a sensitivity better than that of \Herschel\ SPIRE in 
large survey observations (which cover approximately $10\% $ of the sky). 

Figure \ref{fig:telescope_temp} shows the sensitivity of a single \mission\ detector 
as a function of telescope temperature, assuming detector noise equal to photon 
noise, and 25\% bandwidth. At the highest frequencies (above 3 THz, i.e., below 
100 microns), where zodiacal light emission dominates the sky contribution to 
the photon noise, a temperature of 10 K or lower is sufficient to reach the best possible 
detector sensitivity with a sky-dominated background. Below 200 GHz little is gained 
by actively cooling the telescope, as compared to passive cooling to 30-40 K. 
The intermediate frequencies (in particular the 400-2000 GHz range) are where the best sensitivity is 
achieved with the coldest possible mirror temperature. While increased sensitivity is not 
critical for the detection of dusty galaxies (which is background limited), it is important 
for the detection of galaxy clusters via the thermal SZ effect, which is one of the main
mission objectives.

\subsubsection{Cooling chain}

In addition to the cooling of the large telescope of the polarimetric imager to  
below 10 K, the 50 cm telescope of the spectrophotometer will be 
cooled to 2.7 K, and the detectors will operate at sub-Kelvin temperatures.

The cooling chain will rely on a first stage of passive cooling of the payload using a 
set of deployable V-grooves, as well as an inner solid cylindrical shield 
(Fig.~\ref{fig:compared_instr}). The large mirror of the polarimetric imager will be 
cooled using a cryogenic chain that will use as a starting point the study of 
the {\it SPICA} 3.2~meter telescope cooling system, for which a temperature of $5$-$6\,$K can be achieved. 
\mission\ has a mirror about 30\% bigger in area, but the requirement on the telescope 
temperature ($<10\,$K) is less stringent (however with an objective of $4\,$K if possible
for optimal sensitivity).
The focal planes of both instruments will be cooled to $0.1\,$K using a cryogenic system 
adapted from the dilution refrigerator onboard \planck\ but with continuous helium recycling 
for a longer mission duration of 4 years (baseline) or beyond.

\subsubsection{Scan strategy}
The observing strategy must provide: 
(1) full sky coverage for both instruments; 
(2) cross-linked scan paths and observation of all sky pixels in many orientations for all 
detectors of the polarimetric imager;
(3) fast scanning of the imager to avoid low-frequency drifts; 
(4) slow scanning for the spectrophotometer field of view to allow for few seconds 
long interferogram scans with negligible depointing; 
(5) avoiding direct solar radiation on the payload.

These requirements can be satisfied by a spinning spacecraft with the beam of the spectrophotometer 
aligned along the spin axis and the imager field of view offset by $\theta_{\rm spin} 
\approx 30^\circ$ (Fig. \ref{fig:compared_instr}). During each spacecraft rotation 
(with  $\omega_{\rm spin}$ of a few rpm), the field of view of the imager scans 
circles of diameter $\approx 2\theta_{\rm spin}$ while the beam of the spectrophotometer rotates in place.
A slow precession of the spin axis (with a period between a few hours and one 
day) with a precession angle $\theta_{\rm prec} \approx 45^\circ$ results in slow scans of the 
spectrophotometer on large circles of diameter $2\theta_{\rm prec}$. Finally, the precession axis 
evolves by about 1$^\circ $ per day along the ecliptic plane to keep the 
payload away from the Sun, and also slowly moves perpendicular to the ecliptic plane 
so as to map the ecliptic poles. Deployable screens isolate the payload from the 
heat from the Sun, providing a first stage of passive cooling 
to $\approx 40\,$K.

\subsubsection{Polarization modulation}

The precise measurement of polarization imposes stringent requirements on the instrument characterization.
Polarization modulation provides essential redundancy and robustness tests.
The polarization modulation baseline for \mission, similar to the solution proposed in the 
previous SAMPAN and EPIC studies, relies on the scanning strategy and the rotation of the 
entire payload. However alternative strategies such as the use of a half-wave plate in front of 
the focal plane (the receivers being the major source of instrumental polarization) will be 
investigated during a trade-off analysis.

\subsubsection{Detectors} 

The \mission\ mission will observe the sky using a focal plane of 7600 broadband detectors in the millimeter to far infrared frequency
range. It is also planned to include a number of narrow band detectors for 
line monitoring and limited spectroscopy.

Direct detectors (such as TES bolometers, CEBs or KIDs) are the most 
sensitive detectors at millimeter wavelengths. Bolometers have achieved in-flight 
performance with the Planck \citep{2011A&A...536A...4P} and Herschel 
\citep{2010A&A...518L...3G} missions close to the instrinsic photon noise limit. 
Large bolometer arrays with thousands of pixels are 
currently being used on large ground-based telescopes. 

These technologies have not yet been demonstrated to be viable 
for 30 to 70~GHz. Nevertheless, it is likely that their efficiency 
at low frequencies will improve over the next few years. For instance, 
the studies in ref.~\citep{2007stt..conf...93K} have shown that 
70$\,$GHz CEBs could lead to NEPs of (few)$\times 10^{-18}~{\rm W}\cdot {\rm Hz}^{1/2}$. 
As an alternative solution, 
the \mission\ instruments could take advantage of the recent breakthroughs in 
cryogenic HEMT technology, with sensitivities predicted to reach 2-3 times the quantum limit 
up to 150-200$\,$GHz (instead of 4-5 times up to 100$\,$GHz so far). The fact that these 
devices allow for cryogenically cooled miniaturized polarimeter designs will simplify the
thermo-mechanical design. Hence, while a single detector technology throughout the instruments 
would be preferable, the option of using a combination of HEMTs and bolometers remains open 
(Tables \ref{tab:detechno} and \ref{tab:techno}).
While the whole focal plane could be built from pixels based only on direct detectors, the detector 
technology and optical coupling concept will probably vary across the spectral domain from 100 to 6000~GHz. 
Table \ref{tab:detechno} lists potential technology candidates. 

\begin{table}[htdp]
\centering{}%
{
\begin{tabular}{|c|c|c|c|}
\hline
\hline
Frequency (GHz)&Technology&TRL&Used for \\
\hline
30 - 90 & HEMT & 9 & WMAP, Planck-LFI\\
& Bolometers & 2-3 & \\
\hline
90 - 400&Feedhorn coupled TES&7&Ground/balloon projects, SPICA \\
&Antenna coupled TES & 5& PolarBear, SPIDER\\
&KIDs&5&NIKA, MUSE\\
&CEBs&4&Research\\
\hline
400 - 2000 & Antenna coupled TES & 5 & PolarBear, SPIDER\\
& Planar Resistive Bolometers &9& Herschel-PACS \\
\hline
2000 - 6000 & Planar Resistive Bolometers &9& Herschel-PACS\\
& HEB, SIS & 9 &Herschel-HIFI \\
\hline
\hline
\end{tabular}
}
\caption{\small Available detector technologies 
in the \mission\ frequency range with Technology Readiness Level (TRL) 
and experiments that have observed with the corresponding detector
technology.}
\label{tab:detechno}
\end{table}%

\begin{table}[ht]
\centering{}%
{
\begin{tabular}{|c|c|c|c|c|c|c|}
\hline 
$\nu_{c}$ range & Req. NEP & Req. $\tau$ & \multicolumn{4}{c|}{Focal plane technology}\tabularnewline
\hline 
\multirow{2}{*}{$\left[{\rm GHz}\right]$} & \multirow{2}{*}{$\left[10^{-18}\, {\rm W}/\sqrt{\rm Hz}\right]$} 
& \multirow{2}{*}{$\left[ms\right]$} & \multicolumn{2}{c|}{Detector technology} & \multicolumn{2}{c|}{Optical coupling}\tabularnewline
\cline{4-7} 
 &  &  & Baseline & Backup & Baseline & Backup\tabularnewline
\hline 
\hline 
{\small 30 - 75} & 3.3 -- 5.7 & {\small 2.96 -- 1.18} & {\small TES} & {\small HEMT} & {\small MPA/CSA} & {\small HA}\tabularnewline
\hline 
{\small 90 - 320} & 4.6 -- 7 & {\small 1.18 -- 0.4} & {\small TES} & {\small KIDS} & {\small HA+POMT} & {\small MPA}\tabularnewline
\hline 
{\small 395 - 660} & 0.94 -- 3.1 & {\small 0.4 -- 0.13} & {\small TES} & {\small KIDS} & {\small MPA/CSA} & {\small LHA}\tabularnewline
\hline 
{\small 800 - 6000} & 0.011 -- 0.63 & {\small 0.13 -- 0.01} & {\small KIDS} & {\small HEB/CEB} & {\small MPA/CSA} & {\small LHA}\tabularnewline
\hline 
\multicolumn{1}{c}{} & \multicolumn{1}{c}{} & \multicolumn{1}{c}{} & \multicolumn{1}{c}{} & \multicolumn{1}{c}{} & \multicolumn{1}{c}{} & \multicolumn{1}{c}{}\tabularnewline
\end{tabular}
}
\vspace{-4mm}
\caption{\small Required NEP (Noise Effective Power) and time constants for various frequency ranges and 
corresponding baseline and backup focal plane technology. TES: Transition Edge Sensors 
(Technology Readiness Level 5); HEMT: High Electron Mobility Transistor (TRL 5); KID: Kinetic 
Inductance Detector (TRL 5); HEB: Hot Electron Bolometer (TRL 4); CEB: Cold Electron Bolometer 
(TRL 3); HA: Horn Array (TRL 9); LHA: Lithographed Horn Array (TRL 5); MPA: Multichroic Planar 
Antenna (TRL 4); CSA: Crossed Slot Antenna (TRL 5); POMT: Planar Ortho-Mode Transducer (TRL 
5).}
\label{tab:techno}
\end{table}

In order to reduce the size of the focal plane, 
multichroic pixels could be used where one antenna can feed 2 or 3
spectral bands (4 or 6 detectors). For instance, waveguide filtering, which 
could be adapted to split the bands, has been improved lately \citep{Leal2013}, and 
multichroic TES detectors are being developed and tested for PolarBear II 
and LiteBird. Moreover a recent ESA ITT has been released in the last few months 
in order to study large focal plane concepts in Europe.

\subsubsection{Detector time constants} 

The rapid scanning by  \mission\ requires fast 
detector time constants, of order 1$\,$ms at 100 GHz, down to $\sim 10\, \mu$s at 6 THz. These 
time constants are challenging (especially at high frequencies), but have already been 
achieved using recent TESs, KIDs and CEBs.

\subsection{Ancillary spacecraft}

We propose that the mission include a small ancillary spacecraft serving the 
following functions:

\mypar{Telecommunication:} The high resolution mapping of the full sky with the many 
detectors of \mission\ with a lossless compression of 4 gives a total data rate of 
$\sim350\,$Mbit/s (out of which $300\,$Mbit/s is from the channels above 700 GHz). Further 
on-board reduction by a factor $\sim\!10\!-\!20$ can be achieved by averaging the timelines 
of detectors following each other on the same scan path (after automatic removal of spikes 
due to cosmic rays) to yield a total data rate $<40\,$Mbit/s (a few times higher than Euclid or 
Gaia). While a phased-array antenna or counter-rotating antenna on the main spacecraft is a 
possible solution, decoupling the communication function from the main spacecraft 
using an ancillary spacecraft as an intermediate station for data transmission has several 
advantages. Most notably, telecommunications with Earth use power from the 
ancillary spacecraft, allowing power from the main spacecraft to be redeployed 
for payload cooling. 
Using a large pointed 
antenna and the additional power available for telemetry
increases the data rate to Earth.
Finally, because the main spacecraft 
does not have to be pointed toward Earth, 
a more flexible scanning strategy is possible allowing better polarization 
modulation and optimal distribution of the observation time over the sky.

\mypar{In-flight calibration:} The formidable \mission\ design challenge is to ensure that 
performance is limited by detector noise rather than by systematic effects and calibration 
uncertainties. While pre-flight calibration is necessary, an 
ancillary spacecraft
fitted with calibrated, 
polarized sources could be used for precise in-flight calibration of the polarization 
response and polarization angles of the detectors as well as for main beam and far sidelobe 
measurements down to extremely low levels (below -140~dB) at several times during the 
mission lifetime.

While the relative position of the two satellites will have to be monitored precisely to
achieve a goal of $0.05$\% accuracy in relative calibration, this is well within what can be 
achieved with standard ranging techniques. No precise formation flying is required.

 
\section{Competition and complementarity with other observations}
In this last section, we discuss the \mission\ mission in the general framework of relevant current, planned, and proposed
observations.

\subsection{Ground-based and balloon-borne CMB B-mode experiments}

A key scientific 
objective of \mission\ is the detection and characterization of the CMB polarization B-modes and 
corresponding constraints  of inflationary models (Sec.~\ref{sec:CMB-B-modes}). 
Efforts from the ground and balloons are also seeking to discover the B-modes despite considerable obstacles
due to atmospheric interference and an unstable observing environment. 
Observations from suborbital platforms to detect CMB B-modes, both from primordial tensor modes and from lensing, 
are a very active area of current research. Over the last few years nearly two dozen concepts for dedicated experiments 
have been described in talks and in the 
literature, with deployments that are ongoing or envisaged for the forthcoming decade. Some of these 
proposals involve multiple, staged deployments of progressively more advanced and sensitive experimental set-ups. 
Hence it is likely that the status of our knowledge will evolve significantly before the launch of PRISM. 

To date two suborbital experiments QUIET \citep[][]{quiet} and BICEP \citep[][]{bicep} have published analyses of 
their first data sets, setting up upper limits on the B-mode power and $r$ parameter based on the B-mode polarization, $r 
\simlt 0.7$ (95\% c.l.). A few more experiments are currently in operation or in the data analysis stage,
including 
BICEP/Keck~\citep{bicep/keck}, SPTpol~\citep{sptpol}, ABS~\citep{abs}, EBEX~\citep{ebex}, POLARBEAR \citep[][]{polarbear}, 
and a few more scheduled for the deployment within a year or so (e.g., QUIJOTE~\citep{quijote}, ACTpol~\citep{actpol} and 
SPIDER~\citep{spider}). These experiments should start producing results within the next few years, and at the end of that 
period are expected to have sufficient sensitivity to provide a convincing direct first detection of the lensing generated B-mode 
spectrum and start probing primordial B-modes with $r$ below $0.1$ and possibly down to $r \sim 0.05$ (95\% c.l.). 
A limit on $r$ at about the same level can be also expected from {\it Planck} in the coming year. 
Recently, the presence of lensing B-modes has been confirmed by the SPTpol team, with a cross-correlation 
analysis of CMB polarization with CIB measurements at 500 microns from \Herschel\ SPIRE \cite{2013PhRvL.111n1301H}.

On the timescale of a decade or so, more advanced versions of ground-based and balloon-borne experiments as well as some new ones, 
including technologically novel proposals such as the QUBIC
bolometric interferometer~\citep{qubic}, should reach the sensitivities to detect the gravity waves contribution down 
to sub $r \sim 0.01$ levels and potentially as low as $r \sim 0.007$ (e.g., POLARBEAR-EXT~\citep{polarbear}, 
PIPER~\citep{piper}, LSPE~\citep{lspe}), although this prediction is probably optimistic. The performance of 
these last experimental set-ups is typically considered near the limit of what can be achieved with
suborbital observations because of the considerable limitations due to atmospheric opacity and emission, far-side lobe pickup from the 
ground, and unstable observing conditions that make controlling systematic errors very difficult (particularly on  
large angular scales where the B-mode signal is largest). In addition, forecasts of $r$ from ground-based experiments are 
often impressive but assume very simple foreground contamination modelling. 
For this reason a tentative detection of $r$ from the ground would in fact provide a strong motivation for a 
confirmation and more precise characterization from space.

In the light of these forecasts, \mission\ would improve 
by almost two orders of magnitude what can be achieved with 
the most advanced of the above suborbital experiments owing to PRISM's 
full sky coverage, outstanding sensitivity, and immunity against systematics inherent to suborbital experiments.
PRISM would also carry out precise measurements of the B-mode spectrum extending to the largest angular scales, which are inaccessible 
to any single suborbital experiment. Moreover, systematic effects will
play a major role in determining the actual performance of suborbital observations. In 
contrast, a space mission such as \mission\ will allow exquisite control of systematic effects, thanks to the 
combination of benefits of the space environment at L2 and its advanced hardware design. The overall gain in performance from such a 
mission is therefore very likely to exceed the estimates based solely on the statistical considerations quoted earlier.

We note that even if $r$ happens to be large (i.e., $r \simgt 0.01$) so that first hints of a B-mode signal could  
potentially be seen by some of the most sensitive suborbital experiments (which is by no means guaranteed), its precise characterization and scientific exploitation is possible only with a satellite mission. If $r$ is small, a very sensitive space mission with precise  characterization of the instrument and control of systematics, and enough frequency bands to deal with complex foreground emission, is required to make a detection. \mission\ is designed to achieve this objective and give the final word on CMB primordial B-modes. In addition, it will provide the best possible full-sky measurement of lensing B-modes, essential to investigate the distribution of mass in the high redshift universe.

\subsection{Other space CMB projects}

Two US space missions concepts, {\it CMBPol} and {\it PIXIE,} and one Japanese, {\it LiteBird,} have been proposed in the 
past few years. In Europe, the \mission\ proposal follows the previously proposed SAMPAN 
satellite in France, and the {\it BPol} and COrE projects, proposed as M-class missions to ESA in 2007 and 2010 
respectively. None of these missions, however, has been funded or approved yet.
Among the current space mission concepts, \mission\ is the most ambitious and encompasses the broadest science case, B-mode polarization being only one of several main scientific objectives of the mission. 

{\it LiteBird} is a highly-targeted, low-cost Japanese B-mode mission concept, in many respects similar to the {\it BPol} 
mission proposed to ESA in 2007. It is designed 
to detect B-modes at the level of $r \sim10^{-3}$, a sensitivity that it should be able to achieve assuming that the foregrounds are not too complicated. However, {\it LiteBird}
lacks the angular resolution needed to make significant contributions to other key science objectives, and hence is not a competitor for \mission. If {\it LiteBird} is approved, it will probe the B-mode range down to one order of magnitude lower than what can be done from the ground. \mission\ will be about 10 times more sensitive yet, and more robust to contamination by complex foreground emission.

The US {\it EPIC-CS} mission is the most ambitious of the previously proposed mission concepts. 
{\it EPIC-CS} is in some respects similar to the present proposal, also targeting high resolution CMB science (in particular CMB lensing and some galaxy cluster science), 
but has considerably less frequency coverage, fewer frequency bands, and no absolute spectral capability. 

The {\it PIXIE} mission concept is an improved version of the {\it 
FIRAS} spectrometer, which aims at performing absolute spectroscopy and measuring large-scale CMB B-modes simultaneously. One of its major limitations is the effective resolution of only $2.6^\circ$. The absolute spectrophotometer of \mission\ is very similar conceptually to that of {\it PIXIE}, but it will benefit from slightly higher angular resolution and better sensitivity, and is not the instrument used to measure polarization with \mission. It will also, most importantly, benefit from the additional imager observations to clean-up the observed spectra from the contamination due to small scale localized emission (from SZ galaxy clusters, clouds of emission in the ISM, and strong IR and radio point sources).

A small satellite such as {\it LiteBIRD} of {\it PIXIE}, if funded and launched before \mission, could make a first detection of CMB B-modes---or not. \mission\ will then either provide an accurate measurement, or make the ultimate attempt at detecting primordial CMB polarization B-modes and definitively settle the question.

\subsection{Galaxy clusters observations}

By the time \mission\ flies, several large cluster surveys will have been completed, surpassing the first generation of SZ 
surveys from SPT, ACT and \Planck.  These new generation surveys, most notably in the X-ray and optical/near IR bands, will 
complement the \mission\ cluster catalog with invaluable additional information.
In addition, large-aperture ground-based 
sub-millimeter telescopes, such as CCAT,\footnote{\tt http://www.ccatobservatory.org/index.cfm/page/index.htm} will offer 
the exciting possibility of detailed follow-up of \mission\ clusters at high angular resolution, enabling important studies 
of the gas physics of clusters out to high redshifts.

{\it eROSITA} is an X-ray instrument onboard the SRG Russian satellite to be launched in 2014. The mission's principal goal is to explore cosmological 
models using galaxy clusters. Forecasts predict that {\it eROSITA}, 20--30 times more sensitive than ROSAT, 
will detect $\sim10^5$ clusters at more than 100 X-ray 
photon counts, which is sufficient to provide a good detection and in many cases to detect the source as extended in 
X-rays. The main survey provides a good sample of galaxy clusters typically out to $z=1$ with some very massive and 
exceptional clusters at larger distance.

The large majority of these clusters will be re-detected by \mission\ and thus provide an invaluable inter-calibration of 
X-ray and SZ effect cluster cosmology, provide determinations of cluster temperatures by combining the two detection 
techniques, and obtain independent cluster distances for many thousands of clusters whose X-ray temperatures and shape 
parameters can be obtained from the X-ray survey.  With $\sim10^6$ clusters detected with \mission, one can also further exploit 
the {\it eROSITA} survey data by stacking in a way similar to the analysis of the X-ray signals from the {\it ROSAT} 
All-Sky Survey for {\it SDSS} detected clusters \citep{rykoff2008}.

Ambitious imaging surveys in the optical and near-IR will also produce large cluster surveys.  These include the 
ground-based Dark Energy Survey (DES)\footnote{\tt https://www.darkenergysurvey.org/index.shtml}, now operating, and the 
future Large Synoptic Survey Telescope (LSST)\footnote{\tt http://www.lsst.org/lsst/} \citep{lsstbook}, a 8m-class 
telescope to be constructed in Chile and dedicated to a ten-year survey of the Southern Sky starting in 2020.  The LSST 
final survey depth will reach $\sim \! 27$ magnitude in six bands (ugrizY), surpassing the DES depth of $\sim \! 24$ magnitude 
(in grizY bands).  These broad-band imaging surveys will detect clusters out to redshifts of order unity through the 
characteristic colors of their member galaxies, which lie on the so-called ``red sequence" in a color-magnitude diagram.  
The lensing measurements undertaken in these surveys will also produce mass calibrations for cluster scaling relations.

In addition, the European Space Agency's \Euclid\ mission, scheduled for launch in 2020, will extend these cluster catalogs 
to higher redshifts thanks to its space-based IR observations\footnote{\tt http://www.euclid-ec.org} 
\citep{2011arXiv1110.3193L}.  \Euclid\ is estimated to detect between $(5-10)\times 10^4$ clusters out to redshifts well 
beyond unity.  Moreover, \Euclid's exquisite gravitational lensing survey will furnish a high-quality mass calibration for 
the cluster scaling laws.  Comparison of weak lensing and CMB lensing mass calibrations from \mission\ will allow for the 
most robust possible calibration for these scaling relations that are crucial to cluster science. This comparison
is possible in the overlap region at redshifts up to unity.  Beyond $z\approx 1$
only CMB-based lensing by \mission\ will be used to calibrate the scaling 
relations.

\subsection{Other sub-millimeter and far-infrared initiatives} 

 \mission\ will map the full-sky, large-scale continuum emission at higher sensitivities than ground based single-dish telescopes operating in the same frequency range.  Ground-based instruments can observe smaller regions with higher angular resolution than \mission\ and are hence complementary. Existing ({\it APEX, ASTE, IRAM 30m, LMT}) and future ({\it CCAT}) ground-based single-dish sub-millimeter observatories are of interest for ground-based follow-up, although they are not as sensitive above 300 GHz as \mission\ (mainly because of the limitations of observing through the atmosphere).  

{\it CCAT} will initially have two imaging instruments, {\it LWCam} and {\it SWCam}. At low frequencies, {\it LWCam} will be able to 
detect sources below the \mission\ confusion limit relatively quickly. However variations in atmospheric transmissivity and 
thermal radiation from the atmosphere will make it difficult for {\it CCAT} to map large scale structures. At high frequencies, 
{\it SWCam} will have difficulty mapping large areas to the confusion limit of \mission.  Based on the specifications from 
Stacey et al. (2013), {\it CCAT} can map an area of 1 square degree at 857 GHz to a sensitivity of 6 mJy (the \mission\ 
confusion limit) within 1 hour.  To map the entire southern sky to this same depth however requires $\sim900$ days (24h) with 
optimal observing conditions.  Such large scale observations will not be feasible with {\it CCAT.} Only a scanning space mission with a cold telescope such as \mission\ can produce high sensitivity maps over a large area of sky in these frequency bands. 

Interferometers ({\it ALMA, CARMA, PdB Interferometer, SMA}) are ill-suited to observing large fields and insensitive to large-scale structure, but can be used to produce superior maps of selected sky regions in combination with \mission\ data. 
{\it ALMA}, operating in the range 30-1000 GHz, will complement \mission\ with follow-up of sources and clusters to map their structure in total intensity, polarization and to observe spectral lines at high angular and spectral resolution.

Few previous infrared telescopes have performed all-sky surveys in the bands covered by \mission. {\it Akari} was the last 
telescope to perform such observations, but the data is at much lower sensitivity and resolution and is not yet publicly 
available. Several other prior telescopes {\it (Spitzer, Herschel)} as well as the airborne observatory {\it SOFIA} have 
observed or will observe in the 600-4000 GHz range, but only over very limited areas of the sky. Furthermore, except for a 
few deep fields, they observe objects already identified in other bands.  \mission\ will be able to perform observations 
with sensitivities comparable to {\it Herschel} or better, but covering the entire sky in many frequency bands.

\subsection{Other dark energy probes}

The question of Dark Energy will be investigated in 
depth before \mission. Several dark energy experiments as well as a space mission are planned for
the near future, among which DESI, LSST and 
\Euclid\ stand out as among the most competitive. In spite of this vigorous observational program, \mission\ is expected to still shed additional light on this most puzzling piece of our current cosmological model.

Building upon the expertise acquired with SDSS-III/BOSS, DESI is a ground-based dark-energy experiment that will 
study baryon acoustic oscillations and the growth of structure with a wide area (14000 square degrees) spectroscopic survey 
of approximately 20 million galaxies and 600 thousand quasars. The redshift of the targets (luminous red galaxies to $z\! =\! 1$, 
emission line galaxies with $0.7\! < \! z\! < \! 1.7,$ and quasars with $1\! <\! z\! <\! 5$) will be measured with the 4m Mayall telescope at Kitt Peak 
(Arizona). DESI is expected to last from 2018 to 2024 and aims to measure the cosmic distance scale to 1\% in 20 
redshift bins from 0 to 3.5. The galaxy part of the survey will be limited only by cosmic variance.

The Large Synoptic Survey Telescope (LSST) is a ground-based photometric survey 
that will obtain multiple images of the sky with an 8m telescope in Cerro Pachon, Northern Chile. During the 
anticipated 10 years of operation (approximately 2020-2030), LSST will uniformly observe a 18000 square degree region about 
1000 times (summed over the 6 optical bands). The time information will be used to detect and measure light curves of over 
$10^5$ SNIa, while the deep co-added images will provide photometric redshifts of several billion galaxies. These data 
will be used together to provide a combined constrain on dark energy and the growth of structure from weak lensing, baryon 
acoustic oscillations and SNIa.

Unlike the previous two projects, \Euclid\ is a space-based survey, resulting from the merging of a baryon acoustic 
oscillation redshift project (SPACE) and a weak lensing photometric project (DUNE) both proposed to ESA. The resulting 
survey will use these two techniques to constrain the dark universe using the measured redshifts of about 50 million 
galaxies over 15000 square degrees and the photometric information of about 20 billion sources. The launch date of the 
\Euclid\ satellite is currently expected to be around 2020.

A comparison of these projects was recently compiled by Font-Ribera et al. \citep{2013arXiv1308.4164F},
who compared
the expected constraints on the isotropic measurement of the cosmic distance scale from the spectroscopic DESI and \Euclid\ 
BAO projects in addition to constraints from other earlier future or ongoing surveys. Font-Ribera et al.~provide the 
following Dark Energy Task Force figures-of-merit (FoM) for these projects (with the 
normalization ${\rm FoM}=(\sigma_{w}\sigma_{w_a})^{-1}$) including priors 
from \Planck\ in each case: FoM=2 for \Planck\ alone, FoM from 23 to 104 for each of the individual dark energy probes of DESI, 
reaching 318 for an appropriate combination of DESI probes, FoM=27 for Euclid BAO, and FoM=134 for LSST or \Euclid\ weak 
lensing. Combining the two \Euclid\ primary probes yields a FoM of 430 \citep{2011arXiv1110.3193L}.

It is difficult to foresee the state of Dark Energy research when \mission\  will be launched.
However in terms of the usual FoM, the constraints from the galaxy clusters observed 
by \mission\ alone will outperform all the above observations (see Fig.~\ref{fig:cluster-constraints}), and also probe more complex models such as quintessence (Sec.~\ref{sec:clusters}).
Additional constraints will be obtained from \mission\ primary CMB including lensing (Fig.~\ref{fig:highell_params}). Moreover \mission\ will 
significantly improve constraints on other cosmological parameters (Sec.~\ref{sec:CMB-highres}), which can be used to better exploit
galaxy lensing and BAO oscillation data.

\subsection{\SPICA}

The proposed Japanese space mission \SPICA, originally scheduled for 
launch in the first half of the 2020s, addresses a few science objectives closely related to a part of the \mission\ science program. In particular, the SAFARI instrument will carry out imaging and spectroscopic observations of the sky in the 34--200 micron range (instead of 50--10000 microns for \mission) at the focus of a $\sim \! 3\,$m telescope cooled to about 5--6 K. The two mission concepts, however, differ substantially.

\begin{figure}[t]
\begin{center}
\includegraphics[width=10cm]{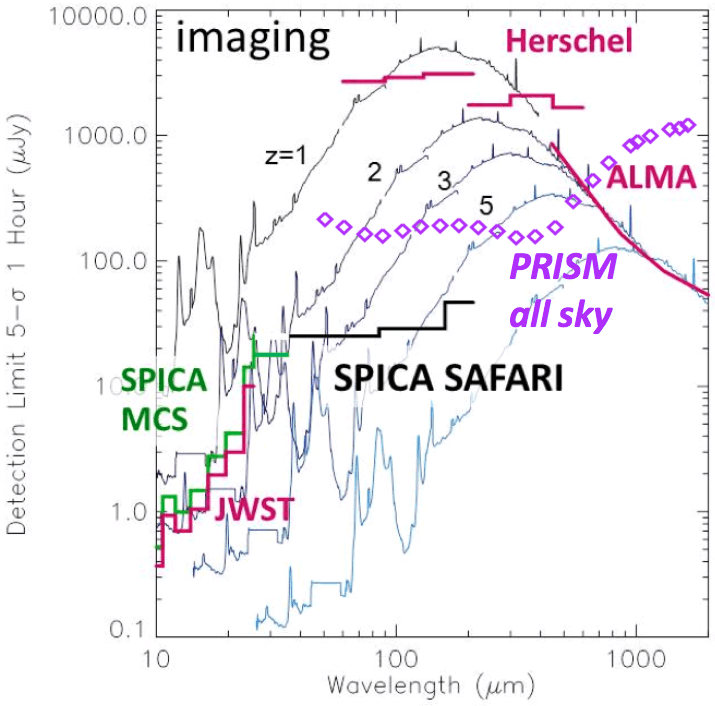}
\caption{{\small The full-sky point source sensitivity of the \mission\ polarimetric imager compared 
to other existing or planned experiments. Each diamond corresponds to the sky-average point source sensitivity 
(excluding confusion) in a single frequency channel of \mission. Note that \mission\ is in fact
confusion limited over the full frequency range, and thus the actual detection limit for both 
\mission\ and \SPICA\ will be similar, slightly higher than the \mission\ limit (Figure adapted from \citep{2012SPIE.8442E..0ON}). For illustrative purposes the SED of the starburst galaxy M82 as redshifted to the values indicated is shown in the background.}}
\label{fig:prism-spica}
\end{center}
\end{figure}

The biggest difference is the fact that while \SPICA\ will perform targeted observations towards a number of selected regions, the
scientific program of \mission\ requires observing the full sky in intensity and polarization. The broad-band sensitivity that can be achieved with a telescope cooled down to below $10\,$K is sufficient to guarantee that the \mission\ survey will be confusion limited in its full frequency range. Therefore, even if SAFARI can in principle achieve better imaging sensitivity by spending more time integrating on a small patch of sky 
(Fig.~\ref{fig:prism-spica}), its actual point source detection performance, limited by confusion, will be comparable to what will be achieved by \mission\ over the full sky. The maps from \mission\ will thus extend to the entire sky the high quality and confusion limited observations obtained with the imaging mode of SAFARI on {\it SPICA}. 

\mission, however, cannot match \SPICA 's spectroscopic capabilities for detecting atomic and molecular lines in distant 
galaxies. The spectral resolution of the PRISM narrow-band imager is of order $R\approx 40$, about 
50 times worse than the SAFARI spectroscopic mode. Hence, both missions will generate complementary 
observations of the distant universe, \mission\ being clearly superior for detecting a large number of objects, \SPICA\ being superior to observe selected objects in detail.

In our own galaxy, \mission\ will map the main atomic and molecular lines to get a complete picture of the ISM on large scales, while \SPICA\ will observe targeted regions in detail with similar angular resolution, but much higher spectral resolution.
A key feature of \mission, matched by no other existing or planned instrument including \SPICA, and of much relevance to galactic astrophysics, is the capability of  \mission\ to observe polarized emission at sub-millimeter wavelengths. Polarization data is essential to untangle the role of the magnetic field for star formation in our Galaxy. 

\subsection{Square Kilometer Array (SKA)}

The Square Kilometer Array (SKA) will span radio frequencies from 0.07 GHz to 10-20 GHz, close to the \mission\ frequency range, but with no overlap.  It will observe more than $10^{9}(f_{sky}/0.5)$ HI 
galaxies over the redshift range $0 < z < 1.5$ and provide maps of the reionization epoch above $z\sim 6$, completing the \mission\ view of structures in our Hubble volume.  The capabilities of SKA will also enhance the range of investigations possible with the \mission\ data set.

Definition of the high frequency reach of SKA Phase 1 is in progress, with a baseline design compatible with 
observations up to 14.5 GHz at the sub-system level, and dish specifications for good performance up to 20 GHz. Capabilities
for observing at the highest frequencies will be implemented around 2025 with the SKA Phase 2, and be fully 
operational at the foreseen \mission\ launch date in 2028 or 2034.

Plans include operations over several decades, which will make follow-up observations of sources identified by 
\mission\ possible, as well as detailed exploration of very faint emission on patches of the sky.  One of the most relevant perspectives offered by the surface brightness sensitivity achievable with SKA's compact core array is the detailed mapping out to 
the virial radius of the majority of the $\sim 10^{6}$ clusters detected by \mission.  Such follow-up observations of the good 
statistical sample provided by \mission\ down to low masses and up to high redshifts will enable the study of magnetic fields 
in clusters through Faraday rotation measurement of background sources and the study of cluster radio halos, as well as 
determinations of the dark matter distribution through lensing of background radio sources.  These capabilities complement and
enhance \mission's, whose higher frequencies covering the zero-crossing of the SZ effect make PRISM more efficient at finding clusters, 
and allow measurement of the kSZ and relativistic SZ effects as well as exploiting CMB lensing for cluster mass determinations.
 
SKA will also complement \mission\ by assessing the SZ effects expected from proto-galactic gas heated at the virial 
temperature and from quasar driven blast-waves, making possible accurate prediction of the corresponding background of 
fluctuations on very small scales (at multipoles of several thousands).

The extreme SKA sensitivity will probe radio source counts down to hundreds or even tens of nJy, thus providing additional 
information about their contribution to the total radio background.  This is relevant for the scientific exploitation of 
\mission\ data to measure the absolute CMB and Galactic emission at the lowest \mission\ frequencies. Furthermore, SKA will 
probe the free-free emission associated with cosmological reionization by looking at single sources and in a
statistical sense, including the global integrated signal, thereby complementing \mission\ spectrum measurements below 30 GHz.

SKA will contribute substantially to understanding galaxy formation and its evolution, 
AGN feedback, and star formation history. These observations complement 
the \mission\ studies of extragalactic sources and the CIB
with information over the radio spectrum.
Similarly, the exquisite mapping of the Galactic radio foreground with SKA in both 
temperature and polarization will complement \mission\ observations of the low-frequency galactic emission, with the 
characterization of the small scales not observed with the space-borne observatory. 



\bibliographystyle{plain}

\small
\bibliography{ref_planck_papers,bibliography,Bprim_refs,zodi,ref_point_sources,macias,bexpts,non_gaussianity,galaxy,Bib_add,clusters,high_res_plus_lensing}
\end{document}